\documentclass[12pt]{article}
\pdfoutput=1
\usepackage{amsmath}
\usepackage{multirow}
\usepackage{amsfonts}
\usepackage{amssymb,graphics,psfrag}
\usepackage{array,epsfig,multirow,graphicx}
\usepackage{comment}
\usepackage{feynmp}
\usepackage[all,cmtip]{xy}
\usepackage{color}
\usepackage{hyperref}
\usepackage{bbold}

\numberwithin{equation}{section}
\numberwithin{table}{section}

\def\hybrid{\topmargin -20pt    \oddsidemargin 0pt
        \headheight 0pt \headsep 0pt
        \textwidth 6.25in      
        \textheight 9 in      
        \marginparwidth .875in
        \parskip 5pt plus 1pt
          \jot = 1.5ex
  }
\hybrid

\newcommand{\beq}{\begin{equation}}
\newcommand{\eeq}{\end{equation}}

\newcommand{\bea}{\begin{eqnarray}}
\newcommand{\eea}{\end{eqnarray}}

\newcommand{\nn}{\nonumber}

\def\P{\mathbb{P}}

\newcommand{\Q}{\mathbb{Q}}

\newcommand{\Z}{\mathbb{Z}}

\newcommand{\suppress}[1]{}

\newcommand{\cref}{{\bf [check ref]}}

\def\blfootnote{\xdef\@thefnmark{}\@footnotetext}
\long\def\symbolfootnote[#1]#2{\begingroup%
\def\thefootnote{\fnsymbol{footnote}}\footnote[#1]{#2}\endgroup}

\begin{document}

\baselineskip=15pt

\begin{titlepage}
\begin{flushright}
\parbox[t]{1.73in}{\flushright UPR-1274-T\\
CERN-PH-TH-2015-157\\
MIT-CTP-4678}
\end{flushright}

\begin{center}

\vspace*{ 1.2cm}

{\Large General U(1)$\times$U(1) F-theory Compactifications and Beyond:\\[.4Em]
Geometry of unHiggsings and novel Matter Structure}

\vskip 0.9cm

\renewcommand{\thefootnote}{}
\begin{center}
 { Mirjam Cveti\v{c}$^{1,2}$,
Denis Klevers$^3$, Hernan Piragua$^1$, Washington Taylor$^4$\ \footnote{cvetic@cvetic.hep.upenn.edu, denis.klevers@cern.ch, hpiragua@sas.upenn.edu, wati@mit.edu} 
 }
\end{center}
\vspace{0.3cm}
{\it \small
$\,^1$ {Department of Physics and Astronomy,
University of Pennsylvania, 209 S. 33rd Street, Philadelphia, PA 19104-6396, USA} \\[.3cm]
$\,^2${Center for Applied Mathematics and Theoretical Physics,
University of Maribor, Krekova Ulica 2,
2000 Maribor, Slovenia}\\[0.3cm]
${}^{3}$Theory Group, Physics Department, CERN, CH-1211, Geneva 23, Switzerland\\[.3cm]
${}^4$Center for Theoretical Physics, Department of Physics,
Massachusetts Institute of Technology, 77 Massachusetts Avenue
Cambridge, MA 02139, USA
}\\[.7cm]

\end{center}

\begin{center} {\bf ABSTRACT } \end{center}

We construct the general form of an F-theory compactification with
two U(1) factors based on 
a general
elliptically fibered Calabi-Yau manifold with Mordell-Weil group of
rank two. This construction produces broad classes of models with
diverse matter spectra, including many that are not realized in earlier F-theory constructions
with U(1)$\times$U(1) gauge symmetry. Generic U(1)$\times$U(1) 
models can be related to a Higgsed non-Abelian model with
gauge group SU(2)$\times$SU(2)$\times$SU(3), SU(2)${}^3\times$SU(3), 
or a subgroup
thereof.  The nonlocal horizontal divisors of the Mordell-Weil
group are replaced with local vertical divisors associated with the
Cartan generators of non-Abelian gauge groups from Kodaira
singularities. We give a global  resolution of codimension two
singularities of the Abelian model; we identify the full anomaly free matter
content, and match it to the unHiggsed non-Abelian model. The non-Abelian 
Weierstrass model exhibits a new algebraic description of the singularities in the fibration
that results in the first explicit construction
of matter in the symmetric representation of SU(3). This matter is realized on
double point singularities of the discriminant locus. 
The construction suggests a generalization to U(1)$^k$ factors with $k >2$, which can be studied by Higgsing theories with larger non-Abelian gauge groups.

\hfill {July, 2015}
\end{titlepage}

\tableofcontents

\section{Introduction}

F-theory \cite{Vafa:1996xn, Morrison:1996na, Morrison:1996pp} provides
a powerful nonperturbative approach 
to understanding large classes of string vacua in four and six
space-time dimensions.  While non-Abelian gauge factors in F-theory
models are classified by the local Kodaira-Tate classification  of
singular fibers in elliptic fibrations, Abelian factors are
represented by elements of the Mordell-Weil group, which are
intrinsically global and more difficult to describe analytically.  In
recent years, progress has been made on constructing general classes
of Weierstrass models that describe F-theory compactifications with
one or more Abelian U(1) factors.  The general form of an F-theory
Weierstrass model with a single U(1) factor was described by
Morrison and Park in \cite{Morrison:2012ei}.\footnote{Certain F-theory 
compactifications with U(1) factors have appeared earlier in  \cite{Aldazabal:1996du,Klemm:1996hh} in the context of heterotic/F-theory duality, see also \cite{Klemm:2004km}.}  
A general class of models
with two U(1) factors was constructed in \cite{Borchmann:2013jwa,Cvetic:2013nia,Cvetic:2013uta,Borchmann:2013hta,Cvetic:2013jta}, and models with
three U(1) factors were studied in \cite{Cvetic:2013qsa}.

Explicit constructions of general F-theory models with multiple U(1)
factors are algebraically quite complex.  A great deal of physical
insight can be gained into systems with Abelian factors by considering
how the Abelian factors can arise from Higgsing of non-Abelian gauge
factors.  For example, a model with a single U(1) can be constructed
by starting with an SU(2) non-Abelian factor tuned on a divisor that
supports an adjoint representation, and then Higgsing the SU(2) by
giving a vacuum expectation value to the adjoint matter, giving a
breaking SU(2) $\rightarrow$ U(1).  A similar consideration allows us
to construct a fairly broad class of models with gauge group 
U(1)$\times$U(1) by Higgsing a rank two gauge group, such as 
SU(2)$\times$SU(2) or SU(3).  Consideration of the resulting spectra, however,
shows that the U(1)$\times$U(1) models constructed in this way
cannot
in general
 be described through the construction of
\cite{Borchmann:2013jwa,Cvetic:2013nia,Cvetic:2013uta,Borchmann:2013hta,Cvetic:2013jta}.\footnote{
On $\P^2$, for example,
only an SU(3) model with one adjoint and 54 fundamentals Higgses to such a U(1)$\times$U(1) model.}  
This observation motivates us to find a more general approach to
constructing the F-theory models with two U(1) factors.

The main result of this paper is the construction of a very general
Weierstrass form for F-theory models with U(1)$\times$U(1) gauge
factors.  This class of models includes all those that can be realized
by Higgsing the rank two groups SU(2)$\times$SU(2) and SU(3) on
adjoint matter.  The most general model also contains more complex
spectra, which can arise from Higgsing more complicated non-Abelian
structures.  In fact, one of
the principal results of this paper is that a large class of generic models
with two U(1) factors can be ``unHiggsed'' to models with the
non-Abelian gauge group $G_{\rm general} =$SU(2)$\times $
SU(2)$\times$SU(3). In various special cases, the model is unHiggsed to a
subgroup of this group, $G \subset G_{\rm general}$.  In this
unHiggsing process, the ``horizontal'' divisors associated with
sections in the Mordell-Weil group become vertical divisors associated
with Kodaira singularity types in the elliptic fibration over the
base.  This generalizes the result found in \cite{Morrison:2012ei, Morrison:2014era} that a single U(1) factor can generally be unHiggsed
to an SU(2) non-Abelian factor (or an increase in rank on an existing
non-Abelian factor), though in some cases the resulting non-Abelian
model may have certain types of singularities.  The form of the
non-Abelian group $G_{\rm general}$ can be understood geometrically by
identifying the general form of the vertical divisors associated with
unHiggsing the two U(1) factors to be $AC, BC$, where $C$ is a
common factor.  The unHiggsing process leads to SU(2) factors on $A$
and $B$, while on $C$ the two elements of the Cartan generators combine in a
nontrivial way to produce the group SU(3). In some cases the divisors $A$, $B$
and $C$ can or must be reducible, thus leading to larger non-Abelian groups such as 
$\text{SU}(2)^3\times$SU(3). 
This general algebraic framework suggests an approach to describing
Weierstrass models with more Abelian factors in terms of analogous,
but more complicated, algebraic structures.

A systematic understanding and classification of F-theory models with
multiple Abel-\\ian factors is an important challenge for F-theory, both
for theoretical and phenomenological reasons.  Much recent work has
focused on various aspects of this problem, largely motivated by
efforts to construct models with Abelian factors related to F-theory
GUT phenomenology  (for a representative list of works, see \cite{Donagi:2008ca,Beasley:2008dc,Donagi:2008kj,Marsano:2009ym,Blumenhagen:2009yv,Grimm:2009yu,Marsano:2009wr,Grimm:2010ez,Dudas:2010zb,Dolan:2011iu,Krause:2011xj,Krause:2012yh,Mayrhofer:2012zy,Braun:2013nqa,Cvetic:2013uta,Borchmann:2013hta,Krippendorf:2014xba,Braun:2014qka}) as well as other particle physics models with Abelian gauge symmetry \cite{Lin:2014qga,Klevers:2014bqa,Cvetic:2015txa}.
On the more theoretical side, a major challenge in constructing a 
completely general F-theory model with Abelian factors is the wide
range of possible spectra that may arise in such a theory.  While for 6D models
anomaly constraints in the low-energy supergravity theory provide some
limits on the set of possibilities 
\cite{Green:1984bx,Sagnotti:1992qw,Erler:1993zy, Park:2011wv}, 
as one considers matter with increasingly large charges under the
U(1) gauge factors, the complexity of the corresponding F-theory
models grows accordingly.  Through the unHiggsing process, these
matter fields with higher charges are related to matter fields with
increasingly complicated transformation properties under the
corresponding non-Abelian gauge factors.

One of the important new results of this paper that signifies the complexity of the representations with multiple U(1) factors  is a construction of a generic class of  U(1)$\times$U(1) models  in which the resolution of the Abelian theory gives matter
fields with specific higher charges that are related to matter
transforming under the symmetric representation of SU(3) in the
associated unHiggsed theory.  As a by-product this is the first explicit realization of the the symmetric matter  representation of the non-Abelian gauge symmetry in F-theory. The associated Weierstrass models have
an intricate and nontrivial algebraic structure that realizes the
non-Abelian gauge theory in a novel way that cannot be understood
directly from the Tate description.
 
While our construction and the geometric results are applicable for Calabi-Yau $n$-folds, in this paper we primarily focus on Calabi-Yau threefolds relevant for 6D F-theory.
However, most of the geometric analyses are also
relevant to 4D F-theory.

The outline of this paper is as follows: In Section
\ref{sec:U1sandUnHiggsing}, we discuss the ways in which models with one or
two U(1) factors can be realized by Higgsing non-Abelian gauge groups
with a variety of structures.  This field-theoretic analysis and
discussion provides a framework for interpreting the rather
complicated algebraic structures that arise in the following sections.
In Section \ref{sec:abcd}, we construct the most general elliptic
fibration with three independent sections, corresponding to a
Weierstrass model with Mordell-Weil group of rank two, and an F-theory model
having two Abelian gauge factors U(1) $\times$ U(1) in the
corresponding low-energy supergravity theory.  In Section
\ref{sec:SingsResMatter}, we give a global resolution of the
singularities in the general U(1) $\times$ U(1) Weierstrass model, and
analyze the resulting matter spectrum.  In Section
\ref{sec:UnhiggsingGeometry}, we describe in general terms the
unHiggsing of the two U(1) model to a non-Abelian theory with
vanishing rank of Mordell-Weil group.  Concrete examples of the unHiggsing
process and the detailed form of the Weierstrass model in different
cases are explored in Section \ref{sec:Examples}.  In Section
\ref{sec:further}, we discuss a natural generalization of our work to more than two U(1) factors  
and  the construction of F-theory models with exotic matter representations  as well as the 
problem  of proving the global equivalence of the space of consistent 6D supergravity theories
and the set of F-theory compactifications. We summarize key results of the paper
in our conclusions in Section \ref{conclusions}.

{\bf Note added:} Our general U(1)$\times$U(1) F-theory construction, 
presented in this paper, is already
employed in the simultaneous work \cite{Krippendorf:2015} for the study of flavor textures in SU(5) GUT's with two U(1)'s. 

\section{U(1)'s and unHiggsing} 
\label{sec:U1sandUnHiggsing}

In this section we describe various ways that two U(1)'s can arise
from the Higgsing of a non-Abelian theory.  This provides a field
theory framework for understanding the wide range of models that can
be produced using the general U(1)$\times$U(1) Weierstrass model
constructed in the following section.  We begin with a review of the
story for a single U(1), and then consider unHiggsing of two U(1)
factors.

The general classes of models considered in this section can be
constructed in the context of F-theory in six or four space-time
dimensions.  For specific examples, we focus here on 6D constructions,
since anomaly cancellation conditions for 6D supergravity theories
provide strong constraints on the allowed spectra of the low-energy
theories \cite{green2012superstring, Sagnotti:1992qw,Kumar:2010ru},
providing in many cases a simple check on the consistency of the
constructions we describe here.  The same classes of theories can also
be realized in 4D F-theory constructions, however, with a richer range
of specific models and applications.

\subsection{Higgsing and a single U(1)}
\label{sec:HiggsingU1}

\subsubsection{Higgsing an SU(2)}
\label{sec:Higgsing-2}

In general, an SU(2) gauge factor  with matter in the adjoint
representation can be constructed in F-theory by tuning a Kodaira type
$I_2$ singularity on a divisor $D$ of the form $-K_B + X$, where $K_B$ is the
canonical class of the F-theory base manifold $B_{n}$, and $X$ is
effective; for such $D$, we write $D \geq - K_B$. 
For a smooth divisor $D$, this condition is necessary and sufficient
for the existence of an adjoint matter representation at the level of
geometry.  For 6D theories, this can be seen directly from the fact
that the genus $g$ of the divisor $D$
is in general given by
\beq \label{eq:genusFormula}
	g=1+\frac{1}{2} D\cdot(D+K_B)\,,
\eeq
which is positive precisely for $D\geq -K_B$.
  As a simple example, if the base is 
 $B_2 =\P^2$, then $- K_B=  3H_B$ where $H_B$ is the hyperplane class in $\mathbb{P}^2$. 
If we tune an SU(2) over a smooth curve of degree $d$
and genus $g =(d -1) (d -2)/2$, in the resulting
low-energy theory there are $g$ matter fields that transform in the
adjoint of SU(2).  
For example, a cubic curve $C$ has genus 1, a quartic has genus 3, etc.~.
From anomaly cancellation or inspection of the explicit Weierstrass
model, {\it cf.}~Appendix \ref{app:SU2SU2WSF}, the number of
fundamental matter fields is $x_2 =6 d^2 +16 (1 - g)$. For the cubic
curve $C$,
this gives $x_2 =54$.

An SU(2) with adjoint matter can be Higgsed by giving a VEV to an
adjoint matter field proportional to the Cartan generator
$\sigma_3$, the third Pauli matrix.  This leaves a residual U(1) gauge symmetry.
  Each matter field that transforms in the fundamental of the original
  SU(2) gives rise to a pair of matter fields with charges $\pm 1$
  under the resulting U(1).  For example, if we start with a cubic
  on $\P^2$, after Higgsing we get a U(1) theory with 108 charged
  matter fields --- the minimum number of charged matter fields
  compatible with anomaly cancellation in a 6D supergravity theory
  with a single U(1) factor \cite{Park:2011wv}.  More generally, if
  we Higgs the SU(2) using one adjoint (whose charged components are eaten up) and there are also
$g-1$ other adjoint fields
  available, each of the $g -1$ 
other adjoints produces a multiplet of charges $(-2, 0, +2)$
  under the resulting U(1).  For example, tuning an SU(2) on a
  quartic on $\P^2$ and then Higgsing gives 128 matter fields with
  charges $\pm 1$ and 4 matter fields with charges $\pm 2$.

\subsubsection{UnHiggsing a U(1)} 

An F-theory compactification on a base manifold $B$ is defined
through a Weierstrass model
$y^2 = x^3 + f\, xz^4 + g\, z^6$,
where $f$ and $g$ are sections $f\in\Gamma({\cal O}(-4K_B))$, $g\in
\Gamma({\cal O}(-6K_B))$.  Such an elliptic fibration has a global section
$z=0$.
We often work in the
coordinate patch
of $\P (2, 3, 1)$ where $z=1$, giving the common
form
\begin{equation}
y^2 = x^3 + f\, x + g\,.
\label{eq:WSF}
\end{equation}

It was shown in \cite{Morrison:2012ei}, using the
elliptic curve in $\text{Bl}_1\mathbb{P}^2(1,1,2)$, that the general form of a
Weierstrass model with a U(1) factor takes the form
\begin{equation}
 y^2 = x^3+ (e_1e_3- \frac{1}{3}e_2^2 -b^2e_0) x
+ (-e_0e_3^2 +\frac{1}{3}e_1e_2e_3 - \frac{2}{27}e_2^3 + \frac{2}{3}b^2e_0e_2
 -\frac{1}{4}b^2e_1^2) \,.
\label{eq:Abelian-1}
\end{equation}
Here, $b$ is a section of a line bundle ${\cal O} (L)$, where $L$ is
effective, and $e_i$ are sections of line bundles $\mathcal{O}((i - 4)
K_B+ (i -2) L)$.  The Weierstrass model \eqref{eq:Abelian-1} has a
nontrivial rational section, so that the {\it Mordell-Weil} group of
rational sections has rank (at least) 1.  As described in
\cite{Morrison:2012ei,Morrison:2014era}, when the parameter $b$ is
taken to vanish, the divisor associated to this rational section is
transformed to a vertical divisor.  In general, assuming that in the
original U(1) model there is no Kodaira singularity on the divisor\footnote{As a short-hand notation, we denote here and in the following {\it e.g.}~the divisor $e_3=0$ simply by $e_3$.}
$e_3$, this leads to a
non-Abelian SU(2) factor on $e_3$, giving an unHiggsing
corresponding to the reverse of the process described above.  Note
that\footnote{We will denote the divisor
class of a section on $B_{n}$ with brackets, {\it e.g.}~the class of $e_3$ by $[e_3]$.} $[e_3] = - K_B + L$  always has a form that allows for an adjoint
representation of the resulting SU(2).

In situations where the original model (\ref{eq:Abelian-1}) already
has some non-Abelian gauge group components, the story can be slightly
more complicated.  In some cases, tuning the SU(2) can lead to $(4,
6)$ singularities at codimension one or two.  If the divisor $[e_3]$
already itself supports a non-Abelian gauge factor, then tuning $b
\rightarrow 0$ leads to an enhancement of the original gauge factor
with an increase in rank, as the horizontal divisor of the rational section is transformed
into vertical form.
The form of (\ref{eq:Abelian-1}) shows that any U(1) can thus be
``unHiggsed'' corresponding to an enhancement of the non-Abelian sector
of the theory, albeit in some cases with resulting singularities.

Another relevant situation can occur when $e_3 = \alpha \beta$, {\it
  i.e.}~$e_3$  
is reducible.
In this case, performing the unHiggsing by taking $b \rightarrow 0$
leads to an SU(2) factor on each component of $e_3$.  For example,
consider the case where the base is $B = \P^2$ and
$[e_3]= 3H, [\alpha] = H,[\beta] = 2H$.  In this case, the unHiggsing
gives a theory with SU(2) factors on a pair of intersecting curves
$\alpha, \beta$
of degrees one and two.  In this case there is no adjoint matter field
to Higgs.  The spectrum of the resulting theory in this example
consists of 22 fundamentals on $\alpha$ and 40 fundamentals on
$\beta$, including two bifundamental hypermultiplets $({\bf 2},\mathbf{2})$ associated with
the two intersection points, {\it cf.}~the general multiplicity formulae \eqref{eq:Multies_SU2SU2}.  The bifundamental fields can be Higgsed
by turning on VEV's of the form $\text{diag}(v_1,v_2)$ for two independent
VEV's $v_1$, $v_2$.  Note that  \emph{two fields} 
must be used for this Higgsing to satisfy the D-flatness conditions
when fields in the fundamental representation are used, unlike in the
case of Higgsing on an adjoint field where only one field is needed
since the D-term constraints are automatically satisfied.
Each bifundamental contributes two fundamentals to the spectrum on
each curve, so after the Higgsing there are $18+36 = 54$ SU(2)
fundamentals that are broken to pairs of $\pm 1$ charges under the
resulting U(1), reproducing the expected number 108  of charged
hypers.  This kind of model can be understood as simply a degeneration of a
single divisor $e_3$ to a combination of divisors $\alpha \beta$.  The
Higgsing can be thought of as occurring in two stages: first a
Higgsing by a VEV proportional to {\it e.g.}~$\mathbb{1}$ that re-combines the two
components, and then a Higgsing by the resulting adjoint.  After the
complete Higgsing, the remaining U(1) can be seen as the  difference
$\sigma^{(1)}_3 - \sigma^{(2)}_3$ of Cartan generators of the original
SU(2) factors.

Note that in the case where $e_3$ is reducible, if either factor
$\alpha, \beta$ can by itself support an adjoint, {\it i.e.}, if
$\alpha \geq - K_B$ or $\beta \geq - K_B$, then the U(1) model can be seen
as coming from an SU(2) on the divisor $\alpha \beta$ in several
different ways.  First, in the fashion just described where both
SU(2)'s are Higgsed simultaneously using a pair of bifundamental
matter fields, and second by first Higgsing an SU(2) adjoint to get
U(1)$\times$SU(2) and then Higgsing a bifundamental with both U(1)
and SU(2) charges.  For example, if $\alpha \geq - K_B$, then the
SU(2) on $\alpha$ already contains an adjoint, and Higgsing that
adjoint will produce a theory with gauge group U(1)$\times$SU(2)
with the SU(2) supported on $\beta$.  This represents a sort of
intermediate stage of the Higgsing process, where the U(1) in the
fully Higgsed model is related to a combination of the U(1) field
associated with $\alpha$ and the Cartan of the SU(2) on $\beta$.
This can be seen algebraically in the F-theory context through the Abelian model
(\ref{eq:Abelian-1}). Here, the residual SU(2) on $\beta$ is introduced
by taking $e_3 = \alpha\beta$ and
having a factor of $\beta$ in $b$ as $b=b'\beta$.  Then, the rational section lives only on $\alpha$ as its coordinates in 
$\text{Bl}_1\mathbb{P}^2(1,1,2)$ are $[- b'\beta:1:\alpha\beta:0]=[-b':1:\alpha:0]$ employing the $\mathbb{C}^*$-action.   
Note also  that in the fully unHiggsed model with $b' = 0$ the
extra factor of $\beta$ can then be moved out of $e_3$ by redefining
$\tilde{e}_3 = e_3/\beta, \tilde{e}_1 = \beta e_1, \tilde{e}_0 =
\beta^2 e_0$. 

We encounter a number of related situations in the later
sections of the paper, where various unHiggsed models can be partially
Higgsed in various sequences to go between a non-Abelian model and an
Abelian model. For the most part, the focus however is on the
completely Higgsed and completely unHiggsed endpoints of these
processes.

\subsection{Higgsing and two U(1)'s: simple constructions}

Now let us consider some simple ways in which a pair of U(1)'s can
be produced from the Higgsing of a   non-Abelian model.
The simplest way in which this can be done is to start with a theory
with a rank two non-Abelian gauge group SU(2)$\times$SU(2) or
SU(3), where each factor has an adjoint representation that can be
Higgsed.  One might also consider a theory with  group $G_2$, but a
Higgsing of $G_2$ on the adjoint can give SU(3), so the resulting
U(1)$\times$U(1) models and spectra are identical to those that
could come from an original SU(3) model.

\subsubsection{Higgsing SU(2)$\times$SU(2)}
\label{sec:HiggsingSU2SU2}

\begin{figure}[ht]
\centering
 \includegraphics[scale=1]{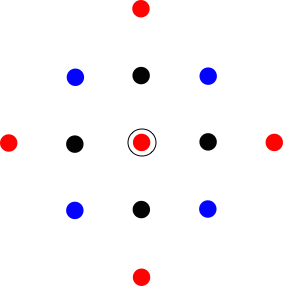}
\caption{Weight lattice of charges in a theory with a
U$(1)\times$U(1)
  gauge group realized by Higgsing a theory with gauge group
SU(2)$\times$SU(2). Dots in red indicate the adjoint, dots in blue the bifundamental 
and black dots are the fundamentals. The origin is indicated by an extra circle.
}
\label{f:2-2-weights}
\end{figure}

If we have two SU(2) factors that are tuned on two independent smooth
divisors $A, B \geq - K_B$, then each of the factors carries an adjoint
representation, and each can be independently Higgsed  just as in the  case of a single
SU(2).  The resulting model has a gauge group U(1)$\times$U(1).
In general, the original model will have  some number of fundamentals
transforming under each of the SU(2) factors, and some number of
bifundamental fields that transform as a fundamental under each of the
factors, in addition to a number of adjoint representations for each
group corresponding to the genera of the two curves.
After the two factors are Higgsed, the fundamentals will become scalar
fields that
carry charges
$(\pm 1, 0)$ and $(0, \pm 1)$ under the two U(1) factors.  The
bifundamental fields will carry charges $(\pm 1, \pm 1)$, and the
extra adjoint fields will carry nonzero  charges $(\pm 2, 0), (0, \pm
2)$.  This gives a characteristic set of spectra that can arise from
this Higgsing structure, depicted in
Figure~\ref{f:2-2-weights}.

As an example, consider a 6D model 
on $\P^2$.  If we tune an SU(2) on each of two divisors
$A, B$ of degrees $a, b \geq 3$, then we will have $g_a =(a -1) (a
-2)/2$ adjoints and $6a^2 +16 (1 - g_a)$ fundamental matter
representations charged under the first SU(2), and similarly for the
second SU(2) with $a \rightarrow b$, see \eqref{eq:Multies_SU2SU2}.  The number of bifundamental
hypermultiplets will be $a b$.  When both SU(2) groups are broken by
Higgsing an adjoint representation, the remaining adjoints become
charge 2 fields, and fundamentals become charge 1 fields under one of
the two U(1) factors (and neutral under the other U(1))
in
the
resulting U(1)$\times$U(1) gauge theory.  In the simplest case, $a =
b = 3$, there are 9 bifundamental fields that carry charges $(\pm 1,
\pm 1)$ after Higgsing, and another 36 fundamental fields for each
factor that carry charges $(\pm 1, 0)$ and $(0, \pm 1)$ each. For each
U(1) the total number of matter fields of charge $\pm 1$ is 108,
satisfying the anomaly cancellation conditions.
Note that the counting of matter multiplets here is in terms of full
6D hypermultiplets, each of which contains two ``half
hypermultiplets'' with opposite charges.

Note that this distribution of charges cannot be realized using the
class of U(1)$\times$U(1) models constructed in \cite{Borchmann:2013jwa,Cvetic:2013nia,Cvetic:2013uta,Borchmann:2013hta,Cvetic:2013jta}.  Thus, a
more general model is needed.

\subsubsection{Higgsing SU(3)}
\label{sec:Higgsing-3}

\begin{figure}[ht]
\centering
 \includegraphics[scale=1]{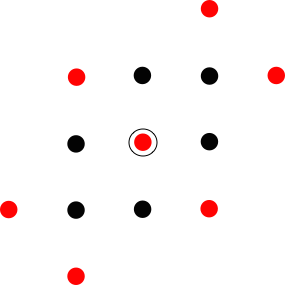}
\caption{Weight lattice of charges in a theory with a
U$(1)\times$U(1)
  gauge group realized by Higgsing a theory with gauge group
SU(3). Dots in red indicate the adjoint
and black dots are the fundamental
and antifundamental. The origin is indicated by an extra circle.
}
\label{f:3-weights}
\end{figure}

In a similar vein, we can tune an SU(3) gauge group on a single
smooth divisor $A\geq - K_B$.  The resulting SU(3) theory will have at
least one adjoint matter representation, and some number of
fundamental matter representations.  Note that in the
six-dimensional theory, with vector-like matter, the ``fundamental''
matter hypermultiplets contain both the fundamental and conjugate
anti-fundamental representations.  On $\P^2$, for example, we again
have $g = (a -1) (a -2)/2$ adjoint representations when the SU(3) is
tuned on a degree $a$ curve, and the corresponding number of
fundamental matter fields is $x_3 = 6a^2+18(1 -g)$,  {\it cf.}~the general multiplicity
formula \eqref{eq:Multies_SU3}.
The  SU(3) can be broken by Higgsing along two Cartan generators,
{\it e.g.}
$\lambda = {\rm diag}(1, -1, 0)$,
$\mu = {\rm diag}(1, 0, -1)$.  The fundamental representation of
SU(3) carries charges $(1, 1), (-1, 0),$ and $(0, -1)$ under the
resulting U(1)$\times$U(1), and the anti-fundamental representation
in the same multiplet carries charges $(-1, -1), (1, 0),$ and $(0,
1)$.  Additional adjoint representations of SU(3) acquire charges of
$\pm (2, 1), \pm (1, 2),$ and $\pm (1, -1)$ (and two singlets) under
the breaking to U(1)$\times$U(1).  This spectrum is shown
graphically in Figure~\ref{f:3-weights}.

Again, this type of spectrum cannot be produced by the U(1)$\times$
U(1) models in \cite{Borchmann:2013jwa,Cvetic:2013nia,Cvetic:2013uta,Borchmann:2013hta,Cvetic:2013jta}.

As a simple example, consider tuning an SU(3) on a cubic on $\P^2$.
The matter content consists of one adjoint and 54 fundamental matter
fields.  Breaking the SU(3) to U(1)$\times$U(1) by Higgsing the
adjoint gives 54 matter fields in each of the representations $(1,
1)$, $(1, 0)$, $(0, 1)$, which appear along with their complex conjugates
$(-1, -1)$, $(-1, 0)$, $(0, -1)$  in each full hypermultiplet.  Note that again we
have 108 charged matter fields
under each U(1) factor, so anomaly cancellation is satisfied.  There
is, however, with this spectrum no way to redefine the basis of U(1)
fields to match the spectrum found in the class of theories realized
by Higgsing an SU(2)$\times$SU(2) model.

It is worth mentioning also here the spectrum that results if an
SU(3) is partially Higgsed on a single Cartan generator to give a
theory with U(1)$\times$SU(2) gauge group.  In this case, Higgsing
for example on $\lambda_8 = {\rm diag}(1, 1, -2)$, the fundamental
representation of SU(3) breaks to the representation content $(+1,
{\bf 2}) + (-2, {\bf 1})$ of U(1)$\times$SU(2), with the
anti-fundamental in the same hypermultiplet carrying the conjugate to
this representation content, and the adjoint breaks to $(+3, {\bf 2})
+ (0, {\bf 3}), (-3, {\bf 2})$.  Note that all fields are invariant
under the diagonal $\Z_2$, so that the precise gauge group in this
case is $(\text{U}(1) \times \text{SU}(2))/\Z_2$.  The presence of this discrete
$\Z_2$ quotient and the difference in spectrum distinguishes this
representation content from that associated with the Higgsing of a
single SU(2) in an SU(2)$\times$SU(2) model such as discussed in
the previous subsection.  Breaking down to U(1)$\times$U(1), it is
straightforward to check that the representation content reproduces
that described above, under an appropriate linear recombination of the
generators.

\subsubsection{Hybrid models from SU(2)$\times$SU(2)$\times$SU(3)}
\label{sec:hybrid}

While the preceding models are the simplest ways of getting  two U(1)
factors from the Higgsing of a non-Abelian theory, by considering the
possibility of reducible divisors supporting the SU(2) factors as in
the single U(1) theories described above, some more interesting
structures can emerge.  
While for a single U(1) factor, the possible
spectra are essentially the same whether the U(1) comes from
Higgsing an SU(2) on an irreducible divisor $A$ or a reducible
divisor $X = A B$,  
when there are two U(1) factors, we can consider
a situation where  one U(1) comes from Higgsing a non-Abelian factor
on $X = A C$, and the other U(1) comes from Higgsing a non-Abelian
factor on $Y = B C$.  
Compared to models with a single U(1) factor, 
much more interesting and
subtle structure is possible when the reduction occurs in this way for
the U(1)$\times$U(1) model. 
In particular, in this situation the divisor $C$ generally will support
a gauge factor SU(3), and we
can get a variety of different charges in the U(1) $\times$ U(1)
theory by starting with different matter content in the unHiggsed theory.

In fact, the general Abelian U$(1)^2$ theory considered in this work
is precisely of this type. It arises from
Higgsing a non-Abelian theory with gauge group 
\begin{equation}
 G = \text{SU(2)} \times \text{SU(2)} \times \text{SU(3)}\,,
\label{eq:GunHiggsed}
\end{equation}
where the three factors are tuned on divisors $A, B, C$.  Most
notably, the gauge group on $C$ is only an SU(3) associated to an
$I_3$ singularity, in contrast to the common expectation of an $I_4$
singularity at the collision of two $I_2$ singularities. In general,
each gauge group factor will have some associated number of
fundamental representations and some number of adjoint
representations, and there can be bifundamental representations
between each pair of gauge group factors.  The multiplicities of these
matter fields are given by the general formulae
\eqref{eq:Multies_SU2SU2SU3}.   In this general context,
the bifundamental fields are primarily used to
Higgs the product group SU(2)$\times$SU(2)$\times$SU(3) (which
requires at least two bifundamentals for D-flatness), the spectrum of
the resulting U(1)$\times$U(1) model will be somewhat different from
either of the rank two Higgsings described above.

As in the case of unHiggsing a single U(1) factor, any of the divisors
$A, B, C$ can be reducible, leading to a larger gauge group after
unHiggsing.  We find one broad class of models where $A$ is
generically reducible, so that the complete gauge group after
unHiggsing is SU(2) $\times$ SU(2) $\times$ SU(2) $\times$ SU(3), and
the Higgsing is again carried out through bifundamental fields.

\begin{figure}[ht]
\centering
 \includegraphics[scale=1]{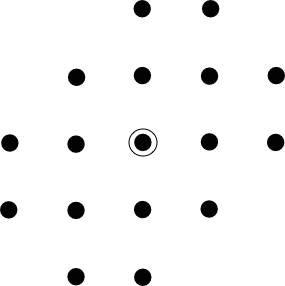}
\caption{Weight lattice of charges in a theory with a
U$(1)\times$U(1)
  gauge group realized by Higgsing a theory with gauge group
SU(2)$\times$SU(2)$\times$SU(3). The origin is indicated by an extra circle.
}
\label{f:223-weights}
\end{figure}

To see how the matter spectrum of the theory follows from Higgsing
the general SU(2) $\times$ SU(2) $\times$ SU(3) model
on bifundamental fields,
consider first Higgsing on a bifundamental $({\bf 1},\mathbf{2},\mathbf{3})$ between the
SU(2)$\times$SU(3) on $B \cap C$, via a vacuum expectation value
proportional to
\begin{equation}
\begin{pmatrix} 
1 & 0 & 0 \\ 
0 & 0 & 0
\end{pmatrix}
\label{eq:VEVbifund}
\end{equation}
This leaves a gauge group SU(2)$\times$ U(1) $\times$SU(2).
We can then Higgs a bifundamental $({\bf 2},\mathbf{1},\mathbf{2})$ on $A \cap C$, leaving a gauge group
U(1)$\times$U(1).  
Carrying this out explicitly, we can write the residual U(1)  
generators $\xi$ and $\zeta$  in terms of the Cartan generators of the factors SU(2),
SU(2), SU(3) as
\begin{equation} \label{eq:embeddingU1s}
\xi  =   \mu-\lambda- \sigma^{(1)}_3\,,\qquad \zeta =  \sigma^{(2)}_3 - \lambda\,.
\end{equation}
A short calculation then shows that the spectra of the different
bifundamentals decompose under U(1)$\times$U(1) as
\begin{eqnarray} \label{eq:branching1}
({\bf 2},\mathbf{2},\mathbf{1}) & \rightarrow &(\pm 1, \pm 1)\,, \\ \nn
({\bf 1},\mathbf{2},\mathbf{3}) & \rightarrow &  (-1, 1) + (-1, -1) + (0, -2) + (1, 0)  + (1,
2) +(0,0)\,,\\ \nn
({\bf 2},\mathbf{1},\mathbf{3}) & \rightarrow &  (-1, -1) + (1,- 1) + (-2, 0) + (0, 1) + (2, 1) +(0,0)\,.
\end{eqnarray}
Similarly, we obtain the following decomposition under U(1) $\times$ U(1) of the fundamentals
\bea \label{eq:branching2}
	(\mathbf{2},\mathbf{1},\mathbf{1})& \rightarrow &(\pm 1,0)\,, \\ \nn
	(\mathbf{1},\mathbf{2},\mathbf{1})& \rightarrow &(0,\pm 1)\,, \\ \nn
		(\mathbf{1},\mathbf{1},\mathbf{3})& \rightarrow &(0,-1)+(1,1)+(1,0)\,.
\eea
This spectrum is shown in Figure~\ref{f:223-weights}, where we emphasize that 
a hypermultiplet in complex representation always contains states of a given charge and their conjugates. If there are
adjoints on any of the divisors $A, B, C$ they will decompose as
\begin{eqnarray} \label{eq:branching3}
({\bf 3},\mathbf{1},\mathbf{1}) & \rightarrow & ( \pm 2,0)+ (0,0)\\ \nn
(\mathbf{1},{\bf 3},\mathbf{1}) & \rightarrow & (0, \pm 2)+ (0,0)\\ \nn
(\mathbf{1},\mathbf{1},{\bf 8}) & \rightarrow & \pm (2, 1) + \pm (1, 2) + \pm (1, -1) +  2(0,0)
\end{eqnarray}
Thus, the Higgsing on bifundamentals of a theory with SU(2)$\times$SU(2)$\times$SU(3) gauge group gives a different, and more general,
spectrum than the rank two Higgsings on SU(2)$\times$SU(2) or
SU(3) above.

\subsubsection{Singular divisors and higher charges}
\label{sec:Higgsing-singular}

In the general U(1)$\times$U(1) Weierstrass model that we derive in
the following section, it turns out that there are even more exotic
classes of spectra, corresponding to Higgsing of models where a gauge
group has been tuned on a singular divisor, associated with a more
complicated matter representation. 

While generic F-theory models in which gauge groups are tuned on
smooth divisors give only simple matter representations for the
low-energy gauge group, such as fundamental and antisymmetric
representations for SU($N$) gauge groups, more exotic representations
can be realized in F-theory when the coefficients in the Weierstrass
model are tuned to realize more complicated codimension two loci in
the elliptic fibration.  While there is not yet a complete map between
codimension two singularities structures and the representation theory
of matter and the corresponding low-energy supergravity theory,
anomaly cancellation in 6D theories gives some insight into this
correspondence, which has partially been explored in \cite{Kumar:2010am,Morrison:2011mb}.  
One of the simplest examples of this kind of
situation arises for representations of SU($N$).  As shown in
\cite{Kumar:2010am}, every  irreducible representation
$R$ of a  simple gauge group
factor $G$ has associated with it a genus contribution $g_R$.  The
value of $g_R$ should
indicate the contribution to the arithmetic genus of a singular curve
in the base surface of an F-theory compactification that carries that
gauge group, where the matter representation $R$
is located at the singularity in question.  All purely antisymmetric
tensor representations of 
SU($N$) ({\it i.e.}, those whose Young diagram has a single column)
have a genus contribution of $g = 0$, and can be realized by
codimension two singularities of the Weierstrass model over a smooth
curve.  The adjoint  and symmetric tensor representations both carry
genus contributions of $g = 1$ for all SU($N$).  The adjoint
representation is the matter representation carried generically by a
smooth divisor of genus $g > 0$.
As described in
\cite{Sadov:1996zm,Morrison:2011mb}, the symmetric tensor representation
arises when the gauge group is supported on a curve $C$ that has an
ordinary double point or cusp singularity.  For SU(2), the adjoint
and symmetric tensor representations are equivalent, so this
distinction does not lead to different spectra in low-energy theories
either of an SU(2) or U(1) theory.  For SU(3), on the other hand,
the adjoint and symmetric tensor representations are distinct.  In
six-dimensional theories, anomaly cancellation conditions treat an
adjoint matter representation as equivalent to the combination of a
symmetric and an anti-symmetric tensor representation of SU(3).
Thus, there are low energy theories of 6D supergravity with an SU(3)
gauge group in which some $g'$ of the total possible $g$ adjoint
representations are replaced by symmetric + antisymmetric
representations.  When such an SU(3) is Higgsed, it gives rise to a
distinctive pattern of charges for the resulting U(1)$\times$U(1)
theory.  In the general model that we describe in the following
section, we find that there is a parameter that determines the number
of such singularities that must be present on the curve carrying an
SU(3) gauge group.  Here we summarize the consequences for the
spectrum after Higgsing.

\begin{figure}[ht]
\centering
 \includegraphics[scale=1]{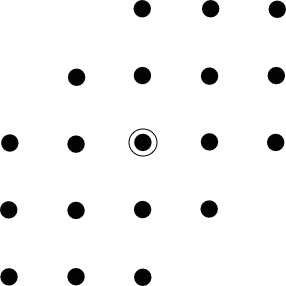}
\caption{Weight lattice of charges in a theory with a
U$(1)\times$U(1)
  gauge group realized by Higgsing a theory with gauge group
SU(2)$\times$SU(2)$\times$SU(3), when the SU(3) carries matter in
  the symmetric representation, associated in F-theory with singular
  points on the curve $C$ carrying the SU(3) factor.
}
\label{f:charges-general}
\end{figure}

We consider the most general model, where the unHiggsed gauge group is
$\text{SU}(2)\times\text{SU}(2)\times\text{SU}(3)$.  We now assume, however, that the
SU(3) carries in addition to $g - g'$ adjoint fields, $g'$ symmetric
and $g'$ antisymmetric representations.  The antisymmetric
representations are just the anti-fundamental, so give rise to the
same U(1) charges as the fundamental representation described in the
previous subsection.  The symmetric representation $\mathbf{6}$ of SU(3), however,
transforms differently and gives rise to charges
\begin{equation} \label{eq:branching4}
 (\mathbf{1},\mathbf{1},\mathbf{6})\rightarrow
(2, 2) + (0, -2) + (-2, 0)+(0,1)+(1,0)+(-1,-1)
\end{equation}
Thus, in the most general situation we expect that we may have a
spectrum of U(1)$\times$U(1) charges such as those shown in
Figure~\ref{f:charges-general}.

\subsection{Examples}

To illustrate some of the different structures just described, we give
some simple examples.  These examples also show how the different
branches of the set of U(1)$\times$U(1) models have different
dimensionalities.

We consider the simplest 6D models on the base $\P^2$ of each of the
types just described.

For the SU(2)$\times$SU(2) model, we have two SU(2) factors tuned
on two cubic curves $A, B$.  The spectrum of this model was described
in \S\ref{sec:HiggsingSU2SU2}.  After Higgsing both SU(2) factors
as described there, using two adjoint fields,
there are a total of $36 + 72 + 72 = 180$ charged matter
fields.  Gravitational anomaly cancellation imposes the condition
$H-V= 273$, so the number of neutral scalar fields parameterizing this
class of models is $H_{\rm uncharged} = 273 + 2-180 = 95$.

For the SU(3) model, we have a single SU(3) factor on a cubic
curve $A$.  
We Higgs using the adjoint field to a U(1)$\times$U(1) model.
The spectrum as discussed above in \S\ref{sec:Higgsing-3}
now contains $ 54 + 54 + 54 = 162$
charged matter fields, so the number of neutral scalars is $H_{\rm
  uncharged} = 275-162 = 113$.  Thus, this branch of the U(1)$\times$U(1) moduli space has many more degrees of freedom than the branch
arising from Higgsing the simplest SU(2)$\times$SU(2) model.

Now we consider a hybrid model, where SU(2)$\times$SU(2)$\times$
SU(3) is tuned on $A, B, C,$ where $A, B$ are degree one (lines) and
$C$ is degree two (conic).  In this case in the non-Abelian theory
there are no adjoint matter fields, and we Higgs the theory by turning
on expectation values for bifundamentals as described in
\S\ref{sec:hybrid}.  The numbers of charged and neutral hypermultiplets lie
in-between those in the two models above: we have 172
charged fields and 103 neutral scalars.  Because there are no adjoints
of the SU(2) factors, and all the SU(2)$\times$SU(3)
bifundamentals are Higgsed,
the set of charges is captured by the subset of charges that appear in
both Figure~\ref{f:2-2-weights} and
Figure~\ref{f:3-weights}; if $A, B$ were higher degree the charge
spectrum would be that of Figure~\ref{f:223-weights}.  Note however
that the detailed multiplicities in the spectrum resulting from this
construction is different from any that could be produced simply by
Higgsing an SU(3).
In particular, in this model the spectrum contains 40 fields with
charge $(1, 1)$, 4 fields with charge $(1, -1)$, and  64 fields each
with charges $(1, 0)$ and $(0, 1)$ (in each case the hypermultiplet
for the field also contains the conjugate charge). Here we used the branching rules \eqref{eq:branching1}-\eqref{eq:branching3} taking into account that two states each with charges $(-2,0)$, $(0,-2)$, $(1,2)$, $(2,1)$ and $(-1,1)$ as well as two neutral singlets are eaten up, respectively. This resulting spectrum
cannot be realized from the other constructions available.

Finally, we consider the simplest apparently consistent supergravity
theory that would correspond to the spectrum depicted in
Figure~\ref{f:charges-general}, as described in \S\ref{sec:Higgsing-singular}.
In principle this could arise from an SU(3) with two adjoint matter
representations, one symmetric representation, and 61 fundamental
representations (of which one comes from the antisymmetric
representation that comes along with the symmetric when replacing an
adjoint).    This model
seems to give a sensible low-energy spectrum,
and one might imagine that this spectrum
could arise from an F-theory
model where an SU(3) is tuned on a quartic curve $C$
with an appropriate double point singularity.  Strangely, however, as
we see later in this paper, the general construction we present here
only gives models of this type associated with curves of degree 5 and
higher.  This puzzle is discussed further throughout the paper.

\section{General elliptic fibrations with three sections}
\label{sec:abcd}

In this section we construct a general class of elliptically fibered
Calabi-Yau manifolds $\pi:X_{n+1}\rightarrow B_n$ over a base $B_{n}$ so that
the general elliptic fiber $\mathcal{E}=\pi^{-1}(p)$ with $p$ a point
in $B_{n}$ has three rational points, {\it i.e.}, there is a rank two
Mordell-Weil group.  F-theory compactified on $X_{n+1}$ then yields a
low-energy effective theory with a U(1)$\times$U(1) gauge
symmetry. First, in Section \ref{sec:TheFiber} we construct the
elliptic curve $\mathcal{E}$ as a specialized cubic in
$\mathbb{P}^2$. Then, we find its Weierstrass form as well as the
Weierstrass coordinates of its rational points in Section
\ref{sec:WSFCubic}. Finally, in Section \ref{sec:cubicFibrations}, we
construct all Calabi-Yau elliptic fibrations $X_{n+1}$ of $\mathcal{E}$ over
$B_{n}$. The construction we describe here is completely generic
  and is valid for an arbitrary F-theory base manifold $B_n$.

The crucial difference between the construction here and the one in
\cite{Borchmann:2013jwa,Cvetic:2013nia,Cvetic:2013uta,Borchmann:2013hta,Cvetic:2013jta,Klevers:2014bqa}
is that the rational points in $\mathcal{E}$ are kept at general
positions and are \textit{not} assumed to be toric points in
$\mathbb{P}^2$
({\it i.e.} the simultaneous vanishings of two
homogeneous coordinates).  This will be the key to constructing a more
general model that can, for example, be unHiggsed to
SU$(2)\times$SU$(2)$ and SU$(3)$

\subsection{A new elliptic curve}
\label{sec:TheFiber}

We consider an elliptic curve $\mathcal{E}$ with three rational points
$P$, $Q$ and $R$. It is well-known due to Deligne \cite{deligne1975courbes}, that these rational
points define a line bundle $\mathcal{M}=\mathcal{O}(P+Q+R)$ that
embeds $\mathcal{E}$ into $\mathbb{P}^2$ with
$\mathcal{M}=\mathcal{O}_{\mathbb{P}^2}(1)\vert_{\mathcal{E}}$,
{\it cf.}~\cite{Borchmann:2013jwa,Cvetic:2013nia,Cvetic:2013uta,Borchmann:2013hta,Cvetic:2013jta}.
In terms of the projective coordinates $[u:v:w]$ on $\mathbb{P}^2$,
this embedding is given by the cubic \beq
	\label{eq:gencubic}
		p:=s_1u^3+s_2 u^2 v+s_3 uv^2+s_4 v^3+(s_5u^2+s_6 uv+s_7 v^2)w+(s_8 u+s_9v)w^2+s_{10}w^3=0\,, 
\eeq
where the coefficients $s_i$ take values in a given field $K$. In an 
elliptic fibration $\pi:\,X_{n+1}\rightarrow B_n$ of $\mathcal{E}$, we identify $K$ with the field of rational 
functions on $B_{n}$.

\begin{figure}[ht]
\centering
 \includegraphics[scale=0.35]{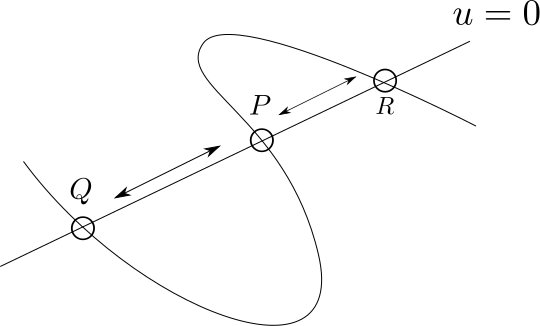}
\caption{Cubic $\mathcal{E}$ with three rational points $P$, $Q$ and 
$R$ contained in the line $u=0$.
}
\label{fig:cubicBU}
\end{figure}

The identification 
$\mathcal{M}=\mathcal{O}_{\mathbb{P}^2}(1)\vert_{\mathcal{E}}$ 
implies that there exists a section $U$ of 
$\mathcal{O}_{\mathbb{P}^2}(1)$ on the ambient space $\mathbb{P}^2$, 
so that $U\vert_\mathcal{E}$ vanishes precisely at the three
points $P$, $Q$ and $R$.\footnote{Indeed the given points
$P$ and $Q$ on $\mathcal{E}$ uniquely define a hyperplane in 
$\mathbb{P}^2$. This hyperplane intersects the 
cubic \eqref{eq:gencubic} in a third point, which automatically is 
rational. Thus, this point has to be $R$, because otherwise 
$\mathcal{E}$ would have a fourth rational point, which is excluded 
by construction.}  By a rotation in the $\mathbb{P}^2$, we  choose
$U=u$, see Figure \ref{fig:cubicBU}.
\footnote{
A general $U$ takes the form $U=Au+Bv+Cw$ with $A$,
$B$, $C$ coefficients in appropriate line bundles on $B_n$. For a 
single elliptic curve, these are constants in $K$ and we can
perform a $\text{Gl}(3)$-transformation so that $U=u$. In an
elliptic fibration,  this is also possible in a general class of models. For a fibration over a
three-dimensional base $B_{3}$,  one might worry about the codimension three locus $A=B=C$ 
in the base where $U\equiv 0$ which would be a point of $I_3$ fiber. 
However, we 
can use the $\mathbb{C}^*$-action on $\mathbb{P}^2$ to set 
{\it e.g.}~$A\equiv 1$ globally. Since $u$ is then a coordinate on 
a $\mathbb{P}^2$-fiber independent of the base $B_{n}$, it follows that $U$ is  
non-vanishing on $B_{n}$, too. Thus, we can globally in the 
$\mathbb{P}^2$-fibration over $B_{n}$ perform the rotation to $U=u$.}
By setting $u=0$ in \eqref{eq:gencubic}, we obtain a cubic  
in $v$, $w$,
\beq \label{eq:cubicvw}
	q_3=s_4v^3+s_7 v^2 w+s_9 v w^2+s_{10}w^3\,,
\eeq
which has to have three distinct roots in $K$ that are the rational points $P$, $Q$ and $R$. 
For elliptic fibrations over $B_{n}$, the field $K$ is in general not algebraically 
closed. Thus, the existence of these roots constrains the 
coefficients in $q_3$ by imposing  it to factorize as
\beq \label{eq:cubicvwfactor}
 q_3=\prod_{i=1}^3 (a_i v + b_i w) \,,
\eeq
where $a_i$, $b_i$ take values in $K$.
Thus,  the coefficients in $q_3$ 
have to be given by the elementary symmetric polynomials in the $a_i$, $b_j$.
In summary, the elliptic fibration \eqref{eq:gencubic} meeting the
requirement of admitting three rational points takes the form
\beq \label{eq:cubicfactorized}
	p=u(s_1u^2+s_2 u v+s_3 v^2+s_5u w+s_6 vw+s_8w^2)+\prod_{i=1}^3
        (a_i v + b_i w)  = 0\,.
\eeq
It is important to note that in an elliptic fibration it is not
generally
possible globally to move two of the roots of \eqref{eq:cubicvwfactor}
to $v=0$ and $w=0$, as the necessary variable transformation will generically
involve denominators that are ill-defined at certain loci on
$B_{n}$. In contrast, for a single elliptic curve
 this is
always possible and leads to the elliptic curve in $dP_2$ studied in
an F-theory context in
\cite{Borchmann:2013jwa,Cvetic:2013nia,Cvetic:2013uta,Borchmann:2013hta,Cvetic:2013jta}.

The three rational points $P$, $Q$ and $R$ of \eqref{eq:cubicfactorized} have the 
general coordinates
\beq \label{eq:coordsPQR}
	P=[0:-b_1:a_1]\,,\qquad Q=[0:-b_2:a_2]\,,\qquad R=[0:-b_3:a_3]\,.
\eeq
We note that these rational points coincide when the two lines 
$a_iv+b_iw=0$ and $a_jv+b_jw=0$ align. This happens when the 
$2\times 2$-matrix of coefficients of these two equations has rank one, which is 
the case if $a_i b_j - a_j b_i=0$, for $i\ne j$. More explicitly, we obtain 
\bea \label{eq:collisionOfPoints}
	&P=Q:\,\, a_1b_2-a_2b_1=0\,, \,\,\quad\,\,  P=R:\,\, a_1b_3-a_3b_1=0\,,\,\,\quad\,\, Q=R :\,\, a_2b_3-a_3b_2=0\,,&\nn\\
	&P=Q=R:\,\, a_1b_2-a_2b_1=a_1b_3-a_3b_1=a_2b_3-a_3b_2=0\,.&
\eea
Note that we have to have $(a_i,b_i)\neq (0,0)$ because otherwise  the elliptic curve 
\eqref{eq:cubicfactorized} would not be smooth and the rational points not well-defined.

\subsubsection{Specialized models}

There are certain elliptic fibrations, for which the model 
\eqref{eq:cubicfactorized} of the elliptic curve $\mathcal{E}$ can be further simplified. 
Indeed, if for a particular $i$ we have that $a_i$ is a constant (this is equivalent to the case of constant 
$b_i$ by exchanging $v$ and $w$), we can perform the \textit{global} variable transformation
\beq
	v\mapsto v-\frac{b_i}{a_i}w\,,
\eeq
which effectively removes $b_i$ so that one root of $q_3$ in \eqref{eq:cubicvwfactor} is at $v=0$.  
Without loss of generality we can assume that $i=1$ and $a_1=1$ so that  the hypersurface equation \eqref{eq:cubicfactorized} takes
the form
\beq \label{eq:abcdModel}
 u( s_1 u^2 + s_2 u v+ s_3 v^2 + s_5  u w + s_6 v w + s_8 w^2) + v( a_2 v + b_2 w)(a_3 v+ b_3 w )=0\,.
\eeq
The coordinate  \eqref{eq:coordsPQR} of the rational points reduce to 
\beq
 P= [0:0:1], \qquad  Q= [0:-b_2:a_2], \qquad  R= [0:-b_3:a_3]\,.
\eeq
Note that in elliptic fibrations of this form the point $P$ is well
defined everywhere on $B_{n}$. For this specialized model, we have also worked out
the presentation as a non-generic quartic in $\mathbb{P}^{2}(1,1,2)$ in Appendix \ref{app:quarticinP112}, which is a specialization of the U(1) model of \cite{Morrison:2012ei}.

We emphasize two main differences between this specialized  model
and the general model \eqref{eq:cubicfactorized}. First of all, the
coefficient $b_1$ is absent from the cubic \eqref{eq:abcdModel}.  As we
demonstrate in Section \ref{sec:cubicFibrations}, this implies that
there is no corresponding effective divisor class associated to $b_1$, 
which gives us more freedom for the degrees of other
divisor classes defining the elliptic fibration. Second, and more
physically, F-theory compactifcations on elliptic fibrations based on
\eqref{eq:abcdModel} do not have matter fields at $a_1=b_1=0$, while
the spectrum is otherwise identical, as we will work out in Section
\ref{sec:Matter}.

Finally, we note that we can have a further specialization if in addition to a constant $a_1$, also $b_2$ is 
a constant. In this case, we can redefine $w$ globally to obtain the cubic 
\beq \label{eq:abModel}
 u( s_1 u^2 + s_2 u v+ s_3 v^2 + s_5  u w + s_6 v w + s_8 w^2) + vw(a_3 v+ b_3 w )=0\,.
\eeq
This is precisely the elliptic curve in $dP_2$ studied in 
\cite{Borchmann:2013jwa,Cvetic:2013nia,Cvetic:2013uta,Borchmann:2013hta,Cvetic:2013jta}. 
$dP_2$-elliptic fibrations in general do not admit rank-preserving unHiggsings as the points at 
$u=v=0$ and $u=w=0$  generally do not collide in the fibration and, consequently, can not be merged by a tuning of the complex structure.
As shown in \cite{Klevers:2014bqa}, the theory can, however, be  
unHiggsed to SU$(3)^3/\mathbb{Z}_3$ and $(\text{SU}(2)^2\times \text{SU}(4))/\mathbb{Z}_2$, 
respectively, where the rational point $u=w=0$ has turned into Mordell-Weil torsion.\footnote{The subset of $dP_2$ models unHiggsing to $\text{SU}(3)\times \text{SU}(2)$ are discussed in Section \ref{sec:dP2comparisondMatter}.} 

In summary, we see that \eqref{eq:cubicfactorized} admits certain
specializations. As the case of constant $a_1$ and $b_2$ reduces to
the elliptic curve in $dP_2$, that has been studied extensively in
recent literature, we will
primarily consider the completely general case and the
novel special case $a_1=1$,
$b_1=0$ with  generic $a_2$, $a_3$, $b_2$, $b_3$.

\subsection{The Weierstrass form}
\label{sec:WSFCubic}

We now proceed with the calculation of the Weierstrass form (WSF) 
\beq \label{eq:WSFcubic}
	 y^2=x^3+fxz^4+gz^6
\eeq
of the elliptic curve \eqref{eq:cubicfactorized}. Choosing $P$ as the
zero point, the birational map to the WSF can be obtained by applying
again the procedure of Deligne and constructing the sections of the
line bundles $\mathcal{O}(nP)$ for $n=1,\ldots, 6$ yielding an
embedding of $\mathcal{E}$ into $\mathbb{P}^{1,2,3}$. This procedure
has been described and applied in an F-theory context recently, see
appendix B of \cite{Morrison:2012ei} and more specifically
\cite{Cvetic:2013nia,Borchmann:2013hta} for the case of
cubics. Alternatively, we can apply Nagell's algorithm to
\eqref{eq:cubicfactorized}, see {\it e.g.}~appendix B of
\cite{Cvetic:2013nia}.

Instead of following these two procedures, we take a shortcut here by using the WSF of the elliptic curve in 
$dP_2$. Indeed, we  can perform the rational variable transformation 
$v\mapsto v-\frac{b_1}{a_1}w$ and $w\mapsto w - \frac{a_1 a_2}{a_1 b_2-a_2 b_1 }v$
on the cubic equation \eqref{eq:cubicfactorized} so that it assumes the standard form of
an $dP_2$-elliptic fibration, that is
\beq \label{eq:dP2constr}
	p_{\text{dP}_2}=u(\tilde{s}_1u^2+\tilde{s}_2uv+\tilde{s}_3v^2+\tilde{s}_5 uw+\tilde{s}_6 vw+\tilde{s}_8w^2)+vw(\tilde{s}_7v+\tilde{s}_9w)\,.
\eeq
Here the coefficients $\tilde{s}_i$ are given by $\tilde{s}_1 = s_1$ and
\bea \label{eq:sTransf}
&\!\!\!\!\!\tilde{s}_2 =  \frac{a_1 (b_2 s_2-a_2 s_5)}{a_1 b_2-a_2 b_1}\,,\,\,\,
\tilde{s}_3 =  \frac{a_1^2 \left(a_2^2 s_8-a_2 b_2 s_6+b_2^2 s_3\right)}{(a_1 b_2-a_2 b_1)^2}\,,\,\,\,
 \tilde{s}_5 =  s_5-\frac{b_1 s_2}{a_1}\,,\,\,\, \tilde{s}_6 = \frac{a_1 (b_2 s_6-2 a_2 s_8)+b_1 (a_2 s_6-2 b_2 s_3)}{a_1 b_2-a_2 b_1}\,,\!\!\!\!\!\!\!\!\!  &\nn\\
&\!\!\!\!\! \tilde{s}_7 =  a_1 (a_3 b_2-a_2 b_3)\,,\,\,\,\,\,\,\tilde{s}_8 =  \frac{b_1 (b_1 s_3-a_1 s_6)}{a_1^2}+s_8\,,\,\,\,\,\,\,\tilde{s}_9 =  \frac{(a_1 b_2-a_2 b_1 ) (a_1 b_3- a_3 b_1 )}{a_1}\,.& 
\eea
Then, we can directly apply the results of \cite{Cvetic:2013nia} for
the functions $f$, $g$ of the Weierstrass model. 
Clearing denominators gives a Weierstrass model where the coefficients
$f, g$ are polynomials in the $s_i, a_j, b_k$.
The resulting
expressions are quite lengthy and can be found in Appendix
\ref{app:WSF}.  We emphasize that both $f$ and $g$ are both manifestly
symmetric under exchange of the rational points, {\it i.e.}~both expressions
\eqref{eq:fullf} and \eqref{eq:fullg} are invariant under exchanging
$(a_i,b_i)\leftrightarrow (a_j,b_j)$ for any pair $i\neq j$.  Note
that neither $f$, $g$ nor the discriminant $\Delta$  admit a
factorization, signaling the absence of codimension one
singularities. The singularities at codimension two and higher will be
discussed in Section \ref{sec:SingsCubic}.

We can use this birational map to WSF in order to compute the
Weierstrass coordinates of the rational points $Q$, $R$. Again, the
obtained formula are lengthy and relegated to Appendix
\ref{app:WSF}. For $Q$ we obtain the coordinates given in
\eqref{eq:WScoordsQ}.  The Weierstrass coordinates of the rational
point $R$, given in \eqref{eq:WScoordsR} and denoted by
$[y_R:x_R:z_R]$, are obtained from the ones for $Q$ by exchanging the
variables $(a_2,b_2)\leftrightarrow (a_3,b_3)$ by symmetry of the
elliptic curve \eqref{eq:cubicfactorized}.

\subsubsection*{Specialized  model}

The WSF of the specialized cubic curve \eqref{eq:abcdModel} is obtained from \eqref{eq:fullf}, 
\eqref{eq:fullg} by setting $a_1=1$ and $b_1=0$. Similarly we obtain the Weierstrass coordinates of the 
rational point $Q$ from \eqref{eq:WScoordsQ} in this limit:
\bea  \label{eq:spWScoordsQ}
z_Q \!\!&\!=\!&\!\! -b_2\,,\nn\\
x_Q\!\!&\!=\!&\!\! \frac{1}{12} (-4 b_2^3 (b_3 s_2 + a_3 s_5) - 12 a_2 b_2 s_6 s_8 + 12 a_2^2 s_8^2 + 
    b_2^2 (8 a_2 b_3 s_5 + s_6^2 + 8 s_3 s_8))\,,\nn \\
y_Q \!\!&\!=\!&\!\! \frac{1}{2} (b_2^4 b_3 (a_3 b_2 - a_2 b_3) s_1\! -\! 
    b_2^2 (a_3 b_2 - a_2 b_3) (b_2 s_2 - a_2 s_5) s_8 \!-\! 
    b_2^2 b_3 s_5 (b_2^2 s_3 - a_2 b_2 s_6 + a_2^2 s_8) \nn \\ && + 
    s_8 (b_2 s_6 - 2 a_2 s_8) (b_2^2 s_3 - a_2 b_2 s_6 + a_2^2 s_8))\,.
\eea
As before, we obtain the coordinates of the point $R$, by symmetry of \eqref{eq:abcdModel}, upon 
exchanging $(a_2,b_2)\leftrightarrow (a_3,b_3)$.

\subsection{Constructing elliptic fibrations}
\label{sec:cubicFibrations}

We proceed with the construction of elliptically fibered Calabi-Yau
manifolds $\pi: X_{n+1} \rightarrow B_n$ as fibrations of the curve
$\mathcal{E}$ defined in \eqref{eq:cubicfactorized} over an arbitrary
base $B_{n}$.  Although we focus in this work on the case of
Calabi-Yau threefolds, the discussion in this section applies to any
complex dimension of $X_{n+1}$.  The procedure outlined here has been
used already,  {\it e.g.},~in
\cite{Cvetic:2013nia,Cvetic:2013qsa,Klevers:2014bqa}, to which
we refer the reader for more details.

An elliptic fibration by the curve $\mathcal{E}$  over a base $B_{n}$ is formally obtained by 
identifying the field $K$ with the field of rational  functions on $B_{n}$. Upon cancelling denominators, the coefficients 
$s_i$, $a_j$ and $b_j$ in  \eqref{eq:cubicfactorized} are identified with sections of appropriate line 
bundles over $B_{n}$. The correct line bundles are determined by imposing the Calabi-Yau condition on the total 
space of the fibration. Clearly, in the obtained elliptic fibration $X_{n+1}$, the rational points in 
\eqref{eq:coordsPQR} lift to rational sections of the fibration, that we denote by 
$\hat{s}_P$, $\hat{s}_Q$ and $\hat{s}_R$. Their homology classes give rise to two U(1) gauge 
symmetries in F-theory compactifications on $X_{n+1}$ \cite{Morrison:1996pp}.

In more detail, we first fiber the two dimensional ambient space 
$\mathbb{P}^2$ of \eqref{eq:cubicfactorized} over $B_{n}$, yielding the projective bundle 
\beq \label{eq:P2fibrationBn}
	\xymatrix{
	\mathbb{P}^2 \ar[r] & 	\mathbb{P}^2_{B}=\mathbb{P}(\mathcal{O}_{B}(D_u)\oplus\mathcal{O}_{B}(D_v)\oplus\mathcal{O}_{B}) \ar[d]\\
	& B\,
	}
\eeq
This is fully specified by the two divisors $D_u$ and $D_v$ on
$B_{n}$, respectively, their associated line bundles being
$\mathcal{O}_B(D_u)$ and $\mathcal{O}_B(D_v)$, which we chose so that  
\beq \label{eq:LBDuDv}
 u \in \mathcal{O}(D_u) \,, \qquad v\in \mathcal{O}(D_v)\,.
\eeq
This implies that the constraint \eqref{eq:cubicfactorized}  becomes
a section of a line bundle over the base $B_{n}$. Consistency,
{\it i.e.}~the requirement that every monomial in \eqref{eq:cubicfactorized} 
transforms as a section of the same line bundle, then
requires that the  coefficients $s_i$, $a_j$ and $b_j$   are 
sections of appropriate line bundles. 
Finally, we require in addition that the constraint 
\eqref{eq:cubicfactorized}  is a  section of the anti-canonical bundle
of \eqref{eq:P2fibrationBn}, {\it i.e.},~defines a Calabi-Yau manifold
$X_{n+1}$. This  condition fixes the line bundles
for the coefficients $s_i$, $a_j$ and $b_j$. 

Computing the anti-canonical bundle of \eqref{eq:P2fibrationBn} by adjunction and solving the constraints 
imposed by the Calabi-Yau condition, we obtain that the coefficients $s_i$, $a_j$, $b_j$
have to be sections in the line bundles 
\beq \label{eq:coeffsSectionsGen}
\qquad \text{
\begin{tabular}{c|c}
\text{Section} & \text{Line bundle}\\
\hline
	$s_1$&$\mathcal{O}(-6K_B-2[a_1]-2[a_2]-2[a_3]-3[s_8])$\rule{0pt}{13pt} \\
	$s_2$&$\mathcal{O}(-4K_B-[a_1]-[a_2]-[a_3]-2[s_8])$\rule{0pt}{12pt} \\
	$s_3$&$\mathcal{O}(-2K_B-[s_8])$\rule{0pt}{12pt} \\
	$s_5$&$\mathcal{O}(-3K_B-[a_1]-[a_2]-[a_3]-[s_8])$\rule{0pt}{12pt} \\
	$s_6$&$\mathcal{O}(-K_B)$\rule{0pt}{12pt} \\
	$s_8$&$\mathcal{O}([s_8])$\rule{0pt}{12pt} \\
	$a_1$&$\mathcal{O}([a_1])$\rule{0pt}{12pt} \\
	$a_2$&$\mathcal{O}([a_2])$\rule{0pt}{12pt} \\
    $a_3$&$\mathcal{O}([a_3])$\rule{0pt}{12pt} \\
	$b_1$&$\mathcal{O}(K_B+[a_1]+[s_8])$ \rule{0pt}{12pt} \\
	$b_2$&$\mathcal{O}(K_B+[a_2]+[s_8])$\rule{0pt}{12pt} \\
	$b_3$&$\mathcal{O}(K_B+[a_3]+[s_8])$\rule{0pt}{12pt} 
\end{tabular}
}
\eeq
Here $-K_B$ is the anti-canonical
divisor of $B_{n}$, and we set 
$D_u=3K_B + [a_1] + [a_2] + [a_3] + 2 [s_8]$,  $D_v = K_B + [s_8]$ so that  all base 
divisors are parametrized by the classes $[a_i]$, $i=1,2,3$, and $[s_8]$, corresponding to the sections $a_i$ and $s_8$, respectively.
This parametrization is convenient as these four classes have to be effective in a generic model. They can chosen freely 
subject  to the condition that all $s_i$ and $a_j$, $b_j$ are effective, which yields a finite discrete set of possible strata in
the moduli space of $X_{n+1}$. 
We note that the line bundles for each of the parameters in
(\ref{eq:coeffsSectionsGen}) can also be determined efficiently
directly from the form of the Weierstrass coefficients $f, g$ in
Appendix~\ref{app:WSF}, where $f \in{\cal O} (-4K_B),
g \in{\cal O} (-6K_B)$.

As immediate consequences of the effectiveness condition of \eqref{eq:coeffsSectionsGen} we infer that 
\beq
[s_8]\leq -2K_B\,,\qquad -K_B\leq[a_i]+[s_8] \,,\qquad [a_1]+[a_2]+[a_3]+[s_8]\leq -3K_B\,,
\eeq
where $-K_B$ must be ($\Q$-) effective for a Calabi-Yau
Weierstrass model to exist.
These conditions will be crucial for the understanding of the unHiggsings of the U(1)-symmetries by tuning of the
parameters in \eqref{eq:cubicfactorized} as discussed below.

In summary, we see that the elliptically fibered Calabi-Yau manifold
$X_{n+1}$ is specified by four discrete parameters.  We note that this
additional freedom is not present in the earlier construction in
\cite{Borchmann:2013jwa,Cvetic:2013nia,Cvetic:2013uta,Borchmann:2013hta,Cvetic:2013jta}, where there are only two 
discrete parameters. The presence of two additional discrete parameters 
is expected for the cubic
in \eqref{eq:cubicvw}: it correspond to the two discrete degrees of freedom 
reflected in the two independent pairwise 
rescalings $((a_i,b_i),(a_j,b_j))\mapsto
(\lambda(a_i,b_i),\lambda^{-1}(a_j,b_j))$  $(i\neq j)$ which leave  
$p=0$ invariant, where $\lambda$ is a section of a line-bundle on $B_n$. 
The additional discrete choices arise from specifying these two line-bundles.

\subsubsection*{The specialized model}

The case of the specialized model \eqref{eq:abcdModel} is covered by the general result \eqref{eq:coeffsSectionsGen} by 
setting $[a_1]=0$ and dropping the coefficient $b_1$, as this is no longer present in the model. We obtain 
\beq \label{eq:coeffsSections2}
\qquad \text{
\begin{tabular}{c|c}
\text{Section} & \text{Line bundle}\\
\hline
	$s_1$&$\mathcal{O}(-6K_B-2[a_2]-2[a_3]-3[s_8])$\rule{0pt}{13pt} \\
	$s_2$&$\mathcal{O}(-4K_B-[a_2]-[a_3]-2[s_8])$\rule{0pt}{12pt} \\
	$s_3$&$\mathcal{O}(-2K_B-[s_8])$\rule{0pt}{12pt} \\
	$s_5$&$\mathcal{O}(-3K_B-[a_2]-[a_3]-[s_8])$\rule{0pt}{12pt} \\
	$s_6$&$\mathcal{O}(-K_B)$\rule{0pt}{12pt} \\
	$s_8$&$\mathcal{O}([s_8])$\rule{0pt}{12pt} \\
	$a_2$&$\mathcal{O}([a_2])$\rule{0pt}{12pt} \\
    $a_3$&$\mathcal{O}([a_3])$\rule{0pt}{12pt} \\
	$b_2$&$\mathcal{O}(K_B+[a_2]+[s_8])$\rule{0pt}{12pt} \\
	$b_3$&$\mathcal{O}(K_B+[a_3]+[s_8])$\rule{0pt}{12pt} 
\end{tabular}
}
\eeq

We emphasize that the absence of the coefficient $b_1$ weakens the
effectiveness conditions imposed by \eqref{eq:coeffsSections2} on the remaining
parameters $[a_2]$, $[a_3]$ and $[s_8]$, compared to
\eqref{eq:coeffsSectionsGen}.
In fact, we do not have the constraint
$-K_B \leq [s_8]$.
This implies that the model
\eqref{eq:abcdModel} has qualitatively different unHiggsings than the
general model \eqref{eq:cubicfactorized}.

\subsection{Comparison with $dP_2$-elliptic fibrations}
\label{sec:dP2comparison}

Before continuing with our discussion, let us pause to compare with the elliptic fibrations with rank two Mordell-Weil group 
constructed in \cite{Borchmann:2013jwa,Cvetic:2013nia,Cvetic:2013uta,Borchmann:2013hta,Cvetic:2013jta}
using the elliptic curve in $dP_2$.

Clearly, for a single elliptic curve over the complex numbers
$\mathbb{C}$, we can always redefine the coordinates $[u:v:w]$ in
\eqref{eq:cubicfactorized} to map to the elliptic curve in $dP_2$
given in \eqref{eq:abModel}.  However, in an elliptic fibration, the
necessary coordinate redefinition is not defined globally unless we
specialize $[a_1]=[b_2]=0$. Then, we can perform a coordinate
redefinition that effectively sets $b_1=a_2=0$.  
Like the specialized model described in (\ref{eq:coeffsSections2}),
the $dP_2$-elliptic fibration gives a class of constructions that are
not captured by the
general model $X_{n+1}$
where all coefficients $a_i, b_i$ are non-vanishing. In
the $dP_2$-elliptic fibration, the coefficients $b_1$ and $a_2$ are
absent.  Consequently, their divisor classes in
\eqref{eq:coeffsSectionsGen} do not have to be effective, which
relaxes the constraints imposed on the remaining two parameters
$[a_3]$ and $[s_8]$, which are now allowed to assume values that would
violate the effectiveness of $b_1$ and $a_2$
if those parameters were non-vanishing. Hence, we obtain new
models based on the $dP_2$-elliptic curve giving rise to different
physics in F-theory, that are not just a specialization of F-theory on
a general $X_{n+1}$.
Note, however, that all the models found in the
specialized model (\ref{eq:coeffsSections2}) and the $dP_2$
construction are still special cases of the Weierstrass model
described in Appendix~\ref{app:WSF}; the distinguishing feature of
these specialized models is that a broader range of classes can be
chosen for the non-vanishing parameters $[a_3], [s_8]$, and $[a_2]$ in
the specialized models.

\begin{figure}[ht]
\centering
 \includegraphics[scale=0.4]{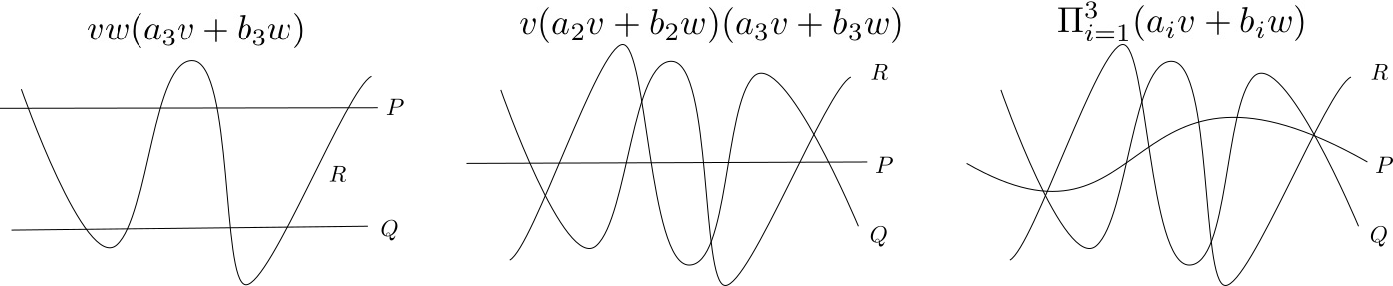}
\caption{Intersection pattern of sections in three different cubic fibers.}
\label{fig:intpattern}
\end{figure}

Another way to formulate the difference between elliptic fibrations
based on \eqref{eq:cubicfactorized}
with generic coefficients and those based on the specialized model \eqref{eq:abcdModel} or the
$dP_2$-curve \eqref{eq:abModel} is to compare the general
intersection patterns of the rational sections in these three models. The intersection
pattern of the sections in elliptic fibrations
based on $dP_2$ is shown in the first figure in Figure
\ref{fig:intpattern}, that of the specialized models in the second figure 
while a generic situation in the general model
$X_{n+1}$ is shown in the third figure.
It is observed that the sections $\hat{s}_P$ and $\hat{s}_Q$ in $dP_2$-models can not
intersect at codimension one in the base $B_{n}$ as they are induced
by the fixed toric points $P=[0:0:1]$ and $Q=[0:1:0]$.  In 
contrast,
in a specialized model, all sections can intersect independently, 
as discussed in \eqref{eq:collisionOfPoints}. Finally,
in a general model all sections can intersect, too, and the zero section is no longer rigid. As we will demonstrate in
Section \ref{sec:Matter}, this gives rise to a different matter spectrum
of the F-theory models constructed as $dP_2$-elliptic 
fibrations, the specialized and
the general Calabi-Yau manifolds $X_{n+1}$; in particular, the
disjoint nature of the sections $P,\, Q$ in the $dP_2$-models indicates
that these sections cannot be brought together to form an unHiggsed
model without Abelian factors in a straightforward fashion.

\section{Singularities, global resolution and the matter spectrum}
\label{sec:SingsResMatter}

In this section, we discuss the singularities of the elliptic
fibration of the Calabi-Yau manifold $X_{n+1}$, the construction of a
global CY-resolution of these singularities and the determination of
the matter spectrum of the corresponding F-theory model.  
As we discuss further in Section \ref{sec:SingsCubic}, 
the relevant singularities of
$X_{n+1}$ are at least codimension two in the base $B_n$.  The global
resolution is represented as a complete intersection in a
$(n+4)$-dimensional ambient space. It describes two blow-ups at two
non-toric points in the ambient space of the elliptic fiber.  In
Section \ref{sec:Matter}, we use these results to obtain the general
matter spectrum of an 6D F-theory compactification on $X_{n+1}$.

Note that in the discussion here we focus on the matter charged under
the U(1) F-theory fields associated with the rank two Mordell-Weil
group, \textit{i.e.}~matter whose existence is imposed by the considered elliptic
fiber of the form \eqref{eq:cubicfactorized}. For simplicity, in this analysis we may assume that we are
considering a generic elliptic fibration over a weak Fano base, so
that there are no non-Abelian gauge group factors.  As described in
\cite{Morrison:2012np,Morrison:2014lca,Halverson:2015jua} for 6D and 4D F-theory
compactifications, most bases $B_n$ are not weak Fano, and have
associated non-Higgsable non-Abelian gauge group factors for generic
elliptic fibrations.  In such cases, or when the Weierstrass model
over a Fano base is specifically tuned, there are codimension one
singularities as well, and a full resolution would involve non-Abelian
gauge groups and associated charged matter as well as the Abelian
factors and Abelian charges we describe here.  More detailed analysis
of such models could be carried out as a further extension of the
Abelian analysis here.

\subsection{Singularities of $X_{n+1}$ and their global resolution}
\label{sec:SingsCubic}

In this section we determine all codimension two singularities of the
elliptically fibered CY-manifold $X_{n+1}$.  In general, the existence
of rational sections in an elliptic fibration implies the presence of
singularities in its Weierstrass model at codimension two in the base
$B_{n}$ of the fibration, see
{\it e.g.}~\cite{Morrison:2012ei,Cvetic:2013nia}.
As stated above, we make the simplifying assumption that there are no
codimension one singularities in the elliptic fibration.
As we will demonstrate below,
$X_{n+1}$ has Kodaira fibers of type $I_2$ at the 
codimension two loci in the base $B_{n}$ given in Table \ref{tab:ChargedHypersLoci}
\begin{table}[ht!]
\begin{center}
 \begin{tabular}{|c|} \hline 
  $I_2$-loci at codimension two in $B_{n}$ \\ \hline
  $V(I_{(1)}):=\{a_1=b_1=0\}$  \\
   $V(I_{(2)}):=\{a_2=b_2=0\}$  \\ 
$V(I_{(3)}):=\{a_3=b_3=0\}$\\   
  $V(I_{(4)}):=\{\Delta_{12}=s_3b_1^2-s_6a_1b_1+s_8a_1^2=0\}\backslash V(I_{(1)})$  \rule{0pt}{13pt}\\
   $V(I_{(5)}):=\{\Delta_{13}=s_3b_1^2-s_6a_1b_1+s_8a_1^2=0\}\backslash V(I_{(1)})$  \\
  $V(I_{(6)}):=\{\Delta_{23}=s_3b_2^2-s_6a_2b_2+s_8a_2^2=0\}\backslash V(I_{(2)})$  \\
  $V(I_{(7)}):=\{y_{Q}=y_R=(z_Q)^4 f+3(x_Q)^2=(z_R)^4 f+3(x_R)^2=0\}\backslash V(I_{(1)})$ \\
$V(I_{(8)}):=\{y_{Q}=(z_Q)^4 f+3(x_Q)^2=0\}\backslash  V(I_{(1)}\cdot I_{(2)}\cdot I_{(4)}\cdot I_{(5)}\cdot I_{(6)}\cdot I_{(7)})$ \\
$V(I_{(9)}):=\{y_{R}=(z_R)^4 f+3(x_R)^2=0\}\backslash  V(I_{(1)}\cdot I_{(3)}\cdot I_{(4)}\cdot I_{(5)}\cdot I_{(6)}\cdot I_{(7)})$ \\
   \hline
 \end{tabular}
 \label{tab:ChargedHypersLoci}
 \caption{Codimension two matter loci of $X_{n+1}$.}
 \end{center}
 \end{table}
Here we have denoted the variety given by the vanishing locus of an
ideal $I_{(k)}$ by $V(I_{(k)})$, $k=1,\ldots,9$;  we have also
used the
Weierstrass coordinates of the rational sections $\hat{s}_{Q}$,
$\hat{s}_R$ given in \eqref{eq:WScoordsQ}, \eqref{eq:WScoordsR}, and the
shorthand notation
\beq \label{eq:Deltaij}
	\Delta_{ij}:=a_ib_j-b_ia_j\,.
\eeq
In addition, we specify the varieties $V(I_{(k)})$, $k=4,\ldots, 9$,
in terms of reducible varieties from which we subtract unwanted
irreducible components, employing $V(I\cdot J)=V(I)\cup
V(J)$ for two ideals $I$, $J$.  The corresponding prime ideals
$I_{(k)}$ can be computed via the primary decomposition of the
respective reducible variety.
As discussed next, this analysis confirms that the list above gives the complete set
of irreducible components of the $I_2$-loci. Further evidence of completeness is provided by 
anomaly cancellation in 6D, which we show in Section \ref{sec:Matter}.

Next, let us comment on the derivation of the codimension two loci in Table \ref{tab:ChargedHypersLoci}. First, we recall
that  the  presence of the two rational sections
$\hat{s}_Q$ and $\hat{s}_R$ implies two factorizations of the Weierstrass form \eqref{eq:WSFcubic} of $X_{n+1}$. In the patch $z=1$,
these factorizations read
\beq
	\Big(y-\frac{y_{Q,R}}{z_{Q,R}^3}\Big)\Big(y+\frac{y_{Q,R}}{z_{Q,R}^3}\Big)=\Big(x-\frac{x_{Q,R}}{z_{Q,R}^2}\Big)\Big(x^2+\frac{x_{Q,R}}{z_{Q,R}^2}x+f+\frac{x_{Q,R}^2}{z_{Q,R}^4}\Big)\,,
\eeq
which implies a singularity at the following codimension two loci in
the base $B_{n}$:
\beq \label{eq:Codim2Sings}
 y_{Q,R} = 0\,, \qquad (z_{Q,R})^4 f + 3(x_{Q,R})^2 =0 \,.
\eeq
The singularity at the loci \eqref{eq:Codim2Sings} is of Kodaira type
$I_2$, as can be read off by the vanishing orders of
$(f,g,\Delta)$. These are precisely the two complete intersections
entering the last two lines in Table \ref{tab:ChargedHypersLoci}.

The locus \eqref{eq:Codim2Sings} is a generically a reducible variety in $B_{n}$. Over each irreducible component 
the behavior of the three sections $\hat{s}_P$, $\hat{s}_Q$ and $\hat{s}_R$ is different, yielding  different matter
representations in F-theory. All irreducible matter loci are found by a decomposition of \eqref{eq:Codim2Sings} into 
its prime ideals as employed in \cite{Cvetic:2013nia,Cvetic:2013uta}.  
We claim that all its irreducible components are given in Table \ref{tab:ChargedHypersLoci}.
As this decomposition is in general computationally 
involved, it is necessary to identify a number of codimension two singularities of the WSF of $X_{n+1}$ by hand, as we demonstrate
next. 

\subsubsection{Singularities in the cubic fibration $X_{n+1}$}

The elliptic fibration  $X_{n+1}$ in the form \eqref{eq:cubicfactorized} is singular and admits conifold singularities. 
These can be made manifest by writing the defining equation \eqref{eq:cubicfactorized} in the more suggestive form
\beq \label{eq:binomial}
 		u f_u(u,v,w) +  v_1 v_2v_3 =0 \,,
\eeq
where we defined
\beq \label{eq:fuv1v2v3}
	f_u(u,v,w):=s_1 u^2+s_2 uv+s_3 v^2+s_5 uw+s_6 vw+s_8w^2\,,\qquad v_i:=a_iv+b_i w\,. 
\eeq
This is a binomial geometry precisely of the form studied in \cite{Esole:2011sm,Marsano:2011hv}. As 
discussed in these works there are three intersecting conifold singularities at 
\beq \label{eq:conifolds}
	u=f_u=0\,,\qquad v_i=v_j=0\,\,\, (i< j)\,,
\eeq
where we only consider the irreducible components of codimension two in $B_{n}$ of this variety. If $(a_i,b_i)\neq (0,0)$
and $(a_j,b_j)\neq 0$, this yields three codimension two loci
\beq \label{eq:conifoldLoci}
	\Delta_{ij}:=a_ib_j-b_ia_j=0\,,\qquad s_3b_i^2-s_6a_ib_i+s_8a_i^2=0\,,\qquad i=1,2,3\,,
\eeq
where the elliptic fiber is singular at $[u:v:w]=[0,-b_i,a_i]$. These are precisely the complete intersections entering the 
third to sixth lines of Table \ref{tab:ChargedHypersLoci}.

For the special values $(a_i,b_i)=(0,0)$ the second term in (\ref{eq:binomial}) vanishes.  Thus.
the elliptic fiber of \eqref{eq:binomial} degenerates into 
the line $u=0$ and the conic $f_u=0$,  which produces a resolved
$I_2$ fiber at the three codimension two loci 
\beq \label{eq:resolvedI21}
	a_i=b_i=0\,,\qquad i=1,2,3\,.
\eeq 
These three loci are the first to third lines in Table \ref{tab:ChargedHypersLoci}.
Note that in these cases, the sections $\hat{s}_P, \hat{s}_Q,$ and $\hat{s}_R$ all intersect
the fiber along the line $u = 0$; it is wrapped by that section whose coordinates
are ill-defined at the respective locus in \eqref{eq:resolvedI21}.
Note  also that while this $I_2$ singularity is resolved in the cubic model
(\ref{eq:cubicfactorized}), the fiber is singular in the Weierstrass presentation.

There are three additional ways to factorize \eqref{eq:binomial} into
a line and a conic at codimension two in $B_n$.
 We write the factorization in the form
\beq \label{eq:resolvedI22}
 	(s_1 u+a_iv+b_iw)q_2(u,v,w)=0\,,\qquad i=1,2,3\,. 
\eeq
Here $q_2(u,v,w)=u^2+\ldots$ is a general conic in $u$, $v$, $w$. For
generic $a_i$, $b_i$ the line $(s_1 u+a_iv+b_iw)= 0$ intersects the
line $u = 0$ at the single point $[u:v:w] =[0, -b_i, a_i]$, and
precisely one of the rational points $P, Q, R$ is on it.  The three different possibilities in \eqref{eq:resolvedI22} 
correspond to the
three different choices of which rational point lies on the line.  The
codimension two locus in $B_{n}$ supporting these $I_2$ fibers can be
computed using elimination ideals, 
see~\cite{Anderson:2014yva,Klevers:2014bqa}, and is expected to yield the
three ideals in the last three lines of Table \ref{tab:ChargedHypersLoci}.
Again, \eqref{eq:resolvedI22} is a resolved $I_2$ fiber in the cubic
presentation of $X_{n+1}$.

\subsubsection{Global resolution of the cubic fibration $X_{n+1}$}

Thus, the only codimension two singularities of $X_{n+1}$ that require  an additional  resolution are the conifold singularities in
\eqref{eq:conifolds}.
As shown in \cite{Esole:2011sm,Marsano:2011hv}, these are resolved by blowing up the ambient space $\mathbb{P}^2_B$ 
of  \eqref{eq:binomial}
defined in 
\eqref{eq:P2fibrationBn}
twice at $u=v_i=0$ and $f_2=v_j=0$, respectively. Introducing 
projective coordinates $[l_1:l_2]$, $[m_1:m_2]$  on $\mathbb{F}_0=\mathbb{P}^1\times \mathbb{P}^1$, these 
two blow-ups are described by the complete intersection \cite{griffiths2014principles}
\beq \label{eq:2blowups}
	l_1u-l_2v_i=0\,,\qquad m_1f_u-m_2v_j=0\,.
\eeq
For these two constraints to be well-defined sections of line bundles, the coordinates on $\mathbb{F}_0$ have to be sections
of the following line bundles:
\beq  \label{eq:linebundles}
	l_1\in \mathcal{O}([v_i]+H_1)\,,\quad l_2\in  \mathcal{O}([u]+H_1)\,,\quad m_1\in \mathcal{O}([v_j]+H_2)\,,\quad m_2\in  \mathcal{O}([f_u]+H_2)\,.\
\eeq
Here $H_1$ and $H_2$ denote the hyperplane classes of the two
$\mathbb{P}^1$'s in $\mathbb{F}_0$ and the divisor class associated to
a section is denoted by $[\cdot]$ as before.  The assignment
\eqref{eq:linebundles} implies that $\mathbb{F}_{0}$ is fibered
non-trivially over $\mathbb{P}^2_B$. Thus, the ambient
space of the complete intersection
\eqref{eq:2blowups} is the following $\mathbb{F}_0$-bundle:
\beq \label{eq:F0fibrationOverP2B}
	\xymatrix{
	\mathbb{F}_{0} \ar[r] & 	\mathcal{W}=\mathbb{P}(\mathcal{O}\oplus \mathcal{O}([u]-[v_i])\oplus\mathcal{O}\oplus \mathcal{O}([f_u]-[v_j])\,.\ar[d]\\
	& \mathbb{P}^2_{B}\,
	}
\eeq
Here the projectivization acts on the fiber coordinates in the four line bundle summands as the $\mathbb{C}^*$-actions in 
$\mathbb{F}_0$. 

The proper transform of \eqref{eq:binomial} in the blow-up \eqref{eq:2blowups} is readily computed using prime ideals. It 
agrees with the naive result one obtains by multiplying  \eqref{eq:binomial} by $l_1m_1$, using the two 
relations \eqref{eq:2blowups} and factoring out $v_iv_j$. In summary, we obtain the complete intersection
\beq \label{eq:CICY}
	l_1u-l_2v_i=0\,,\qquad
	m_1f_u-m_2v_j=0\,,\qquad l_2m_2+v_kl_1m_1=0\,,\,\, k\neq i,j\,
\eeq
in the ambient space \eqref{eq:F0fibrationOverP2B} as the global resolution of \eqref{eq:binomial}. We denote this global 
resolution of $X_{n+1}$ as $\hat{X}_{n+1}$ in the following. A complete intersection
resolution as \eqref{eq:CICY} has also been used recently in \cite{Cvetic:2015moa}
for the study of the $\mathbb{Z}_3$ Tate-Shafarevich group.

The Calabi-Yau condition of the complete intersection \eqref{eq:CICY} is readily checked as follows. The first Chern class
of \eqref{eq:F0fibrationOverP2B} is computed as the sum of the first Chern class $c_1(\mathbb{P}^2_{B})$ of its base 
$\mathbb{P}^2_{B}$ and the four fiber line bundles,
\bea
	c_1(\mathcal{W})&=&c_1(\mathbb{P}^2_{B})+[u]+[f_u]-[v_i]-[v_j]+2(H_1+H_2)\nn\\
	&=&2([u]+[f_u]+H_1+H_2)-[v_i]-[v_j]\,,
\eea
where we have imposed the condition $c_1(\mathbb{P}^2_{B})=[u]+[f_u]$ which is the Calabi-Yau condition of \eqref{eq:binomial} in the ambient space \eqref{eq:P2fibrationBn} before resolution. Employing \eqref{eq:linebundles}, we see that this is precisely the sum of the classes of the three constraints
in \eqref{eq:CICY}, {\it i.e.}~the complete intersection is indeed  Calabi-Yau.

Next, we compute the coordinates of the rational points $P$, $Q$ and $R$ in the resolution $\hat{X}_{n+1}$. 
By plugging in their coordinates \eqref{eq:coordsPQR} into  the complete intersection \eqref{eq:CICY} we obtain
their new coordinates in $[u:v:w:l_1:l_2:m_1:m_2]$ as
\bea \label{eq:coordsPQRCICY}
	&P:&[0:-b_1:a_1:f_u(0,-b_1,a_1) :\Delta_{k1}\Delta_{1j}:\Delta_{j1}:f_u(0,-b_1:a_1)]\nn\\
	&Q:&[0:-b_2:a_2:f_u(0,-b_2,a_2):\Delta_{k2}\Delta_{2j}:\Delta_{j2}:f_u(0,-b_2,a_2)]\nn\\
	&R:&[0:-b_3:a_3:f_u(0,-b_3,a_3):\Delta_{k3}\Delta_{3j}:\Delta_{j3}:f_u(0,-b_3,a_3)]
\eea
where we used the shorthand notation $\Delta_{ij}=a_ib_j-b_i a_j$ defined in \eqref{eq:conifoldLoci}.

We note that we have obtained different resolutions, corresponding to the different choices of the 
labels $(i,j,k)$ in \eqref{eq:CICY}.  For the purpose of computing the charged matter spectrum of F-theory,
we only need one particular resolution of $X_{n+1}$. For convenience we choose  $(i,j,k)=(1,2,3)$ in 
\eqref{eq:CICY}  yielding the resolution
\beq \label{eq:CYCI123}
	l_1u-l_2v_1=0\,,\qquad m_1 f_u-m_2v_2=0\,,\qquad l_2m_2+v_3l_1m_1=0
\eeq
and the coordinates  \eqref{eq:coordsPQRCICY}
\bea \label{eq:coordsPQRCICY123}
	&P=[0:-b_1:a_1:f_u(0,-b_1,a_1) :\Delta_{31}\Delta_{12}:\Delta_{21}:f_u(0,-b_1:a_1)]\,,&\\ &Q=[0:-b_2:a_2:1:0:0:1]
\,,\quad R=[0:-b_3:a_3:1:0:\Delta_{23}:f_u(0,-b_3,a_3)]\,.&\nn
\eea
For latest results on the geometrical construction of  different geometrical resolution phases, we refer {\it e.g.}~to the 
recent works \cite{Esole:2014hya,Braun:2014kla}.

We emphasize  that the conifold singularities \eqref{eq:conifolds} are replaced by $\mathbb{P}^1$'s, which are wrapped by 
rational points. The particular wrapping and the $\mathbb{P}^1$ inside $\mathbb{F}_0$ constituting the exceptional curve in $\hat{X}_{n+1}$ can be inferred from 
\eqref{eq:CYCI123} and the coordinates \eqref{eq:coordsPQRCICY123} of the rational points. 
For example, at the original conifold locus $u=f_u=v_1=v_2=0$ in \eqref{eq:conifolds}
the first and second equation in \eqref{eq:CYCI123} are trivial so that $(l_1,l_2)$ 
and $(m_1,m_2)$ remain unconstrained. The third equation, that is of degree $(1,1)$ determines 
a diagonal $\mathbb{P}^1$ in $\mathbb{F}_0$ replacing the singularity. By \eqref{eq:coordsPQRCICY123} we see that it is 
wrapped by $P$, whose coordinates are ill-defined as 
$\Delta_{12}=f_u(0,-a_1,b_1)=0$.

We conclude the discussion of the resolution by noting that by Theorem 4.1 in  \cite{Esole:2014hya} the complete intersection \eqref{eq:CICY} is 
equivalent to 
\beq \label{eq:blowupatQR}
l_1u-l_2v_i=0\,,\qquad
	 m_1v_k-m_2l_2=0\,,\qquad m_1f_u+l_1m_2v_j=0\,,\,\, k\neq i,j\,.
\eeq
This describes two blow-ups, one at $u=v_i=0$, which is precisely the location of one rational point in $X_{n+1}$, and another 
at $v_k=l_2=0$, which is exactly the proper transform of a second rational point. The latter statement is clear since the first 
constraint in \eqref{eq:CICY} implies 
that the set $\{l_2=v_k=0\}$  is equivalent to $\{u=v_k=0\}$ (except when $v_k\equiv v_i=0$ or $v_k\equiv v_j=0$ and 
$f_u\vert_{u=v_k=0}=0$, where $l_2$ is unconstrained as we are at one conifold singularity in \eqref{eq:conifolds}). Thus, we 
see that the resolution  \eqref{eq:CICY} is nothing but an equivalent description of the blow-up at two rational points in 
$X_{n+1}$ given in \eqref{eq:blowupatQR}.

\subsection{ The matter spectrum}
\label{sec:Matter}

In this section, we proceed to investigate the codimension two fibers in the resolution $\hat{X}_{n+1}$
constructed above. The charges of the hypermultiplets
related to the isolated curves of the $I_2$ fibers are calculated in general from
the intersection with the Shioda map of the sections,
$\sigma(\hat{s}_Q)$ and $\sigma(\hat{s}_R)$, as
\cite{Park:2011ji,Morrison:2012ei} 
\beq
 q_{1,2} = c \cdot \sigma(\hat{s}_{Q,R})=c\cdot(S_{Q,R}-S_{P})\,,
\label{eq:Shioda-intersection}
\eeq
where $S_{P,Q,R}$ denotes the divisor class of the section $\hat{s}_{P,Q,R}$, respectively.
We employ this formula and the particular resolution phase in \eqref{eq:CYCI123} to compute the charged
matter spectrum of the F-theory compactification on $X_{n+1}$.

\subsubsection{General model}

We begin by analyzing the resolution $\hat{X}_{n+1}$ of the elliptic fibration of the general model
 \eqref{eq:cubicfactorized} at the nine codimension
two loci in Table \ref{tab:ChargedHypersLoci}. In all cases, we obtain
$I_2$ fibers, as discussed above, that enjoy different intersections
with the rational sections $\hat{s}_{Q}$, $\hat{s}_R$ and the zero section $\hat{s}_P$. 
The resulting
matter spectrum including 6D multiplicities $x_{(q_1,q_2)}$ of
hypermultiplets with charges $(q_1,q_2)$ and a representation of the
$I_2$ fibers in $\hat{X}_{n+1}$ in the phase \eqref{eq:CYCI123} is shown in
Table  \ref{tab:MultiesGenModel}. The behaviour of the fiber is obtained by
analyzing the complete intersection presentation \eqref{eq:CYCI123} for $\hat{X}_{n+1}$
at all loci in Table \ref{tab:ChargedHypersLoci}.
In the figure, fiber components completely contained in one of the
rational sections are shaded.  We note that all fibers we find  are compatible with the survey of possible $I_2$ fibers in 
\cite{Lawrie:2015hia}.

The charges can be
read off from the location of the rational sections on the fiber
and (\ref{eq:Shioda-intersection}), noting that the U(1) generators are associated
with $Q-P$ and $R-P$.  The fiber component not containing the zero point $P$ is the
curve $c$ whose charges we compute.
For example, in the first entry in Table
  \ref{tab:MultiesGenModel},
the curve $c$ intersects twice with $P$ and not at all with $Q, R$, so
the charges are $-2, -2$. 
\begin{table}
\begin{tabular}{|c|c|c|} \hline
Charges &Multiplicity  & Fiber  \rule{0pt}{14pt} \\ \hline
$(-2,-2)$ &   $x_{(-2,-2)}=[a_1]\cdot([a_1]+[s_8]+K_B)$ \rule{0pt}{14pt} & \rule{0pt}{1.3cm}\parbox[c]{2.8cm}{\includegraphics[scale=0.75]{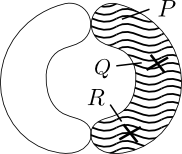}} \\[.95cm] \hline
$(2,0)$ &  $x_{(2,0)}=[a_2]\cdot ([a_2]+[s_8]+K_B)$ \rule{0pt}{14pt}& \rule{0pt}{1.3cm}\parbox[c]{2.8cm}{\includegraphics[scale=0.75]{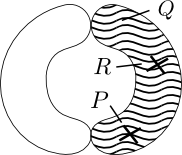}} \\[.95cm] \hline
$(0,2)$ &  $x_{(0,2)}=[a_3]\cdot ([a_3]+[s_8]+K_B)$ \rule{0pt}{14pt} & \rule{0pt}{1.3cm}\parbox[c]{2.8cm}{\includegraphics[scale=0.75]{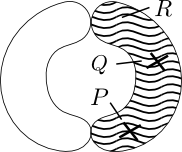}} \\[.95cm] \hline
$(-2,-1)$ & $x_{(-2,-1)}=[s_8]\cdot  ([s_8]+[a_1]+[a_2]+K_B)+2[a_1] \cdot [a_2]$ \rule{0pt}{14pt} & \rule{0pt}{1.3cm}\parbox[c]{2.8cm}{\includegraphics[scale=0.75]{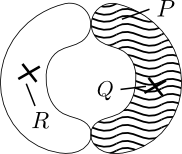}} \\[.95cm] \hline
$(-1,-2)$ & $x_{(-1,-2)}=[s_8]\cdot  ([s_8]+[a_1]+[a_3]+K_B)+2[a_1]\cdot [a_3]$  \rule{0pt}{14pt} & \rule{0pt}{1.3cm}\parbox[c]{2.8cm}{\includegraphics[scale=0.75]{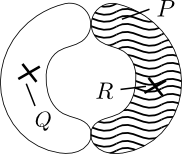}} \\[.95cm]  \hline
$(-1,1)$ & $x_{(-1,1)}=[s_8]\cdot  ([s_8]+[a_2]+[a_3]+K_B)+2[a_2] \cdot  [a_3] $ \rule{0pt}{14pt}& \rule{0pt}{1.3cm}\parbox[c]{2.8cm}{\includegraphics[scale=0.75]{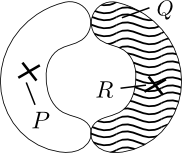}}  \\[.95cm] \hline
$(1,1)$ & $\begin{array}{c}x_{(1,1)}=-K_B\cdot  (16 [a_1]+9 [s_8])-4 [a_1] \cdot ([a_1]+[a_2])\\
 +2  [a_3]\cdot ([a_2]-2 [a_1])-(8 [a_1]+[a_2]+[a_3])\cdot  [s_8]-3 [s_8]^2\end{array}$ \rule{0pt}{20pt} & \rule{0pt}{1.3cm}\parbox[c]{2.8cm}{\includegraphics[scale=0.75]{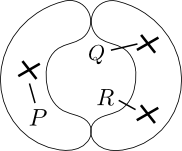}} \\[.95cm] \hline
$(1,0)$ & $\begin{array}{c}x_{(1,0)}= -K_B\cdot (16 [a_2]+9 [s_8])-4 [a_2] \cdot ([a_2]+[a_3])\\ 
+2 [a_1]\cdot ([a_3]-2 [a_2])-([a_1]+8 [a_2]+[a_3])\cdot  [s_8]-3 [s_8]^2 \end{array}$ \rule{0pt}{20pt} & \rule{0pt}{1.3cm}\parbox[c]{2.8cm}{\includegraphics[scale=0.75]{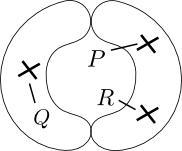}} \\[.95cm] \hline
$(0,1)$ & $\begin{array}{c} x_{(0,1)}=  -K_B\cdot (16 [a_3]+9 [s_8])-4 [a_3]\cdot  ([a_2]+[a_3])\\ +2 [a_1] \cdot ([a_2]-2 [a_3])-([a_1]+[a_2]+8 [a_3]) \cdot [s_8]-3 [s_8]^2\end{array}$ \rule{0pt}{20pt} & \rule{0pt}{1.3cm}\parbox[c]{2.8cm}{\includegraphics[scale=0.75]{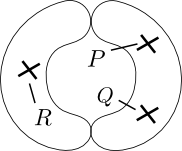}} \\[.95cm]
 \hline
\end{tabular}
\caption{Matter spectrum and corresponding $I_2$-fibers in $\hat{X}_{n+1}$.}
\label{tab:MultiesGenModel}
\end{table}
We note that the matter spectrum is completely symmetric under
exchange of the three points $P$, $Q$ and $R$ as expected by the
symmetries of $X_{n+1}$ in the form \eqref{eq:cubicfactorized}.
(Note that this symmetry is not manifest in
Table \ref{tab:ChargedHypersLoci}, though it is still there and can be seen
by a different choice of generators of the respective ideals.)

The representation content in Table \ref{tab:MultiesGenModel} can be
conveniently summarized by drawing the charge lattice of the
theory. Since a 6D hypermultiplet contains fields both in a
representation and its complex conjugate, we draw all the charges in
Table \ref{tab:MultiesGenModel} including their negative, yielding
precisely 
the charge lattice that was depicted in Figure
\ref{f:charges-general}.
Thus, all the generic models have spectra that appear compatible with
a Higgsed $\text{SU}(2)  \times \text{SU}(2) \times \text{SU}(3)$ model, with symmetric
$\text{SU}(3)$ matter included; we see in the
following section how this works in detail.

The multiplicities of the matter representations  in Table \ref{tab:MultiesGenModel} are given by the number of
points in the base $B_{n}$ supporting a given $I_2$ fiber in $\hat{X}_{n+1}$.
The multiplicities of the hypermultiplets in the first three lines of Table \ref{tab:MultiesGenModel} 
readily follow from 
Table \ref{tab:ChargedHypersLoci} and the line 
bundles in \eqref{eq:coeffsSectionsGen}. The multiplicities in the fourth to sixth lines are computed from 
Table \ref{tab:ChargedHypersLoci} starting with the corresponding reducible complete intersection and subtracting the 
respective irreducible component $a_i=b_i=0$ which is contained in it 
with the appropriate multiplicity, that is calculated to be two using the resultant
(see \cite{Cvetic:2013nia} for more details on this method). For example, for the matter
with charge $(-2,-1)$, we compute the resultant of $\Delta_{12}$ and 
$s_3b_1^2-s_6a_1b_1+s_8a_1^2$ w.r.t.~$a_1$. It has a factor of $b_1^2$, indicating that 
$a_1=b_1=0$ is a zero of order two as claimed. 
Thus we obtain
\bea
	x_{(-2,-1)}&=&(2[a_1]+[s_8])\cdot([a_1]+[a_2]+[s_8]-[K_B^{-1}])-2x_{(2,2)} \,,\\
	x_{(-1,-2)}&=&(2[a_3]+[s_8])\cdot([a_1]+[a_3]+[s_8]-[K_B^{-1}])-2x_{(0,2)}\,,\\  x_{(-1,1)}&=&(2[a_2]+[s_8])\cdot([a_2]+[a_3]+[s_8]-[K_B^{-1}])-2x_{(2,0)} \,,
\eea
which precisely reproduces the fourth to sixth line of Table \ref{tab:MultiesGenModel} employing \eqref{eq:coeffsSectionsGen}.

For the computation of the multiplicity $x_{(1,1)}$ of the
hypermultiplets with charges $(1,1)$, we instead use the ideal (4.19)
in \cite{Cvetic:2013nia}, that is $\langle g^{\prime}_{9}, \delta g_6
\rangle$, with the coefficients $s_i\rightarrow \tilde{s}_i$ according
to \eqref{eq:sTransf}.\footnote{It is advantageous to use this ideal because it
  already has some components subtracted.}  We calculate $x_{(1,1)}$
by subtracting unwanted irreducible components with their appropriate
multiplicities. These are computed using the resultant as outlined before. Due to computational limitations 
we have to use random numbers for those coefficients $s_i$, $a_j$, $b_j$ which are not the 
variables on which the resultant depends.
We obtain
\bea \label{eq:mult11}
 x_{(1,1)} &=& ([g_9^\prime]\cdot[ \delta g_6 ])_{s\rightarrow \tilde{s}} - 2 x_{(2,0)} - 8 x_{(-2,-1)} - 4 x_{(-1,-2)} - 20 x_{(-2,-2)}\,, \\
 &=& ( [a_1^4 a_2^2 s_8^3])\cdot  ( [a_1^4 a_2 b_3 s_8^2 ] )- 2 x_{(2,0)} - 8 x_{(-2,-1)} - 4 x_{(-1,-2)} - 20 x_{(-2,-2)}\,, \nn
\eea
which yields precisely line seven in Table \ref{tab:MultiesGenModel} using \eqref{eq:coeffsSectionsGen}.

Finally, the multiplicities of the hypermultiplets with charges
$(1,0)$ and $(0,1)$ are calculated from the ideals in the last two
lines of Table \ref{tab:MultiesGenModel} after subtracting, with the
right degrees, the components corresponding to the other charged
hypermultiplets.  The multiplicities of the hypermultiplets with
charges $(1,0)$ and $(0,1)$ are
\beq \label{eq:mult10}
 x_{(1,0)} = [y_Q]\cdot[ z_Q^4f+3x^2_Q ] - 16 x_{(2,0)} - 16 x_{(-2,-1)} -   x_{(-1,-2)} - 16 x_{(-2,-2)} - x_{(-1,1)} - x_{(1,1)} \,
\eeq
and 
\beq \label{eq:mult01}
 x_{(0,1)} = [y_R]\cdot[ z_R^4 f+3x^2_R ]  -  x_{(-2,-1)} - 16  x_{(0,2)} - 16 x_{(-1,-2)} - 16 x_{(-2,-2)} - x_{(-1,1)} - x_{(1,1)} \,, 
\eeq
where the latter can also be obtained from $x_{(1,0)}$ using the symmetry $Q \leftrightarrow R$, as expected.
Employing  \eqref{eq:coeffsSectionsGen}, we obtain the last two lines of Table \ref{tab:MultiesGenModel}.

We conclude by discussing anomaly cancellation in 6D.  Using the
charges and their respective multiplicities in Table
\ref{tab:MultiesGenModel}, we proceed to calculate the anomalies
of the Abelian theory following \cite{Erler:1993zy,Park:2011wv} (we omit the details of this calculation). 
The only piece of information missing are 
the coefficients $b_{mn}$ of the Green-Schwarz counterterms. They are calculated by the Neron-Tate height pairing as 
\beq
b_{mn} = -\pi( \sigma(\hat{s}_m) \cdot \sigma(\hat{s}_n) )
\eeq
where $\sigma$ is the Shioda map which maps sections to elements in 
$H^{(1,1)}(\hat{X}_{n+1})$ \cite{shioda1990mordell,Park:2011ji,Morrison:2012ei}, and the map $\pi$ is 
the projection operator to homology $H^{(1,1)}(B_n)$ of the base.
The Shioda map for a section $\hat{s}_m$ with divisor class $S_m$ reads  explicitly
\beq
\sigma(\hat{s}_m) = S_m - S_P - [K_B^{-1}] - \pi(S_m\cdot S_P  )\,,
\eeq
thus, the Green-Schwarz counterterms are
\beq \label{eq:bmnCoeff}
 b_{mn} =
 \left( \begin{array}{cc}
2( \pi(S_Q \cdot S_P)  - K_B) & \!\!\!\!\pi(S_Q \cdot S_P + S_R \cdot S_P - S_Q \cdot S_R) - K_B \\
\!\!\pi(S_Q \cdot S_P + S_R \cdot S_P - S_Q \cdot S_R)  - K_B & 2( \pi(S_R \cdot S_P)  - K_B)\!\!  \end{array} \right)_{mn}
\eeq
The divisor classes  $\pi(S_m \cdot S_n)$ are precisely the classes of the respective constraint
in \eqref{eq:collisionOfPoints}, which are just the classes of $\Delta_{ij}$ defined in
\eqref{eq:Deltaij}. Using \eqref{eq:coeffsSectionsGen}, we obtain the following Neron-Tate height pairing matrix:
\bea \label{eq:bmnCoeff2}
 b_{mn} &=& 
 \left( \begin{array}{cc}
2( [a_1]+[b_2] + [s_8]) & 2[a_1]+[s_8]  \\
2 [a_1]+[s_8]  &  2( [a_1]+[a_3] + [s_8])  \end{array} \right)_{mn} \,.
\eea
Finally, using the full spectrum and the coefficients $b_{mn}$, we confirm that all 
anomalies are canceled by a generalized Green-Schwarz mechanism in 6D (refer {\it e.g.}~to 
\cite{Erler:1993zy,Park:2011wv}  for a review on Abelian anomalies).

\subsubsection{Specialized model}

Next, we turn to the specialized model defined by \eqref{eq:abcdModel}.
Its $I_2$ fibers and 
spectrum of charged matter fields are obtained directly from Table \ref{tab:ChargedHypersLoci} and
Table \ref{tab:MultiesGenModel}, respectively, by setting $a_1=1$,
$b_1=0$. We obtain
the results summarized in Table~\ref{tab:Multies_Simpl}.
\begin{table}[ht!]
\begin{center}
\begin{tabular}{|c|c|} \hline
Charges &Multiplicity    \rule{0pt}{14pt} \\ \hline
$(2,0)$ &  $x_{(2,0)}=[a_2]\cdot ([a_2]+[s_8]+K_B)$ \rule{0pt}{14pt} \\ \hline
$(0,2)$ &  $x_{(0,2)}=[a_3]\cdot ([a_3]+[s_8]+K_B)$ \rule{0pt}{14pt} \\ \hline
$(-2,-1)$ & $x_{(-2,-1)}=[s_8]\cdot  ([a_2]+[s_8]+K_B)$ \rule{0pt}{14pt} \\ \hline
$(-1,-2)$ & $x_{(-1,-2)}=[s_8]\cdot  ([a_3]+[s_8]+K_B)$  \rule{0pt}{14pt} \\  \hline
$(-1,1)$ & $x_{(-1,1)}=[s_8]\cdot  ([a_3]+[s_8]+K_B)+[a_2] \cdot (2 [a_3]+[s_8]) $ \rule{0pt}{14pt} \\ \hline
$(1,1)$ & $\begin{array}{c}x_{(1,1)}=-9K_B\cdot   [s_8]+2  [a_2]\cdot [a_3]\\
 -([a_2]+[a_3])\cdot  [s_8]-3 [s_8]^2\end{array}$ \rule{0pt}{20pt} \\ \hline
$(1,0)$ & $\begin{array}{c}x_{(1,0)}= -K_B\cdot (16 [a_2]+9 [s_8])-4 [a_2] \cdot ([a_2]+[a_3])\\ 
-(8 [a_2]+[a_3])\cdot  [s_8]-3 [s_8]^2 \end{array}$ \rule{0pt}{20pt} \\ \hline
$(0,1)$ & $\begin{array}{c} x_{(0,1)}=  -K_B\cdot (16 [a_3]+9 [s_8])-4 [a_3]\cdot  ([a_2]+[a_3])\\ -([a_2]+8 [a_3]) \cdot [s_8]-3 [s_8]^2\end{array}$ \rule{0pt}{20pt} \\
 \hline
\end{tabular}
\caption{Matter spectrum of the simplified model $\hat{X}_{n+1}$ with $a_1=1$ and $b_1=0$.}
\label{tab:Multies_Simpl}
\end{center}
\end{table}
We note that the matter fields with charge $(-2,-2)$ in Table \ref{tab:MultiesGenModel} are not
present due to $[a_1]=0$.
Again, we can summarize the spectrum in Table \ref{tab:Multies_Simpl} by drawing the charge lattice of the 
theory, giving the spectrum shown in Figure \ref{f:223-weights},
corresponding to a Higgsed $\text{SU}(2) \times \text{SU}(2) \times
\text{SU}(3)$ model without symmetric $\text{SU}(3)$ matter.

Anomaly cancellation can be readily checked using the spectrum in Table \ref{tab:Multies_Simpl} and
using the anomaly coefficients $b_{mn}$ in \eqref{eq:bmnCoeff} for $[a_1]=0$.

\subsection{Comparison with matter in $dP_2$-elliptic fibrations}
\label{sec:dP2comparisondMatter}

We conclude by continuing the discussion of Section \ref{sec:dP2comparison} on the connection between the models $X_{n+1}$  and $dP_2$-elliptic fibrations, now at the level of the matter spectrum.

\begin{figure}[ht]
\centering
 \includegraphics[scale=0.8]{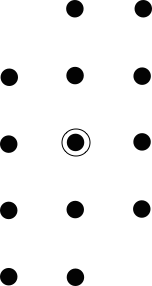}
\caption{Weight lattice of charges of the 
U$(1)\times$U(1) theory obtained from $dP_2$-elliptic fibrations. 
The origin of the lattice is indicated by a circle.
}
\label{f:charges-dp2}
\end{figure}

First, we recall that $a_1$ and $b_2$ have to be non-vanishing constants in a $dP_2$-elliptic fibration. Thus,  
the representations with charges $(2,0)$, $(-2,-1)$ and $(-2,-2)$ in Table \ref{tab:MultiesGenModel} are not realized. 
These are the representations related to the collision of the sections $\hat{s}_Q$ and $\hat{s}_P$, which, however, is 
impossible for an elliptic fiber in $dP_2$. Drawing all the charges in a two-dimensional  
lattice yields Figure \ref{f:charges-dp2}. A quick glance at this picture makes it clear why a 
complete unHiggsing of the $dP_2$-model to the gauge groups $\text{SU}(2) \times \text{SU}(2)$ and $\text{SU}(3)$ with more than one adjoint is not feasible: the representations $\pm(2,0)$ and $\pm(2,1)$, which come from the SU(2) or SU(3) adjoints, respectively, are not present.

As mentioned in section \ref{sec:dP2comparison}, there are
$dP_2$-models which are not accessible from specializations of $X_{n+1}$, as the effectiveness constraints in the latter models 
are more constraining than in the former. To see this, let us set the classes
$[a_1]=0$ and $[b_2]=0$ in \eqref{eq:coeffsSections2}. Once these
constraints are imposed, the effectiveness condition $[b_1]\ge 0$ and
$[a_2]\ge 0$ automatically forces the constraints $[s_8] = -K_B$,
$[a_2]=0$ and $[b_1]=0$, restricting the set of possible models to a
one-parameter family parametrized by the degree of $[a_3]$.
However, it has been shown in 
\cite{Cvetic:2013nia,Cvetic:2013uta,Borchmann:2013hta,Cvetic:2013jta} that a $dP_2$-elliptic 
fibration is parametrized by two divisor classes. Thus, we only obtain a subset of models by 
specializing $X_{n+1}$. These models, however, admit unHiggsings to $\text{SU}(3)\times \text{SU}(2)$ on $s_8=0$ and $a_3=0$, respectively, along the lines of Section \ref{sec:UnhiggsingGeometry}.

\begin{figure}[ht!]
\centering
\includegraphics[scale=0.4]{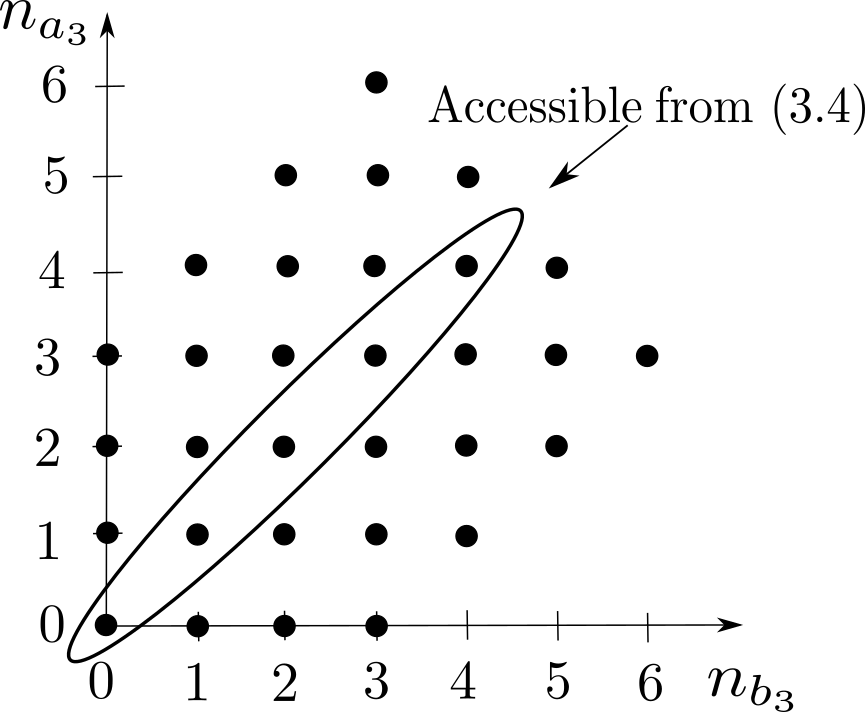}
\caption{Region of allowed $dP_2$ fibrations over $\mathbb{P}^2$. The set of models obtainable by specializing
\eqref{eq:cubicfactorized} is given by the five encircled points.}
\label{fig:dP2Region}
\end{figure}

More concretely, specific examples of $dP_2$-fibrations over 
$\mathbb{P}^2$ and $\mathbb{P}^3$ 
were listed in Figure 2 in \cite{Cvetic:2013uta} and Table
2.3 in \cite{Borchmann:2013hta}. In those cases
it was shown that the complex structure moduli space is stratified, with integral points
in a two-dimensional polygon corresponding to the different strata.
Let us  compare the models obtained from $X_{n+1}$
through the specialization $[a_1]=0$ and $[b_2]=0$, with the models
obtained from $dP_2$-fibrations over $\mathbb{P}^2$. Imposing these
specializations in the Calabi-Yau manifolds $X_{n+1}$, we obtain only five different models 
labelled by the different values of $[a_3]$:
\beq\label{eq:P2withDP2spec}
\begin{tabular}{|c|c||c|c|c|c|c|c|c|c|} \hline
$[a_2]$ & $[a_3]$& $b_2$ & $b_3$ & $s_1$ & $s_2$ & $s_3$ & $s_5$ & $s_6$ & $s_8$ \rule{0pt}{1Em}\\ \hline
 0& 0& 0& 0& 9& 6& 3& 6& 3& 3 \\
 0& 1& 0& 1& 7& 5& 3& 5& 3& 3 \\
 0& 2& 0& 2& 5& 4& 3& 4& 3& 3 \\
 0& 3& 0& 3& 3& 3& 3& 3& 3& 3 \\
 0& 4& 0& 4& 1& 2& 3& 2& 3& 3 \\
 \hline
\end{tabular}
\eeq
In contrast, there are 31 $dP_2$-elliptic fibrations over $\mathbb{P}^2$. We
can draw all the models in a two-dimensional diagram, with axes given by the values 
of $n_{a_3}$ and $n_{b_3}$ where we 
expanded $[a_3]:=n_{a_3} H_B$ and $[b_3]=n_{b_3} H_B$ with $H_B$ the hyperplane in 
$\mathbb{P}^2$.\footnote{The 
parameters $s_7$ and $s_9$ in \cite{Cvetic:2013nia,Cvetic:2013uta} are identified with $a_3$, $b_3$ in $X_{n+1}$, respectively.} 
We obtain Figure \ref{fig:dP2Region}. As indicated here, only the subset
of models on the diagonal is described by models $X_{n+1}$ with $[a_1]=[b_2]=0$.

\section{UnHigssing two U$(1)$'s in F-theory }
\label{sec:UnhiggsingGeometry}

In this section, we discuss the unHiggsing of Abelian to 
non-Abelian gauge symmetries for the case of two U(1) factors.
We distinguish between rank-preserving unHiggsings to non-Abelian groups with adjoints and unHiggsing with rank 
enhancement to non-Abelian groups that do not necessarily have adjoints.  
In the latter case, there are situations when the additional 
non-Abelian groups are spectators to a  rank-preserving unHiggsing of the U(1)'s and other situations when the additional 
groups are involved in the unHiggsing of the Abelian theory.

We first discuss the general geometrical unHiggsing procedure in Section \ref{sec:Unhiggsing}. 
Then we apply this to the particular case of our Abelian model $X_{n+1}$ in Sections 
\ref{sec:TuningModel}, \ref{sec:unHiggsingBothU1s} and \ref{sec:specialunHiggsing} 
and discover an unHiggsing to a model with SU(2)$\times$SU(2)$\times$SU(3) gauge group 
and no U(1)'s. In Section 
\ref{sec:WSFunHiggsed} we analyze its Weierstrass form which we use in Section 
\ref{sec:Matter} to derive the full matter spectrum. We put special emphasis on the
emergence of the symmetric representation $\mathbf{6}$ of SU(3). In Section 
\ref{sec:NewWSFs} we relate the appearance of this new matter representation to an ordinary 
double point singularity on the SU(3) divisor which can not be deformed,

\subsection{UnHiggsing in the complex structure moduli space}
\label{sec:Unhiggsing}

We begin with a discussion of the general geometrical procedure underlying the unHiggsing of a U(1) gauge symmetry in 
F-theory.  We put special emphasis on rank-preserving unHiggsings because they have a clear  
geometric interpretation as a transition of a rational section to a Cartan divisor.

The unHiggsing process of an F-theory compactification with U$(1)^k$ gauge symmetry can be understood geometrically as
a transition from one Calabi-Yau manifold $X_{n+1}^{(k)}$ with a rank $k$ MW-group to another
Calabi-Yau manifold $X_{n+1}^{(k')}$ with a lower rank MW-group, $k'< k$. The manifold 
$X_{n+1}^{(k')}$ is reached via a 
tuning of the complex structure of $X_{n+1}^{(k)}$ so that two rational sections
are placed on top of each other,  {\it i.e.}~define maps from the base $B_n$ to the \textit{same} point in the elliptic fiber 
$\mathcal{E}$. In this process certain codimension two singularities, typically of Kodaira type $I_2$, 
that the fibration of $X_{n+1}^{(k)}$ exhibits as a consequence of the presence of U(1)'s  and that give rise to U$(1)$-charged 
matter in F-theory ({\it cf.}~Section \ref{sec:SingsResMatter}) are promoted to codimension one singularities of $X_{n+1}^{(k-1)}$.
Repeating such tunings of the complex structure until all sections are placed on top of each other we eventually obtain
a Calabi-Yau manifold $X_{n+1}^{(0)}$ with trivial MW-group,  {\it i.e.}~no abelian gauge group factors. 

The structure of the codimension one singularities of the elliptic
fibration of $X_{n+1}^{(0)}$ encodes the original Abelian gauge theory
of F-theory on $X_{n+1}^{(k)}$ in terms of a non-Abelian gauge theory
with gauge group $G$ and a specific matter spectrum. The case of a
single U(1) has been analyzed in \cite{Morrison:2014era} and is
reviewed in Section \ref{sec:HiggsingU1}. It has been shown that every
model with a U(1) can be unHiggsed to a
model with SU(2) or larger non-Abelian
gauge group, although in some cases, particularly
when there are additional non-Abelian factors present before the
unHiggsing, the unHiggsed model can develop singularities. 
UnHiggsing of toric models with up to
three U(1)'s to $(\text{SU}(2)^2\times\text{SU}(4))/\mathbb{Z}_2$ and
$\text{SU(3)}^3/\mathbb{Z}_3$ have been found in
\cite{Klevers:2014bqa}. In general, one naturally expects that the
minimal rank of the gauge group $G$ is at least $k$, so that by
Higgsing {\it e.g.}~via adjoints we can recover the full Abelian gauge group
$\text{U}(1)^k$. However, for some abelian theories the rank of $G$
has to be larger depending on the matter spectrum of the original
Abelian theory. Theories with a sufficiently simple spectrum unHiggs
to a group $G$ of rank $k$, where the Higgs is in the adjoint, whereas
theories with a more complex spectrum including matter of higher
charges require $G$ of rank greater than $k$ and Higgses in other
representations, {\it e.g.}~the (bi-)fundamental.

The preceding discussion is general and thus should apply to elliptic
fibrations with an arbitrary number of rational sections.  For the
rest of this section, however, we work explicitly in the specific
context of the three section fibration $X_{n+1}$ defined by 
\eqref{eq:cubicfactorized}, \textit{i.e.}~we identify $X_{n+1}^{(2)}\equiv X_{n+1}$.
We establish the following picture for unHiggsing F-theory models with
U$(1)^2$ gauge symmetry obtained from these general elliptic fibrations
$X_{n+1}$.  In agreement with the expectation from
field theory, {\it cf.}~Section \ref{sec:U1sandUnHiggsing}, we
demonstrate that, indeed, for sufficiently simple Abelian models the
minimal rank of $G$ is $k=2$.  This leads to two natural classes of
models that admit a rank-preserving unHiggsing of $\text{U}(1)^2$
either to $\text{SU}(2)\times\text{SU}(2)$ or to $\text{SU}(3)$ on
base divisors supporting adjoints, {\it i.e.}~divisors in classes of
the form $-K_B+Z$ for an effective divisor $Z$ in $B_{n}$.  There are,
however, many models with U$(1)^2$ gauge group that due to the
complexity of their matter spectrum require the introduction of
additional non-Abelian groups in the unHiggsed geometry; the resulting
non-Abelian models, typically have rank-reducing Higgses in
(bi-)fundamental representations.

We demonstrate furthermore, in accord with the discussion of
Section \ref{sec:U1sandUnHiggsing}, that the non-Abelian gauge symmetry
underlying
a large class of 
models with U$(1)^2$ gauge symmetry realized via $X^{(2)}_{n+1}$ via a complete unHiggsing 
is given by
\beq \label{eq:unHiggsedGG}
	G_{\text{uni}}=\text{SU}(2)\times\text{SU}(2)\times\text{SU}(3)\,.
\eeq
The minimal non-Abelian gauge group can be even larger, as in the case
of a single U(1), in degenerate cases where the non-Abelian factors
can be thought of as living on reducible divisors.  We also show that
in some cases the unHiggsing is not unique and multiple different
gauge groups $G$ can be obtained, some of which can exceed
$G_{\text{uni}}$ even if others do not.

\subsection{Geometrical unHiggsing of one of the U(1)'s}
\label{sec:TuningModel}

Following the discussion of the previous subsection, we  wish to
unHiggs the
$\text{U}(1)^2$ gauge group of F-theory compactified on
$X_{n+1}$ by reducing its Mordell-Weil group through an
appropriate tuning of its complex structure.

\begin{figure}[ht]
\centering
\includegraphics[scale=0.35]{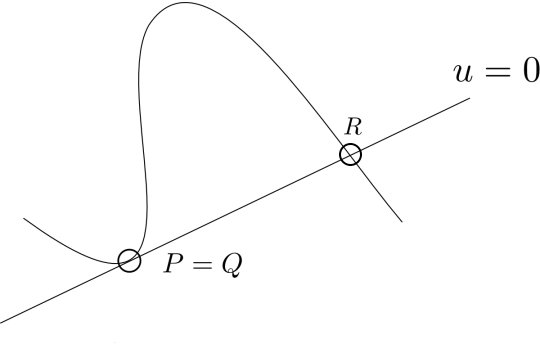}
\caption{Tuned elliptic curve $\mathcal{E}$ with $P=Q$ yielding an elliptic fibration with rank one Mordell-Weil 
group.}
\label{fig:TuningP=Q}
\end{figure}

Given the elliptically fibered CY-manifold $X_{n+1}$ in the form
\eqref{eq:cubicfactorized}, we recall that two of its rational
sections coincide if the coefficients $a_i$, $b_i$ are chosen so that
one of the constraints in the first line of
\eqref{eq:collisionOfPoints} is obeyed.  For instance, in order to
achieve $\hat{s}_P=\hat{s}_Q$ we have to demand $P=Q$ in the elliptic
fiber $\mathcal{E}$ as shown in Figure \ref{fig:TuningP=Q}.
This is achieved by choosing the complex structure of $X_{n+1}$ such that
\beq \label{eq:Delta12=0}
	\Delta_{12}=a_1b_2-a_2b_1\equiv 0\,,
\eeq
where we have used the definition \eqref{eq:Deltaij}.

There are a number of ways in which (\ref{eq:Delta12=0}) may be
realized.  We list some of the main possibilities:
\vspace*{0.1in}

\noindent
{\bf A)} One simple way in which  (\ref{eq:Delta12=0})
may be satisfied is if the divisor class $[a_2]-[a_1]$ is effective.
In this case, we can introduce a section $\lambda_1$ of the 
line bundle
\beq \label{eq:classLambda1}
	\lambda_1\in \mathcal{O}([a_2]-[a_1])\,,
\eeq
and impose the condition
\beq \label{eq:tuningSU2}
	(a_2v+b_2w)\stackrel{!}{=}\lambda_1 (a_1v+b_1w)\,.
\eeq   
We then have $a_2 = \lambda_1a_1, b_2 = \lambda_1 b_1$ and
(\ref{eq:Delta12=0}) is satisfied.  This possibility is available in
many of the simplest cases, such as when $[a_1] = 0$ (then we can also shift $v$ so that $b_1=0$; the solution to \eqref{eq:Delta12=0} is then $b_2$=0), or for example
on the base $\P^2$, where all divisors are multiples of the line $H_B$
so we can simply order $[a_2] >[a_1]$.  As we show explicitly in the
next section, this choice of tuning leads to an unHiggsing of the
U(1) associated to $\hat{s}_Q$ to a non-Abelian group
SU(2)$\times$SU(2).
This leads
directly to a general unHiggsing of the two U(1) model that matches
nicely with the field-theoretic discussion of
\S\ref{sec:U1sandUnHiggsing}

\vspace*{0.1in}

\noindent
{\bf B)} In some situations, neither $[a_2]-[a_1]$
or $[a_1]-[a_2]$ is effective, but the ring of functions/sections is a
UFD, so that we can solve 
 (\ref{eq:Delta12=0}) by taking
\begin{equation}
a_1 = AB, \;
b_2 = CD, \;
a_2 = AC, \;
b_1 = BD \,.
\label{eq:UFD}
\end{equation}
In this situation we again get an unHiggsing of one U(1) to SU(2)$\times$SU(2),
on the divisors $B, C$.  We do not explore the details of this
unHiggsing here.

\vspace*{0.1in}

\noindent
{\bf C)} Even if the ring is not a UFD, there can be situations where
 (\ref{eq:Delta12=0}) can again be solved for nonzero functions $a_1,
a_2, b_1, b_2$.
This unHiggsing gives only a single SU(2)
on $f_u(0,-b_1,a_1)=0$  with $f_u$ defined in \eqref{eq:fuv1v2v3}.

\vspace*{0.1in}

\noindent
{\bf D)} Finally, there are situations where there are no solutions of
 (\ref{eq:Delta12=0}) for  $a_1,
a_2, b_1, b_2$
all nonzero.  For example, if the base is $dP_3$, where the cone of
effective curves is generated by three exceptional curves $E_1, E_2, E_3$
from blowing up three points on $\P^2$, and three lines $L_{12},
L_{23}, L_{13}$ that connect the points, we may have
$[a_1] = E_1,[b_2] = L_{12},[a_2] = L_{23},[b_1] = E_3$.  In this case
even though the divisor classes are equivalent, $[a_1 b_2] =[a_2b_1]$,
there is no way to solve  (\ref{eq:Delta12=0}), since {\it e.g.} $a_1$
must vanish on the rigid curve $E_1$, but $a_2, b_1$ cannot.

Even in this type of
situation, however, there is a solution of
 (\ref{eq:Delta12=0}), which is in fact available for all possible
bases and choices of $[a_i],[b_i]$.  We can set {\it e.g.}
$b_2 = b_1 = 0$, which immediately leads to a solution of
 (\ref{eq:Delta12=0}).  While this tuning can be carried out in any of
the possible $\text{U}(1)^2$ models, even when the tuning of type (A) above
is not available, we can relate this to a special case of the tuning
(A) by taking the tuning in two  steps.  First, we can tune $b_1 =
0$.  This sets $v_1 = a_1 v$, so that the equation
\begin{equation}
uf_u = a_1 v v_2 v_3
\label{eq:b1=0}
\end{equation}
develops  an $I_2$ singularity on the locus $a_1 = 0$.  We may then,
however, move the factor of $a_1$ into $v_2$, defining a new model with $\tilde{a}_2 =
a_1 a_2$, $\tilde{b_2} = a_1 b_2$ and $\tilde{a}_1=1$.  This then becomes a special case
of a geometry where (A) is available, since $\tilde{a}_1 = 1$, though
the UnHiggsing through (A) leads to an extra SU(2) factor since $\tilde{a}_2$ is a
 reducible polynomial.
\vspace*{0.2in}

In the remainder of this section we focus on approach (A) to
unHiggsing, since not only is it relevant in the clearest and simplest
cases, but through the mechanism just mentioned it also covers the 
completely general case (D), albeit at the cost of producing an
additional SU(2) factor due to reduciblity of $\tilde{a}_2$.

\subsection{Unhiggsing both U(1)'s}
\label{sec:unHiggsingBothU1s}

We now describe the process of unHiggsing both U(1) factors in a
situation where approach (A) can be done in both cases.  This is the
situation, for example, on the base $\P^2$, or if one of the $a_i$ is
in the trivial class $[a_1] = 0$.  More generally, it is possible to
do this whenever $[a_2]-[a_1]$ and $[a_3]-[a_1]$ are both effective.
In a general situation where a direct application of (A) is not
possible in both cases, as discussed above we can unHiggs by first
setting $b_1 = 0$ and then considering two applications of (A), using
the redefinitions
\begin{equation}
\tilde{a}_1 = 1, \;
\tilde{a}_2 = a_1 a_2
\label{eq:a-redefinitions}
\end{equation}
where $\tilde{a}_2$ is a reducible divisor.  In the discussion here we assume that the
direct application of (A) can proceed in both cases; other situations
such as those where only the unHiggsing approach (D) is available
are special cases where one of the divisors $a_i$ becomes degenerate.

As discussed above, we begin by solving
 (\ref{eq:Delta12=0})
by setting $v_2 = \lambda_1 v_1$.
Plugging this into the CY-equation \eqref{eq:cubicfactorized} for $X_{n+1}$ we obtain
\beq \label{eq:firsttuning}
p = u f_u(u,v,w) +\lambda_1 (a_1 v + b_1 w)^2 (a_3 v+ b_3 w )\,,
\eeq
where the polynomial $f_u$ is defined in \eqref{eq:fuv1v2v3}.

Clearly, this tuned elliptic fibration is singular at codimension
one. We immediately observe the $I_2$ fiber $uf_u=0$, corresponding to
an SU(2) gauge group in F-theory, at the divisor
$\lambda_1=0$ in $B_{n}$, which we simply denote by $\lambda_1$. It is in the class
\beq
\label{eq:firstSU2}
   \text{SU}(2)\,:\quad [\lambda_1]= [a_2]-[a_1]\,.
\eeq
This $I_2$ fiber is already resolved in the $\P^2$ model
 (\ref{eq:cubicfactorized}).
In addition, \eqref{eq:firsttuning} is the equation of an
(unresolved) $A_1$-singularity at $u=f_u=a_1v+b_1w=0$, 
 {\it i.e.}~a Kodaira singularity of type $I_2$, corresponding to
 another SU(2) gauge group in F-theory. It is localized along the
 codimension one locus in $B_{n}$ given by
\beq \label{eq:tDivisor}
 t:=s_3b_1^2-s_6a_1b_1+s_8a_1^2=0\,,
\eeq
which is simply $f_u(0,-b_1,a_1)=0$.
Denoting the divisor $t=0$ by abuse of notation by $t$  we note that its divisor class is
\beq
\label{eq:SU2fromA1}
\text{SU}(2)\,:\quad [t]= [s_8]+2[a_1]=-K_B+[a_1]+[b_1]\,,
\eeq
where we used \eqref{eq:coeffsSectionsGen}. We emphasize, that the
divisor $t$ has an ordinary double point singularity at $a_1=b_1=0$
that can give rise, as we will see below, to symmetric matter
representations in an F-theory model.

Thus, in summary we have found the following  F-theory gauge group  for
the tuned model where we have unHiggsed the first U(1) associated to $\hat{s}_Q$:
\beq
	\label{eq:GGafterfirstUnHiggs}
	G=\text{SU}(2)\times \text{SU}(2)\times \text{U}(1)\,.
\eeq
We note that the U(1) associated to the section $\hat{s}_R$
remains. Thus, we denote the tuned geometry \eqref{eq:firsttuning} by $X_{n+1}^{(1)}$,
indicating the rank of its remaining MW-group.

The complete unHiggsing of the theory is achieved by tuning the complex structure of $X^{(1)}_{n+1}$  
further so that the rational point $R$ is placed on top of the points $P=Q$ in the fiber $\mathcal{E}$. 
To this end, we have to impose according to \eqref{eq:collisionOfPoints}
\beq\label{eq:FullTuning}
	\Delta_{13}=a_1b_3-a_3b_1\equiv 0\,,
\eeq
where we again used the definition \eqref{eq:Deltaij}. 
Assuming again that we can apply procedure (A), where
$[a_3]-[a_1]$ is effective, we take
\beq \label{eq:secondtuning}
	a_3v+b_3w\stackrel{!}{=}\lambda_2(a_1v+b_1 w)\,,
\eeq
so that \eqref{eq:cubicfactorized} reduces to
\beq \label{eq:fullyTunedModel}
	p = u f_u(u,v,w) +\lambda_1\lambda_2 (a_1 v + b_1 w)^3\,.
\eeq
As before, we denote the divisor $\lambda_2=0$ simply by $\lambda_2$.
The line bundles associated to the sections $\lambda_1$, {\it cf.}~\eqref{eq:classLambda1}, and 
$\lambda_2$ in summary read
\beq
	\lambda_1\in\ \mathcal{O}([a_2]-[a_1])\,\qquad \lambda_2\in  \mathcal{O}([a_3]-[a_1])\,.
\eeq

The tuned model has trivial MW-group and is denoted by
$X^{(0)}_{n+1}$. Its elliptic fibration has Kodaira fibers of type
$I_2$ of the form $uf_u=0$ at the two codimension one loci
$\lambda_1=0$ and $\lambda_2=0$ in
$B_n$. In addition, the $I_2$-singularity of \eqref{eq:firsttuning} on
the divisor $t=0$ in \eqref{eq:SU2fromA1} is enhanced to an
$I_3$ singularity since \eqref{eq:fullyTunedModel} has an
$A_2$-singularity at $u=f_u=a_1v+b_1w=0$. We will show later in
Section \ref{sec:NewWSFs} that this is an $I_3^{\text{s}}$-singularity,
yielding an SU(3) gauge group in F-theory. We note that the presence
of the ordinary double point singularity on $t$ will be
crucial. In summary, the divisors of all three codimension one
singularities are
\bea \label{eq:DivsUnHiggsed}
	&\text{SU}(2)\,:& [\lambda_1]=[a_2]-[a_1]\,,\nn\\
	&\text{SU}(2)\,: & [\lambda_2]=[a_3]-[a_1] \,,\\
	&\text{SU}(3)\,:& [t]= [s_8]+2[a_1]\,.\nn
\eea

In summary, the total gauge group of the unHiggsed theory is
\beq
\label{eq:GGUnHiggsed}
	G_{\text{uni}}=\text{SU}(2)\times \text{SU}(2)\times \text{SU}(3)\,.
\eeq
We see that, indeed, all abelian factors have been enhanced to a
non-Abelian gauge symmetry. We emphasize that this is the generic
maximal gauge group that can be present after imposing the tunings
\eqref{eq:Delta12=0} and \eqref{eq:FullTuning}, modulo the possibility
that one or more of the divisors $\lambda_1, \lambda_2, t$
may be reducible as mentioned at the end of Section \ref{sec:TuningModel}. For a particular gauge
group factor in $G_{\text{uni}}$ to be actually present in a given
model requires that the corresponding divisor class in
\eqref{eq:DivsUnHiggsed} is non-trivial. Clearly, this depends on the
choices of divisor classes $[a_i]$, $i=1,2,3$, and $[s_8]$ in
\eqref{eq:coeffsSectionsGen} entering the definition of the
CY-manifold $X_{n+1}$. Furthermore, the particular embedding of
the two U(1)-factors of F-theory on $X_{n+1}$ into
$G_{\text{uni}}$ depends on this choice as well. We will analyze the
different cases in the following
section, starting with the simplest cases
where the two U(1)'s are embedded as the Cartan generators of
$\text{SU}(2)\times \text{SU}(2) $ or $\text{SU}(3)$, before we focus
on more complicated cases, where the U(1)'s can also be embedded into
a linear combination of Cartan generators of different group factors
in $G_{\text{uni}}$.

\subsubsection*{The specialized model}

Before continuing with the discussion of the details of the unHiggsed model $X_{n+1}^{(0)}$, let us pause by 
analyzing the important specialized model \eqref{eq:abcdModel} with $a_1=1$ and $b_1=0$ in some
detail.
The tunings \eqref{eq:Delta12=0} and \eqref{eq:FullTuning} of the general model simplify  in this case to
\beq \label{eq:tuningSpecialized}
 b_2=b_3=0\,
\eeq
yielding the elliptic curve
\beq \label{eq:fullyTunedModelSpecial}
	p = u f_u(u,v,w) +a_2a_3  v^3\,.
\eeq
Clearly, this is identical to \eqref{eq:fullyTunedModel} with $\lambda_1\equiv a_2$, $\lambda_2\equiv a_3$ (and $a_1=1$, $b_1=0$) 
in the above discussion.
Note that for any model $X_{n+1}$, there is a tuning of this type where we have
fixed $b_1= 0$ and performed the field redefinitions
(\ref{eq:a-redefinitions}).
The  gauge group of the low energy effective theory is again SU(2)$\times$SU(2)$\times$SU(3) 
supported on the following divisors
\beq\label{eq:DivsUnHiggsedSpecial}
	\text{SU}(2)\,:\,\, [a_2]\,, \qquad
	\text{SU}(2)\,:\,\,  [a_3] \,,\qquad
	\text{SU}(3)\,:\,\, [t]=[s_8]\,,
\eeq
where used that $t$ in \eqref{eq:tDivisor} reduces to $s_8$ in the limit $a_1=1$, $b_1=0$.

\subsection{Weierstrass model from unHiggsing the general U$(1)^2$ model}
\label{sec:WSFunHiggsed}

We obtain the Weierstrass form of the Calabi-Yau manifold $X^{(0)}_{n+1}$ of the unHiggsed 
theory by inserting the tunings \eqref{eq:tuningSU2} and \eqref{eq:secondtuning} into the general 
Weierstrass form of $X_{n+1}$ in Appendix \ref{app:WSF}. It reads
\bea \label{eq:WSFSU2SU2SU3}
	y^2\!&\!=\!&\!x^3+\left[-\tfrac{1}{48} \left(s_6^2-4 s_3 s_8\right)^2+  \left(a_1 \left( s_2 s_8-\tfrac{1}{2}s_5 s_6\right)+b_1 \left( s_3 s_5-\tfrac{1}{2}s_2 s_6\right)\right)\lambda_1 \lambda_2 t\right]xz^4\nn\\
	&+&\left[\tfrac{1}{864} \left(s_6^2-4 s_3 s_8\right)^3-\tfrac{1}{12} \left(s_6^2-4 s_3 s_8\right)  \left(a_1 \left(s_2 s_8-\tfrac{1}{2}s_5 s_6\right)+b_1 
	\left(s_3 s_5-\tfrac{1}{2}s_2 s_6\right)\right)\lambda _1 \lambda _2 t\right.\nn\\&+&\left.\big(\tfrac{1}{4}  \left(b_1 s_2-a_1 s_5\right)^2-s_1 t\big)\lambda _1^2 \lambda _2^2t^2 \right]z^6\,,
\eea
where we recall the definitions of the variables $\lambda_1$, $\lambda_2$ and $t$ in 
\eqref{eq:firsttuning}, \eqref{eq:secondtuning} and \eqref{eq:tDivisor}, respectively.

We note that $f$ and $g$ do not vanish at any common codimension one locus, while the discriminant vanishes to order 
two and three at $\lambda_1=0$, $\lambda_2= 0$ and $t=0$, respectively, indicating two $I_2$ and 
one $I_3$ singularities. This can be seen by computing  $\Delta=4f^3+27g^2$ for
the WSF \eqref{eq:WSFSU2SU2SU3} and reducing modulo
(\ref{eq:tDivisor}),
\beq \label{eq:discSU2SU2SU3}
	\Delta=\lambda_1^2\lambda_2^2t^3\Delta'\,,
\eeq
where $\Delta'$ denotes the $I_1$ locus of the discriminant.

The Weierstrass model \eqref{eq:WSFSU2SU2SU3} matches precisely the general form 
\eqref{eq:WSFSU2SU2} of a Weierstrass model with two $I_2$ singularities at 
$\lambda_1=0$, $\lambda_2=0$ and the leading Tate coefficients
$\mathfrak{a_i}=(\lambda_1\lambda_2)^k\mathfrak{a}_i^{(k)}$ of the form
\beq\label{eq:SU2SU2SU3Tate}
	\mathfrak{a}_1^{(0)}\!\!= s_6\,,\,\,\,\mathfrak{a}_2^{(0)}\!\!=\!-s_3s_8\,,\,\,\,\mathfrak{a}_3^{(1)}\!\!=(b_1 s_2 - a_1 s_5) t\,,\,\,\, \mathfrak{a}_4^{(1)}\!\!= (b_1 (s_3 s_5 - s_2 s_6) + a_1 s_2 s_8) t\,,\,\,\, \mathfrak{a}_6^{(2)}\!\!=\!-s_1t^3\,.
\eeq
Thus, we confirm the presence of the gauge algebra $\text{su}(2)\oplus\text{su}(2)$. 

Clearly, the Tate coefficients \eqref{eq:SU2SU2SU3Tate} do not have the correct orders
of vanishing at $t=0$, namely $(0,1,1,2,3)$, required for an $I_3^{s}$ singularity in Tate's algorithm
\cite{tate1975algorithm}  In addition, the leading terms of the $t$-expansion of both $f$ and $g$
in the Weierstrass form \eqref{eq:WSFSU2SU2SU3} deviate from
that of a standard $I_3^{\text{s}}$ singularity at $t=0$ given in \eqref{eq:WSFSU3}. 
This naively seems to imply that there is a  monodromy acting on the fiber so that we have an $I_3^{\text{ns}}$ singularity at 
$t=0$, corresponding to an su(2) gauge algebra. 
However, due to the singularity 
structure of the divisor $t=0$ and the subtle link to the form of $f$, $g$, the singularity is indeed an 
$I_3^{\text{s}}$ 
corresponding to an $\text{su}(3)$ gauge algebra. We will elaborate on this further in Section 
\ref{sec:NewWSFs}. We just emphasize here that even for factoring out $t^3$ in $\Delta$ 
we had to exploit the special form of $t$ in terms of the sections $s_i$, $a_1$ 
and $b_1$ given in \eqref{eq:tDivisor}. In fact, if we had just used the general form 
\eqref{eq:WSFSU2SU2SU3} of the Weierstrass model 
for  generic $t$ we would have obtained only an order two vanishing of $\Delta$ at $t=0$.  

\subsubsection*{The specialized model}

We obtain the Weierstrass form of the unHiggsed specialized model in  \eqref{eq:abcdModel} directly 
from the general Weierstrass model \eqref{eq:WSFSU2SU2SU3} by setting $a_1=1$, $b_1=0$. In this case, we have  $\lambda_1\equiv a_2$, $\lambda_2\equiv a_3$ and $t\equiv s_8$, 
{\it cf.}~\eqref{eq:firsttuning}, \eqref{eq:secondtuning} and \eqref{eq:tDivisor},  yielding the Weierstrass form
\bea \label{eq:WSFSU2SU2SU3special}
	y^2&=&x^3+\left[-\tfrac{1}{48} \left(s_6^2-4 s_3 t\right)^2+  \left( s_2 t-\tfrac{1}{2}s_5 s_6\right)a_2 a_3 t\right]xz^4\\
	&&+\left[\tfrac{1}{864} \left(s_6^2-4 s_3 t\right)^3-\tfrac{1}{12} \left(s_6^2-4 s_3 t\right)  \left(  s_2 t-\tfrac{1}{2}s_5 s_6\right)a_2a_3 t+ \big(\tfrac{1}{4}   s_5^2-s_1 t\big)a_2^2a_3^2t^2\right]z^6\,.\nn
\eea

We emphasize that \eqref{eq:WSFSU2SU2SU3special} is
\textit{simultaneously} of the general Weierstrass forms
\eqref{eq:WSFSU2SU2SU3gen} with two $I_2$-singularities at $a_2=0$ and
$a_3=0$ and one $I_3^{\text{s}}$-singularity at $t=0$.  The leading Tate
coefficients $\mathfrak{a}_i=\mathfrak{a}_i^{(k,l)}t^k(\lambda_1\lambda_2)^l$ are
\beq \label{eq:replsSU22}
 \mathfrak{a}_1^{(0,0)}= s_6\,,\qquad\mathfrak{a}_2^{(1,0)}=-s_3\,,\qquad \mathfrak{a}_3^{(1,1)}= -  s_5\,,\qquad \mathfrak{a}_4^{(2,1)}=  s_2 \,,\qquad \mathfrak{a}_6^{(3,2)}=-s_1\,,
\eeq
This confirms the presence of an $\text{su}(2)\oplus\text{su}(2)\oplus\text{su}(3)$ gauge algebra as 
claimed in \eqref{eq:GGUnHiggsed}. This also ensures that the unHiggsed model $X^{(0)}_{n+1}$
obtained by tuning 
the complex structure of $X_{n+1}$ reproduces all models with 
$\text{su}(2)\oplus\text{su}(2)\oplus\text{su}(3)$ gauge algebra
where the gauge factors are tuned on smooth divisors,
 including the special cases where
only an $\text{su}(2)\oplus\text{su}(2)$ or su(3) gauge algebra are realized, as we will 
demonstrate in more detail in Sections \ref{sec:SU2SU2unHiggsing}, \ref{sec:SU3unHiggsing} and
\ref{sec:SU2SU2SU3Unhiggsing}.

\subsection{The matter spectrum}
\label{eq:MatterUnHiggsed}

We proceed with the determination of the charged matter spectrum of F-theory on the Calabi-Yau 
manifold $X^{(0)}_{n+1}$.  The following analysis is based
on the Weierstrass models \eqref{eq:WSFSU2SU2SU3} for the general and \eqref{eq:WSFSU2SU2SU3special} for the 
specialized model.

Clearly, we have bi-fundamental matter at the mutual intersections of $\lambda_1$, $\lambda_2$ and 
$t$. The respective multiplicities are given as the product of the degrees of the two sections that 
vanish at the relevant intersection loci.
\begin{table}[ht!]
\begin{center}
\begin{tabular}{|c|c|} \hline
Representation &Multiplicity    \rule{0pt}{14pt} \\ \hline
$(\mathbf{1},\mathbf{1},\mathbf{6})$ & $ x_{(\mathbf{1},\mathbf{1},\mathbf{6})}=  [a_1]\cdot ([t]+K_B-[a_1])$ \rule{0pt}{14pt} \\
\hline
$(\mathbf{2},\mathbf{2},\mathbf{1})$ &  $x_{(\mathbf{2},\mathbf{2},\mathbf{1})}=[\lambda_1]\cdot [\lambda_2]$ \rule{0pt}{14pt} \\ \hline
$(\mathbf{2},\mathbf{1},\mathbf{3})$ &  $x_{(\mathbf{2},\mathbf{1},\mathbf{3})}=[\lambda_1]\cdot [t]$ \rule{0pt}{14pt} \\ \hline
$(\mathbf{1},\mathbf{2},\mathbf{3})$ & $x_{(\mathbf{1},\mathbf{2},\mathbf{3})}=[\lambda_2]\cdot [t]$ \rule{0pt}{14pt} \\ \hline
$(\mathbf{2},\mathbf{1},\mathbf{1})$ & $x_{(\mathbf{2},\mathbf{1},\mathbf{1})}=[\lambda_1]\cdot  (-8K_B-2[\lambda_1]-2[\lambda_2]-3[t])$  \rule{0pt}{14pt} \\  \hline
$(\mathbf{1},\mathbf{2},\mathbf{1})$ & $x_{(\mathbf{1},\mathbf{2},\mathbf{1})}=[\lambda_2]\cdot  (-8K_B-2[\lambda_1]-2[\lambda_2]-3[t]) $ \rule{0pt}{14pt} \\ \hline
$(\mathbf{1},\mathbf{1},\mathbf{3})$ & $x_{(\mathbf{1},\mathbf{1},\mathbf{3})}=[t]\cdot (-9K_B-2[\lambda_1]-2[\lambda_2]-3[t])+x_{(\mathbf{1},\mathbf{1},\mathbf{6})}$ \rule{0pt}{14pt} \\ \hline
$(\mathbf{3},\mathbf{1},\mathbf{1})$ & $x_{(\mathbf{3},\mathbf{1},\mathbf{1})}=\frac12 [\lambda_1] \cdot ([\lambda_1]+K_B)+1$ \rule{0pt}{14pt} \\ \hline
$(\mathbf{1},\mathbf{3},\mathbf{1})$ & $x_{(\mathbf{1},\mathbf{3},\mathbf{1})}=\frac12 [\lambda_2] \cdot ([\lambda_2]+K_B)+1$ \rule{0pt}{14pt} \\
\hline
$(\mathbf{1},\mathbf{1},\mathbf{8})$ & $x_{(\mathbf{1},\mathbf{1},\mathbf{8})}=\frac12 [t] \cdot ([t]+K_B)+1-x_{(\mathbf{1},\mathbf{1},\mathbf{6})}  $ \rule{0pt}{14pt} \\[0.05Em]
 \hline
\end{tabular}
\caption{Matter spectrum of the unHiggsed model $X_{n+1}^{(0)}$. The multiplicity of adjoints in 6D is included for completeness.}
\label{tab:spectrumUnHiggsed}
\end{center}
\end{table}
In addition, there is fundamental matter under each factor of the gauge group in 
\eqref{eq:GGUnHiggsed}. The multiplicity of the fundamental matter of the two SU$(2)$'s is computed
analogous to the multiplicities in \eqref{eq:Multies_SU2SU2}, taking into account that the class of 
$\Delta'$ in \eqref{eq:discSU2SU2SU3} is  changed to
\beq
	[\Delta']=-12K_B-2[\lambda_1]-2[\lambda_2]-3[t]\,,
\eeq
which follows as $[\Delta]=-12K_B$. In addition, we have to subtract the order two loci 
$  4 s_3 s_8-s_6^2=\lambda_1=0$ and $4 s_3 s_8-s_6^2=\lambda_2=0$, respectively, 
that support singularities of type $III$ which do not yield additional matter representations.

The calculation of the multiplicity of the representation 
$\mathbf{3}$ of the
the SU$(3)$ is performed as follows. We find that $t=\Delta'=0$  has two minimal associated prime 
ideals, one of which given by
\beq \label{eq:pideal}
	\mathfrak{p}=\{s_6b_1-2s_8a_1,  2s_3b_1-s_6a_1,s_6^2-4s_3s_1\}\,.
\eeq
and the second one given by a large ideal $\mathfrak{J}$ with seven generators. 
Along the variety $V(\mathfrak{J})$, neither $f$ and $g$ vanish while $\Delta$ vanishes to order four, 
indicating an  $I_4$-singularity, which signals the presence of fundamental matter. 
On $ V(\mathfrak{p})$, the sections $f$ and $g$ both vanish to order two, while $\Delta$ vanishes 
to order four, corresponding to a Kodaira fibre of type $IV$ not giving rise  to matter fields in F-theory. 
In addition, we check using the resultant technique of \cite{Cvetic:2013nia} that $V(\mathfrak{p})$ is 
of multiplicity three inside $\Delta'$. This allows us to compute the class $[V(\mathfrak{J})]$, 
 {\it i.e.}~the contribution $x'_{(\mathbf{1},\mathbf{1},\mathbf{3})}$ of $I_4$-singularities
to the number  of matter fields in the representation  $\mathbf{3}$. 
We obtain the multiplicity 
\beq \label{eq:fundsSU3}	
	x'_{(\mathbf{1},\mathbf{1},\mathbf{3})}=[\Delta']-3[V(\mathfrak{p})]=[t]\cdot (-9K_B-2[\lambda_1]-2[\lambda_2]-3[t])\,,
\eeq
where we employed
\beq \label{eq:[V(p)]}
	[V(\mathfrak{p})]=([a_1]-K_B)\cdot([s_8]+[a_1])-[a_1]\cdot [b_1]=-K_B\cdot[t]\,,
\eeq
using \eqref{eq:coeffsSectionsGen} and the class of $t$ in \eqref{eq:SU2fromA1}.
Here, the first equality  is obtained by multiplying the degrees of the first two 
generators of $\mathfrak{p}$, yielding the first summand, and 
subtracting the solution $a_1=b_1=0$  which does not satisfy the third generator of $\mathfrak{p}$.

In addition, we obtain adjoints for  each gauge group. In 6D these are given by the topological genus of 
the curve supporting the respective gauge group factor. For the two SU(2)'s, it agrees with  the 
arithmetic genus \eqref{eq:arithmeticGenus} of the generically smooth curves $\lambda_1=0$ and 
$\lambda_2=0$. There is one subtlety regarding the divisor
$t=0$. Its arithmetic genus is the sum of the topological genus and the number $[a_1]\cdot [b_1]$
of double points.  The matter supported at these double points $p_g$ can be either in the adjoint or
symmetric plus anti-symmetric representations, depending on the global properties of the resolution 
\cite{Morrison:2011mb}. As we will argue in Section \ref{sec:NewWSFs}, in the case at hand we have 
symmetrics plus anti-symmetrics. Thus, we  have a number of $[a_1]\cdot[b_1]$ additional 
matter fields in the representations $(\mathbf{1},\mathbf{1},\mathbf{3})+
(\mathbf{1},\mathbf{1},\mathbf{6})$. 
Accordingly, the number of adjoints $x_{(\mathbf{1},\mathbf{1},\mathbf{8})}$ 
of the SU(3) in 6D is given by
\beq
	x_{(\mathbf{1},\mathbf{1},\mathbf{8})}=p_g=\frac{1}{2}[t]\cdot([t]+K_B)+1-[a_1]\cdot [b_1]\,.
\eeq

We summarize the full spectrum of the unHiggsed model
\eqref{eq:fullyTunedModel} in Table \ref{tab:spectrumUnHiggsed} where we used the divisor classes of $t$, $\lambda_1$ 
and $\lambda_2$ given in \eqref{eq:DivsUnHiggsed} and denote matter multiplicities in a representation $\mathbf{R}$ by 
$x_{\mathbf{R}}$. 

Next, we readily check that all 6D anomalies are canceled for the spectrum in Table \ref{tab:spectrumUnHiggsed}, see
{\it e.g.}~\cite{Taylor:2011wt} for a review of 6D anomalies from an F-theory perspective. This is a 
crosscheck of the geometrical analysis of the unHiggsed model $X_{(n+1)}^{(0)}$.

Finally, we observe that the matter multiplicities in Table \ref{tab:spectrumUnHiggsed} of the unHiggsed 
theory are related to those in Table \ref{tab:MultiesGenModel} of the abelian theory. 
We  readily obtain the following relations between the 
multiplicities of the non-Abelian and the abelian theories:
\begin{align} \label{eq:relationsUnHiggsing}
	x_{(-2,-2)}&\!=\!x_{(\mathbf{1},\mathbf{1},\mathbf{6})}\,,\qquad \qquad\qquad\qquad\quad\quad\,\,\,\,\,\,\, x_{(2,0)}\!=\!x_{(\mathbf{2},\mathbf{1},\mathbf{3})}+2x_{(\mathbf{3},\mathbf{1},\mathbf{1})}+x_{(\mathbf{1},\mathbf{1},\mathbf{6})}-2\,,\nn\\  
	x_{(0,2)}&\!=\!x_{(\mathbf{1},\mathbf{2},\mathbf{3})}\!+\!2x_{(\mathbf{1},\mathbf{3},\mathbf{1})}\!+\!x_{(\mathbf{1},\mathbf{1},\mathbf{6})}\!-\!2\,,\,\,\,  x_{(-2,-1)}=x_{(\mathbf{2},\mathbf{1},\mathbf{3})}+2x_{(\mathbf{1},\mathbf{1},\mathbf{8})}\!-\!2\,,\nn\\
	x_{(-1,-2)}&\!=\!x_{(\mathbf{1},\mathbf{2},\mathbf{3})}\!+\!2x_{(\mathbf{1},\mathbf{1},\mathbf{8})}\!-\!2\,,\qquad \qquad  \,\,\,\,\,\,x_{(-1,1)}\!=\!2x_{(\mathbf{2},\mathbf{2},\mathbf{1})}\!+\!x_{(\mathbf{2},\mathbf{1},\mathbf{3})}\!+\!x_{(\mathbf{1},\mathbf{2},\mathbf{3})}\!+\!2x_{(\mathbf{1},\mathbf{1},\mathbf{8})}\!-\!2\,,\nn\\
	&\qquad \qquad\,\, x_{(1,0)}=x_{(\mathbf{1},\mathbf{2},\mathbf{3})}+2x_{(\mathbf{2},\mathbf{1},\mathbf{1})}+x_{(\mathbf{1},\mathbf{1},\mathbf{3})}+x_{(\mathbf{1},\mathbf{1},\mathbf{6})}\,,\nn\\	
	&\qquad \qquad\,\, x_{(0,1)}=x_{(\mathbf{2},\mathbf{1},\mathbf{3})}+2x_{(\mathbf{1},\mathbf{2},\mathbf{1})}+x_{(\mathbf{1},\mathbf{1},\mathbf{3})}+x_{(\mathbf{1},\mathbf{1},\mathbf{6})}\,,\nn\\
	&\qquad \qquad\,\,  x_{(1,1)}=2x_{(\mathbf{2},\mathbf{2},\mathbf{1})}+x_{(\mathbf{2},\mathbf{1},\mathbf{3})}+x_{(\mathbf{1},\mathbf{2},\mathbf{3})}+x_{(\mathbf{1},\mathbf{1},\mathbf{3})}+x_{(\mathbf{1},\mathbf{1},\mathbf{6})}\,.
\end{align}
Here we used \eqref{eq:DivsUnHiggsed} to express the divisor classes of $\lambda_1$, $\lambda_2$ and $t$ in terms of the
classes $[a_i]$ with $i=1,2,3$  and $[s_8]$. 
These relations have a clear interpretation if we Higgs the non-Abelian theory back to the abelian
one defined by $X_{n+1}$. Indeed, we can view these relations in terms of the Higgsing by
at least two bi-fundamentals $(\mathbf{1},\mathbf{2},\mathbf{3})$ and 
$(\mathbf{2},\mathbf{1},\mathbf{3})$ each 
as discussed in Section \ref{sec:Examples}. Then we 
employ the branchings of representations in \eqref{eq:branching1}, \eqref{eq:branching2}, 
\eqref{eq:branching3} and
\eqref{eq:branching4} to obtain the spectrum  the relations \eqref{eq:relationsUnHiggsing}.  Note that
we have to subtract $2$ as two states with charges $(2,1)$, $(1,2)$ and $(-1,1)$,
respectively, inside the two bi-fundamentals are eaten up by the massive gauge bosons. 
Alternatively, we can view  the relations \eqref{eq:relationsUnHiggsing} as encoding
the combination of Higgsings on adjoints of SU$(2)\times$SU(2) and at least two fundamentals
of an unbroken SU(3) or adjoints of SU(3) and at least two fundamentals
of the unbroken SU$(2)\times$SU(2). Both point of views will be crucial below in Section
\ref{sec:Examples}.

\subsubsection*{The specialized model}

The matter spectrum of the specialized Calabi-Yau manifold after unHiggsing as given in  \eqref{eq:fullyTunedModelSpecial}
follows directly from that of the general model in Table \ref{tab:spectrumUnHiggsed} by setting 
$[a_1]=0$. This yields the spectrum in
Table \ref{tab:spectrumUnHiggsedSpecial}, where we note that $[\lambda_1]=[a_2]$, $[\lambda_2]=[a_3]$ and $[t]=[s_8]$
according to \eqref{eq:DivsUnHiggsed}.
\begin{table}[ht!]
\begin{center}
\begin{tabular}{|c|c|} \hline
Representation &Multiplicity    \rule{0pt}{14pt} \\ \hline
$(\mathbf{2},\mathbf{2},\mathbf{1})$ &  $x_{(\mathbf{2},\mathbf{2},\mathbf{1})}=[a_2]\cdot [a_3]$ \rule{0pt}{14pt} \\ \hline
$(\mathbf{2},\mathbf{1},\mathbf{3})$ &  $x_{(\mathbf{2},\mathbf{1},\mathbf{3})}=[a_2]\cdot [s_8]$ \rule{0pt}{14pt} \\ \hline
$(\mathbf{1},\mathbf{2},\mathbf{3})$ & $x_{(\mathbf{1},\mathbf{2},\mathbf{3})}=[a_3]\cdot [s_8]$ \rule{0pt}{14pt} \\ \hline
$(\mathbf{2},\mathbf{1},\mathbf{1})$ & $x_{(\mathbf{2},\mathbf{1},\mathbf{1})}=[a_2]\cdot  (-8K_B-2[a_2]-2[a_3]-3[s_8])$  \rule{0pt}{14pt} \\  \hline
$(\mathbf{1},\mathbf{2},\mathbf{1})$ & $x_{(\mathbf{1},\mathbf{2},\mathbf{1})}=[a_3]\cdot  (-8K_B-2[a_2]-2[a_3]-3[s_8]) $ \rule{0pt}{14pt} \\ \hline
$(\mathbf{1},\mathbf{1},\mathbf{3})$ & $x_{(\mathbf{1},\mathbf{1},\mathbf{3})}=[s_8]\cdot (-9K_B-2[a_2]-2[a_3]-3[s_8])$ \rule{0pt}{14pt} \\ \hline
$(\mathbf{3},\mathbf{1},\mathbf{1})$ & $x_{(\mathbf{3},\mathbf{1},\mathbf{1})}=\frac12 [a_2] \cdot ([a_2]+K_B)+1$ \rule{0pt}{14pt} \\ \hline
$(\mathbf{1},\mathbf{3},\mathbf{1})$ & $x_{(\mathbf{1},\mathbf{3},\mathbf{1})}=\frac12 [a_3] \cdot ([a_3]+K_B)+1$ \rule{0pt}{14pt} \\
\hline
$(\mathbf{1},\mathbf{1},\mathbf{8})$ & $x_{(\mathbf{1},\mathbf{1},\mathbf{8})}=\frac12 [s_8] \cdot ([s_8]+K_B)+1  $ \rule{0pt}{14pt} \\[0.05Em]
 \hline
\end{tabular}
\caption{Matter spectrum of the unHiggsed specialized model $X_{n+1}^{(0)}$. The multiplicity of adjoints in 6D is included for completeness.}
\label{tab:spectrumUnHiggsedSpecial}
\end{center}
\end{table}

We emphasize that this agrees precisely with the spectrum of a general 
Weierstrass model with two $I_2$  and one $I_3$ singularity found in Appendix 
\ref{app:SU2SU2SU3WSF}. This is not surprising since we already observed in \eqref{eq:replsSU22} 
that the unHiggsed specialized model  reproduces precisely
the most general Weierstrass model \eqref{eq:WSFSU2SU2SU3gen} with these singularities.

Comparing with the spectrum in Table  \ref{tab:spectrumUnHiggsed}, we note that the main difference 
of the spectrum in Table  \ref{tab:spectrumUnHiggsedSpecial} is the absence of the symmetric 
representation $(\mathbf{1},\mathbf{1},\mathbf{6})$ of 
the SU(3). This is due to the fact that the divisor $s_8=0$ supporting the SU(3) gauge theory is 
smooth. Furthermore, 
the range of allowed values for $[a_i]$, $i=2,3$, and $[s_8]$ is larger in the specialized 
compared to the general model, due to different
effectiveness constraints. This yields different matter multiplicities in the unHiggsed specialized model.

We conclude by noting that the relations \eqref{eq:relationsUnHiggsing} also hold for the specialized
models,  {\it i.e.}~we can use Table  \ref{tab:spectrumUnHiggsedSpecial} to precisely reproduce
the spectrum of the specialized abelian model in Table \ref{tab:Multies_Simpl}. As before
these relations encode the possibility of Higgsing the non-Abelian model in different ways back to
the abelian model.

\subsection{Special unHiggsings: $b_i\rightarrow 0$}
\label{sec:specialunHiggsing}

We have discussed in Section \ref{sec:TuningModel} the possibility of situations where the 
only possible solution to $\Delta_{12}=\Delta_{13}=0$ is given by $b_i\rightarrow 0$, for 
$i=1,2,3$. As further explained at the beginning of Section \ref{sec:unHiggsingBothU1s}, this
can be viewed as a specialized model \eqref{eq:abcdModel} with one reducible coefficient
$\tilde{a}_2=a_1a_2$ as in \eqref{eq:a-redefinitions}. This implies that this special unHiggsing, 
its matter spectrum and Higgsings back to the Abelian theory
$X_{n+1}$ is already covered by the discussion in Section \ref{eq:MatterUnHiggsed}. 
However it is instructive to spell out some details of this analysis.

We begin by noting that the fully unHiggsed model $X^{(0)}_{n+1}$ with $b_i\rightarrow 0$
is described by
\beq
	uf_u+a_1a_2a_3v^3=0\,,
\eeq
which has gauge group $\text{SU(2)}^3\times \text{SU}(3)$, as already seen above. 
Its Weierstrass form is given by
\eqref{eq:WSFSU2SU2SU3gen}, after setting  $a_1=1$, $b_1=0$ and $t=s_8$ and then 
identifying  $\lambda_1=a_2a_1$, $\lambda_3=a_3$. This confirms the presence of three 
$I_2$ and one $I_3^{\text{s}}$ singularities. 

The matter spectrum of the theory can be 
computed from analysis of this Weierstrass form (or 6D anomaly cancellation). It is given in Table
\ref{tab:spectrumUnHiggsedbi=0}, which is completely symmetric under an exchange of 
SU(2)-factors.
\begin{table}[ht!]
\begin{center}
\begin{tabular}{|c|c|} \hline
Representation &Multiplicity    \rule{0pt}{14pt} \\ \hline
$(\mathbf{2},\mathbf{2},\mathbf{1},\mathbf{1})$ &  $x_{(\mathbf{2},\mathbf{2},\mathbf{1},\mathbf{1})}=[a_1]\cdot [a_2]$ \rule{0pt}{14pt} \\ \hline
$(\mathbf{2},\mathbf{1},\mathbf{2},\mathbf{1})$ &  $x_{(\mathbf{2},\mathbf{1},\mathbf{2},\mathbf{1})}=[a_1]\cdot [a_3]$ \rule{0pt}{14pt} \\ \hline \rule{0pt}{14pt} 
$(\mathbf{1},\mathbf{2},\mathbf{2},\mathbf{1})$ &  $x_{(\mathbf{1},\mathbf{2},\mathbf{2},\mathbf{1})}=[a_2]\cdot [a_3]$ \rule{0pt}{14pt} \\ \hline \rule{0pt}{14pt} 
$(\mathbf{2},\mathbf{1},\mathbf{1},\mathbf{3})$ &  $x_{(\mathbf{2},\mathbf{1},\mathbf{1},\mathbf{3})}=[a_1]\cdot [s_8]$ \rule{0pt}{14pt} \\ \hline
$(\mathbf{1},\mathbf{2},\mathbf{1},\mathbf{3})$ &  $x_{(\mathbf{1},\mathbf{2},\mathbf{1},\mathbf{3})}=[a_2]\cdot [s_8]$ \rule{0pt}{14pt} \\ \hline
$(\mathbf{1},\mathbf{1},\mathbf{2},\mathbf{3})$ &  $x_{(\mathbf{1},\mathbf{1},\mathbf{2},\mathbf{3})}=[a_3]\cdot [s_8]$ \rule{0pt}{14pt} \\ \hline
$(\mathbf{2},\mathbf{1},\mathbf{1},\mathbf{1})$ & $x_{(\mathbf{2},\mathbf{1},\mathbf{1},\mathbf{1})}=[a_1]\cdot  (-8K_B-2([a_1]+[a_2]+[a_3])-3[s_8])$  \rule{0pt}{14pt} \\  \hline
$(\mathbf{1},\mathbf{2},\mathbf{1},\mathbf{1})$ & $x_{(\mathbf{1},\mathbf{2},\mathbf{1},\mathbf{1})}=[a_2]\cdot  (-8K_B-2([a_1]+[a_2]+[a_3])-3[s_8])$  \rule{0pt}{14pt} \\  \hline
$(\mathbf{1},\mathbf{1},\mathbf{2},\mathbf{1})$ & $x_{(\mathbf{1},\mathbf{1},\mathbf{2},\mathbf{1})}=[a_3]\cdot  (-8K_B-2([a_1]+[a_2]+[a_3])-3[s_8])$  \rule{0pt}{14pt} \\  \hline
$(\mathbf{1},\mathbf{1},\mathbf{1},\mathbf{3})$ & $x_{(\mathbf{1},\mathbf{1},\mathbf{1},\mathbf{3})}=[s_8]\cdot (-9K_B-2([a_1]+[a_2]+[a_3])-3[s_8])$ \rule{0pt}{14pt} \\ \hline
$(\mathbf{3},\mathbf{1},\mathbf{1},\mathbf{1})$ & $x_{(\mathbf{3},\mathbf{1},\mathbf{1},\mathbf{1})}=\frac12 [a_1] \cdot ([a_1]+K_B)+1$ \rule{0pt}{14pt} \\ \hline
$(\mathbf{1},\mathbf{3},\mathbf{1},\mathbf{1})$ & $x_{(\mathbf{1},\mathbf{3},\mathbf{1},\mathbf{1})}=\frac12 [a_2] \cdot ([a_2]+K_B)+1$ \rule{0pt}{14pt} \\ \hline
$(\mathbf{1},\mathbf{1},\mathbf{3},\mathbf{1})$ & $x_{(\mathbf{1},\mathbf{1},\mathbf{3},\mathbf{1})}=\frac12 [a_3] \cdot ([a_3]+K_B)+1$ \rule{0pt}{14pt} \\
\hline
$(\mathbf{1},\mathbf{1},\mathbf{1},\mathbf{8})$ & $x_{(\mathbf{1},\mathbf{1},\mathbf{1},\mathbf{8})}=\frac12 [s_8] \cdot ([s_8]+K_B)+1  $ \rule{0pt}{14pt} \\[0.05Em]
 \hline
\end{tabular}
\caption{Matter spectrum of the specialized unHiggsins $b_i\rightarrow 0$, $i=1,2,3$. The multiplicity of adjoints in 6D is included for completeness.}
\label{tab:spectrumUnHiggsedbi=0}
\end{center}
\end{table}
The multipliticies of bifundamentals and adjoints follow as before. For the multiplicities of 
fundamentals, we account for the third SU(2) by modifying the fourth to sixth lines in Table 
\ref{tab:spectrumUnHiggsedSpecial} as follows. We just have to take into account in the 
parenthesis that we have a reducible $a_2$, 
\textit{i.e.}~$a_2\rightarrow a_1a_2$ so that $-2[a_2]\rightarrow -2([a_1]+[a_2])$. 

The Higgsing to the Abelian model goes in steps. Basically the Higgses are certain components in 
all three possible bifundamental representations of one SU(2) and the  SU(3). We obtain
\bea
		&\text{SU}(2)_{a_1}\times \text{SU}(2)_{a_2}\times \text{SU}(2)_{a_3}\times \text{SU}(3)_{s_8} \longrightarrow \text{SU}(2)_{a_1}\times \text{SU}(2)_{a_2}\times \text{U}(1)'\times \text{SU}(2)_{s_8} \nn&\\  &\longrightarrow  \text{SU}(2)_{a_1}\times \text{U}(1)'\times \text{U}(1)''\rightarrow  \text{U}(1)\times \text{U}(1)\,, &
\eea
where we have indicated the codimension one loci of the various gauge group factors
by a subscript. Note also that the U(1)'s after the second  and third Higgsing are different.
The first Higgsing is performed by the state with Dynkin labels $(0;0;1;1,0)$ inside 
$(\mathbf{1},\mathbf{1},\mathbf{2},\mathbf{3})$, the second Higgsing by the  state
$(0;-1;0;0,-1)$ inside the representation $(\mathbf{1},\mathbf{2},\mathbf{2})_{(0)}$ of the original $(\mathbf{1},\mathbf{2},\mathbf{1},\mathbf{3})$, and the final Higgsing by the state $ (1;0;0;-1,1)$ inside the  $\mathbf{2}_{(-1,1)}$ of the original $(\mathbf{2},\mathbf{1},\mathbf{1},\mathbf{3})$.  Then the final unbroken U(1) generators are
\beq
\xi
=\sigma_3^{(1)}-\sigma_3^{(3)}+\lambda\,,\qquad 
\zeta
=\sigma_3^{(1)}+\sigma_3^{(2)}-(\lambda-\mu)\,,
\eeq
where $\sigma_3^{(i)}$ denotes the third Cartan matrix of  the i-th SU(2) and 
$\lambda=\text{diag}(1,-1,0)$, $\mu=\text{diag}(1,0,-1)$\footnote{Note that the Dynkin labels have been computed here w.r.t.~$\lambda$ and $\lambda-\mu$, as usual.}. For these U(1) generators, 
we see that the states $(-1; 0; 0; -1, 1)$ inside 
$(\mathbf{2},\mathbf{1},\mathbf{1},\mathbf{3})$ and the states $\pm(2;0;0;0,0)$ inside 
$(\mathbf{3},\mathbf{1},\mathbf{1},\mathbf{1})$ will have charges $(-2,-2)$, \textit{i.e.}~the $(-2,-2)$ matter arises as some sort of “diagonal adjoint” and the first SU(2)$\times$ SU(3) 
bifundamental. It does not appear as an ordinary double point singularity of the SU(3) divisor.

The multiplicities of the Higgsed Abelian theory $X_{(n+1)}$ in Table \ref{tab:MultiesGenModel}
are related to those of the unHiggsed non-Abelian theory in Table 
\ref{tab:spectrumUnHiggsedbi=0} by the following relations:
\bea
x_{(0,2)}=x_{(\mathbf{1},\mathbf{1},\mathbf{2},\mathbf{3})}+2x_{(\mathbf{1},\mathbf{1},\mathbf{3},\mathbf{1})}-2\,,&\nn\\
x_{(2,0)}=x_{(\mathbf{1},\mathbf{2},\mathbf{1},\mathbf{3})}+2x_{(\mathbf{1},\mathbf{3},\mathbf{1},\mathbf{1})}-2\,,&\nn\\
x_{(-2,-2)}=x_{(\mathbf{2},\mathbf{1},\mathbf{1},\mathbf{3})}+2x_{(\mathbf{3},\mathbf{1},\mathbf{1},\mathbf{1})}-2\,,&\nn\\
x_{(-2,-1)}=x_{(\mathbf{2},\mathbf{1},\mathbf{1},\mathbf{3})}+x_{(\mathbf{1},\mathbf{2},\mathbf{1},\mathbf{3})}+2x_{(\mathbf{1},\mathbf{1},\mathbf{1},\mathbf{8})}+2x_{(\mathbf{2},\mathbf{2},\mathbf{1},\mathbf{1})}-2\,,&\nn\\
x_{(-1,-2)}=x_{(\mathbf{2},\mathbf{1},\mathbf{1},\mathbf{3})}+x_{(\mathbf{1},\mathbf{1},\mathbf{2},\mathbf{3})}+2x_{(\mathbf{1},\mathbf{1},\mathbf{1},\mathbf{8})}+2x_{(\mathbf{2},\mathbf{1},\mathbf{2},\mathbf{1})}-2\,,&\nn\\
x_{(-1,1)}=x_{(\mathbf{1},\mathbf{2},\mathbf{1},\mathbf{3})}+x_{(\mathbf{1},\mathbf{1},\mathbf{2},\mathbf{3})}+2x_{(\mathbf{1},\mathbf{1},\mathbf{1},\mathbf{8})}+2x_{(\mathbf{1},\mathbf{2},\mathbf{2},\mathbf{1})}-2\,,&\nn
\eea
\bea
x_{(0,1)}=2x_{(\mathbf{2},\mathbf{2},\mathbf{1},\mathbf{1})}+x_{(\mathbf{2},\mathbf{1},\mathbf{1},\mathbf{3})}+x_{(\mathbf{1},\mathbf{2},\mathbf{1},\mathbf{3})}+2x_{(\mathbf{1},\mathbf{1},\mathbf{2},\mathbf{1})}+x_{(\mathbf{1},\mathbf{1},\mathbf{1},\mathbf{3})}\,,&\nn\\
x_{(1,0)}=2x_{(\mathbf{2},\mathbf{1},\mathbf{2},\mathbf{1})}+x_{(\mathbf{1},\mathbf{1},\mathbf{2},\mathbf{3})}+x_{(\mathbf{2},\mathbf{1},\mathbf{1},\mathbf{3})}+2x_{(\mathbf{1},\mathbf{2},\mathbf{1},\mathbf{1})}+x_{(\mathbf{1},\mathbf{1},\mathbf{1},\mathbf{3})}\,,&\nn\\
x_{(1,1)}=2x_{(\mathbf{1},\mathbf{2},\mathbf{2},\mathbf{1})}+x_{(\mathbf{1},\mathbf{1},\mathbf{2},\mathbf{3})}+x_{(\mathbf{1},\mathbf{2},\mathbf{1},\mathbf{3})}+2x_{(\mathbf{2},\mathbf{1},\mathbf{1},\mathbf{1})}+x_{(\mathbf{1},\mathbf{1},\mathbf{1},\mathbf{3})}\,.&
\eea
This shows that the spectrum of the Abelian model $X_{n+1}$ will indeed fall into representations of $\text{SU}(2)^3\times \text{SU}(3)$ after unHiggsing. 

\subsection{Novel Weierstrass models with $I_3^{\text{s}}$ singularities}
\label{sec:NewWSFs}

We conclude the discussion of the tuned Calabi-Yau manifold $X_{n+1}^{(0)}$ by a detailed 
analysis of  the global structure of its Weierstrass model and the interplay with the singularities of the
divisor $t=0$ supporting the $I_3$-singularity.

The central object to distinguish between the split and non-split form of a Kodaira singularity of type
$I_3$ is the monodromy cover \cite{Grassi:2011hq}
\beq \label{eq:I3monoCover}
	\Psi^2+\left.\frac{9g}{2f}\right\vert_{t=0}=0\,.
\eeq
The standard story in F-theory is that we have a split $I_3$,
corresponding to an su(3) gauge algebra, if $\frac{9g}{2f}$ evaluated
at $t=0$ is a perfect square and otherwise a non-split $I_3$, which
yields just an $\text{sp}(1)\cong\text{su}(2)$.  
This condition is determined locally from the structure of the
Weierstrass model around $t = 0$.  While in most applications
considered in the literature the monodromy can be determined in a
straightforward fashion from the geometry using global considerations,
we demonstrate here an example of the subtlety that may arise
if $t=0$ is singular and we
are not in a unique factorization domain.  In the situation we
consider here it turns out that $\frac{9g}{2f}$ is a square
\textit{locally} around $t=0$, though this is encoded in a nontrivial
fashion in the algebra.

\begin{figure}[ht]
\begin{center}
\parbox[c]{0.4\textwidth}{\includegraphics[scale=0.27]{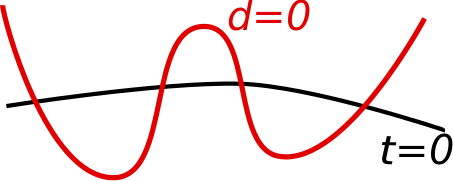}} \parbox[c]{0.3\textwidth}{\includegraphics[scale=0.27]{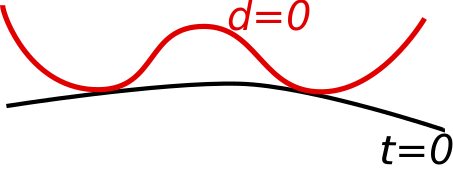}}
\caption{Generic (left) and non-generic (right) $I_3^{\text{ns}}$ along the divisor $t=0$ with normal and tangential crossing of $t=0$ and $d=0$, respectively.}
\label{fig:I3ns}
\end{center}
\end{figure}

We begin by  evaluating the monodromy cover \eqref{eq:I3monoCover} for the Weierstrass form 
\eqref{eq:WSFSU2SU2SU3}. We obtain
\beq \label{eq:g/f}
	\left.\frac{9g}{2f}\right\vert_{t=0}=\tfrac14(s_6^2-4s_3s_8)=:\tfrac14 d\,.
\eeq
It is one key property of the Weierstrass form that this $d$ is at the same time the discriminant
of the divisor $t=s_8a_1^2-s_6a_1b_1+s_3b_1^2=0$  viewed as a conic in
the variables $a_1$, $b_1$.  The consequences of this coincidence are most obvious by contrasting
the geometry at hand with a generic situation. In general, the intersection points of $t=0$ and 
$\frac{9g}{2f}=0$ are the branch points around which, in a resolution, two nodes in the 
reducible fiber over $t=0$ are exchanged by a $\mathbb{Z}_2$-monodromy, {\it cf.}~the left figure in 
Figure  in \ref{fig:I3ns}.
This reduces the gauge algebra from $\text{su}(3)$ to $\text{su}(2)$ \cite{Bershadsky:1996nh}.
In contrast, for the geometry at hand we have a \textit{non-generic} 
divisor $t=0$ with double-points at $a_1=b_1=0$ and discriminant $d=0$, which agrees with 
\eqref{eq:g/f}. Thus,  $t=0$ intersects $d$ tangentially,  {\it i.e.}~$t=d=0$ have only double zeros, see the right picture in
Figure \ref{fig:I3ns}. 
This double zero arises since two zeros of multiplicity one have
merged. Furthermore, as the zeros of $d=0$ along $t=0$ are also the
branch points of the monodromy cover we see that its branch points
have come together in pairs. As the monodromy around a pair of branch
points is trivial, no nodes in the resolution are not interchanged,
leaving a full $\text{su}(3)$ gauge algebra.

The fact that the divisors $d$ and $t$ intersect only at points of
tangency can be seen in a simple fashion by considering the normal
vector to each of the divisors at the intersection points.  Where $t =
d = 0$ we have 
\begin{equation}
 (s_8a_1^2 + s_3b_1^2)^2=s_6^2 a_1^2 b_1^2 = 4s_3s_8 a_1^2 b_1^2 \,,
\label{eq:normal1}
\end{equation}
from which it follows that $s_8a_1^2 = s_3b_1^2 = s_6a_1 b_1/2$, and
\begin{equation}
2s_8a_1 = s_6b_1, \;\;\;\;\;
2s_3b_1 = s_6a_1 \,.
\label{eq:normal2}
\end{equation}
We can then relate the differentials of $d$ and $t$,
\begin{eqnarray}
a_1 b_1 d (d) & = &  (a_1 b_1) (2s_6ds_6-4s_3ds_8-4s_8ds_3)\\
 & = & (-2s_6) d (t) \,,
\end{eqnarray}
 which shows that the normals are proportional and hence the
 intersection is at a point of tangency.

We emphasize that it is also this interplay between the structure of
$f$, $g$ and the divisor $t=0$ that does not permit us to deform $t$
so that its ordinary double singularities disappear. In fact, if we
changed the structure of $t$ without changing the Weierstrass form
\eqref{eq:WSFSU2SU2SU3}, we would only have an order two vanishing of
$\Delta$ at $t=0$ and just an $I_2$-singularity with an su(2) gauge
algebra, as noted above. If we changed only the Weierstrass form, {\it
  e.g.}~by modifying the leading coefficient of $f$ and $g$, while
keeping $t$ unchanged, we might get an $I_3^{\text{ns}}$ and again
only an su(2) gauge algebra. Thus, the ordinary double points of $t=0$
are able to support symmetric plus anti-symmetric matter representations,
in such a way that the geometry cannot be deformed to a smooth
discriminant locus which would support only adjoint representations
 \cite{Morrison:2011mb}.
It is an interesting open question whether  symmetric matter
representations can only be formed in this nontrivial algebraic
fashion, or may also arise when smooth divisors are deformed into
singular divisors with double point singularities.

Let us conclude by showing in another way
 that the zeros of $t=d=0$ are indeed double-points. To this end, we define 
the ideal
\beq
	\mathcal{I}:=\{t,d\}=\{s_8a_1^2-s_6a_1b_1+s_3b_1^2,s_6^2-4s_3s_1\}\,,
\eeq
Generically, there would be $\text{deg}(t)\cdot\text{deg}(d)=-2K_B\cdot([s_8]+2[a_1])$ points of 
multiplicity one in the vanishing set $V(\mathcal{I})$. Computing the primary decomposition 
of $\mathcal{I}$ we obtain only one prime ideal $\mathfrak{p}$ that is given in \eqref{eq:pideal},  
showing the irreducibility of
$V(\mathcal{I})$. However, the corresponding variety  $V(\mathfrak{p})$ has multiplicity two inside 
$V(\mathcal{I})$ as can be seen using the resultant technique \cite{Cvetic:2013nia}. In other words 
$V(\mathcal{I})$ consistent only of points of multiplicity two,  {\it i.e.}~double zeros. Their number is 
computed by the class
\beq 
[V(\mathfrak{p})]=-K_B\cdot([s_8]+2[a_1])=\tfrac12\text{deg}(t)\cdot\text{deg}(d)\,
\eeq
according to \eqref{eq:[V(p)]}, which is half the product of the degrees of $t$ and $d$ as expected.

Performing a local computation around an intersection point in $V(\mathfrak{p})$, we can show explicitly that the tangents of
$t=0$ and $d=0$ become parallel when they intersect. This implies that locally around intersection points with $t=0$, the 
constraint $d$  becomes a perfect square to leading order, so that the monodromy cover \eqref{eq:I3monoCover}
is reducible, corresponding to a Kodaira singularity of type $I_3^{\text{s}}$.

\section{UnHiggsing of two U(1)'s: concrete examples}
\label{sec:Examples}

In this section we systematically describe examples of the unHiggsing
of two U(1) factors.  Following the spirit of Section
\ref{sec:U1sandUnHiggsing}, we
discuss first the rank-preserving unHiggsings to SU(2) $\times$ SU(2)
and SU(3),  and
then those to a fully fledged SU(2) $\times$ SU(2)
$\times$ SU(3) hybrid model.
The discussion here is based on the unHiggsings of type A described in
the previous section; in the completely general cases where approach A
cannot be used directly, as discussed earlier we can unHiggs using
approach D, giving a model which is equivalent to one with unHiggsings
of type A but with a reducible divisor $\tilde{a}_2$.

\subsection{UnHiggsing of two U(1)'s to SU$(2)\times$SU(2)}
\label{sec:SU2SU2unHiggsing}

We begin with the discussion of Abelian F-theory models that admit an unHiggsing to SU$(2)\times$SU(2) with adjoints. The 
two U(1)-factors are embedded as the Cartan generators of SU$(2)\times$SU(2) in the adjoint Higgsing
\beq \label{eq:U1xU1->SU2SU2}
		\xymatrixcolsep{4pc}\xymatrix{\text{U}(1)\times \text{U}(1)  \ar[r]^{\hspace{-0.35cm}\text{adjoints}}_{\hspace{-0.3cm}b_2,\,b_3\rightarrow 0}& \ar[l] \text{SU}(2)\times \text{SU}(2)\,.}
\eeq
As we will show below, these Higgsings occur only for certain
Calabi-Yau manifolds $X_{n+1}$ with $b_1=0$ and $[a_1]=0$, {\it
  i.e.}~cases that are included in the specialized model of Section
\ref{sec:TheFiber}, if we tune $b_2,\,b_3\rightarrow 0$ as indicated.
In addition, we have to have $[a_2],\,[a_3]\geq -K_B$, which ensures
the presence of adjoints of the two SU(2)'s on these divisors,
and $[s_8]=0$, which is
required to guarantee the absence of the additional SU(3) gauge group.

\subsubsection{Geometric discussion}

To proceed more explicitly, we rewrite the effectiveness condition
for the section $s_1$ in \eqref{eq:coeffsSectionsGen} in terms of the classes of $\lambda_1$, 
$\lambda_2$ and $t$ defined in \eqref{eq:DivsUnHiggsed} as
\beq \label{eq:masterCond}
	-6K_B-2[\lambda_1]-2[\lambda_2]-3[t]\geq 0\,.
\eeq
From this it is clear that the two SU(2)'s can only both carry adjoints, which is the case if 
$[\lambda_1]$, $[\lambda_2]$ are of the form $-K_B+Z$ for an effective class $Z$, if the class of $t$ 
is strictly smaller than $-K_B$. As $[t]\geq -K_B$ by 
\eqref{eq:SU2fromA1} unless $b_1\equiv 0$ (the choice $a_1\equiv 0$ is 
equivalent under $v\leftrightarrow w$), which also imposes $[a_1]=0$ by the assumption of a 
smooth Abelian model\footnote{Recall  the presence of an $I_2$ singularity at $a_1=0$ if $b_1\equiv 0$.}, 
we conclude that only the specialized models \eqref{eq:abcdModel} can admit an unHiggsing 
to two SU(2)'s with adjoints. In this case we have $\lambda_1\equiv a_2$, $\lambda_2\equiv a_3$ and
the Calabi-Yau manifold $X^{(0)}_{n+1}$ describing the unHiggsed theory  is given by \eqref{eq:fullyTunedModelSpecial}.

In order to have a clear unHiggsing to two SU(2)'s we have to require $[s_8]=0$ so that the SU(3) in 
$G_{\text{uni}}$ is  absent. Then the two parameters $b_2$ and $b_3$
that are switched off in the  tuning \eqref{eq:tuningSpecialized} triggering the unHiggsing obey the relation
\beq \label{eq:dofmatchingSU2SU2}
	[b_i]=[a_i]+K_B\,,\quad i=2,3\,
\eeq
as follows from \eqref{eq:coeffsSections2}. Thus, the number of
independent VEVs
of the $1+\tfrac12[a_i]\cdot([a_i]+K_B)$ adjoint Higgses agrees
precisely with the number of monomials in the respective $b_i$ that is
tuned for the unHiggsing, as can be seen on a concrete base, {\it
  e.g.}~$B_n=\mathbb{P}^2$.

Geometrically, the unHiggsing \eqref{eq:U1xU1->SU2SU2} corresponds to a transition of the generators of the Mordell-Weil
group of $X_{n+1}$ corresponding to the U(1)'s into the Cartan divisors of the two $I_2$-singularities supporting the 
SU(2) gauge groups. In other words,
a horizontal divisor given by a section of the elliptic fibration of $X_{n+1}$ is transformed into a vertical divisor
in $X^{(0)}_{n+1}$ localized at a codimension one locus in the base $B_n$ .

In the Weierstrass form, the unHiggsing $b_2$, $b_3\rightarrow 0$ of the Abelian theory
produces the  model \eqref{eq:WSFSU2SU2SU3special}. As pointed out 
earlier it precisely agrees with the general Weierstrass model \eqref{eq:WSFSU2SU2SU3gen} with
two $I_2$- and one $I_3^{\text{s}}$-singularities.  If we set $[s_8]=0$ the $I_3^s$-singularity is absent and we 
simply reproduce the  most general model with two $I_2$-singularities, {\it cf.}~\eqref{eq:WSFSU2SU2}, 
with the identification
\beq
	\sigma=a_2\,,\qquad \tau=a_3\,.
\eeq
This direct matching of Weierstrass models with two $I_2$ singularities ensures that any 
such model can be obtained by unHiggsing our Abelian model defined via $X_{n+1}$.

\subsubsection{UnHiggsing from matching of spectra}

Next, we note that the identification of Abelian models that unHiggs to two  SU(2)'s 
can performed directly from investigation of the spectrum of the general Abelian model.

Recalling the discussion of Section \ref{sec:HiggsingSU2SU2}, an Abelian  
model with two U(1)'s arising from a broken SU(2)$\times$SU(2) has to have a charge lattice as shown 
in Figure \ref{f:2-2-weights}, {\it i.e.}~the matter with charges $(-2,-2)$, $(-2,-1)$ and $(-1,-2)$ has to be 
absent while matter with charges  $(2,0)$ and $(0,2)$ has to be generically present.
It follows directly from Table \ref{tab:MultiesGenModel} that the unique solution to this is precisely 
$[a_1]=[s_8]=0$, consistent with the above geometric arguments. 

Furthermore,
the spectrum of the Abelian model of Table \ref{tab:MultiesGenModel}  then automatically assumes
the form of  an  SU(2)$\times$SU(2) gauge theory. Indeed, we find that  the non-zero multiplicities
of the Abelian theory obey the relations
\begin{align} \label{eq:relationsHiggsingSU2SU2}
	x_{(2,0)}&=[a_2]\!\cdot\! (K_B + [a_2])= 2x_{(\mathbf{3},\mathbf{1})}-2\,,\quad\qquad\,
	x_{(0,2)}=[a_3]\!\cdot\! (K_B + [a_3])= 2x_{(\mathbf{1},\mathbf{3})}-2\,,\nn\\
	x_{(1,0)}&= \!-4 [a_2] \!\cdot\!(4K_B  + [a_2] + [a_3])= 2x_{(\mathbf{2},\mathbf{1})}\,,\,\,\, x_{(0,1)}=\!-4 [a_3] \!\cdot\!(4K_B + [a_2] + [a_3])= 2x_{(\mathbf{1},\mathbf{2})}\,,\nn\\
	&\qquad \qquad\qquad \qquad	x_{(-1,1)}=x_{(1,1)}=2 [a_2]\!\cdot\! [a_3]= 2x_{(\mathbf{2},\mathbf{2})}\,,
\end{align}
with the multiplicities  $x_{\mathbf{R}}$ of matter in a 
representation $\mathbf{R}$  of the SU(2)$\times$SU(2) gauge group. 
These relations ensure that the spectrum of the theory
falls into representations of SU(2)$\times$SU(2) in the limit $b_2$, $b_3\rightarrow 0$. In fact, 
\eqref{eq:relationsHiggsingSU2SU2} follows directly from the branching rules of 
SU(2)$\times$SU(2) representations in an adjoint Higgsing  as discussed in Section 
\ref{sec:HiggsingSU2SU2} and a comparison of Tables \ref{tab:spectrumUnHiggsedSpecial} and
\ref{tab:Multies_Simpl}. Note that we have to subtract 
$1$ from the multiplicities
$x_{(\mathbf{3},\mathbf{1})}$, $x_{(\mathbf{1},\mathbf{3})}$ of the adjoints as one field is eaten 
up by the massive gauge bosons, respectively. 

We conclude by  noting that the relations \eqref{eq:relationsHiggsingSU2SU2} formally follow also 
from the general relations \eqref{eq:relationsUnHiggsing}. To this end, we have to set all multiplicities 
to zero that involve non-trivial representations of SU(3), which includes the $-2$ in  $x_{(-1,1)}$.

\subsubsection{Examples: models on $\mathbb{P}^2$}

We conclude by demonstrating the general observations made in this section
for a simple and concrete example. 
We consider a Calabi-Yau threefold $X_3$ with base $B_2=\mathbb{P}^2$. We solve the 
effectiveness constraints \eqref{eq:coeffsSections2} and perform the tunings 
\eqref{eq:tuningSpecialized} of the complex structure of $X_3$ yielding $X^{(0)}_3$. We then  
obtain the following list of models with $[s_8]=0$:
\beq\label{eq:P2withSU2SU2}
\begin{tabular}{|c|c||c|c|c|c|c|c|c|c|} \hline
$[a_2]$ & $[a_3]$& $b_2$ & $b_3$ & $s_1$ & $s_2$ & $s_3$ & $s_5$ & $s_6$ & $s_8$ \rule{0pt}{1Em}\\ \hline
 6& 3& 3& 0& 0& 3& 6& 0& 3& 0 \\
 5& 4& 2& 1& 0& 3& 6& 0& 3& 0 \\
 5& 3& 2& 0& 2& 4& 6& 1& 3& 0 \\
 4& 5& 1& 2& 0& 3& 6& 0& 3& 0 \\
 4& 4& 1& 1& 2& 4& 6& 1& 3& 0 \\
 4& 3& 1& 0& 4& 5& 6& 2& 3& 0 \\
 3& 6& 0& 3& 0& 3& 6& 0& 3& 0 \\
 3& 5& 0& 2& 2& 4& 6& 1& 3& 0 \\
 3& 4& 0& 1& 4& 5& 6& 2& 3& 0 \\
 3& 3& 0& 0& 0& 6& 6& 3& 3& 0 \\
 \hline
\end{tabular}
\eeq
As expected, this precisely reproduces the list of models \eqref{eq:HiggsableSU2OnP2} based on 
the completely general WSF \eqref{eq:WSFSU2SU2}. In addition, we check the necessary relation 
between the degrees of the polynomials, $[a_i]-[b_i]=3$, $i=2,3$. This ensures that the number
of adjoint VEVs agrees precisely with the monomials in $b_2$, $b_3$, that have been switched
off in the unHiggsing procedure.

We emphasize again that the Abelian theory resulting from Higgsing SU$(2)\times$SU(2) via adjoints 
can not be described by F-theory on the $dP_2$-elliptic fibrations in \cite{Borchmann:2013jwa,Cvetic:2013nia,Cvetic:2013uta,Borchmann:2013hta,Cvetic:2013jta}.
Thus, we see that the Calabi-Yau manifolds $X_{n+1}$ are crucial for describing the branch in the moduli space
of F-theory with two Abelian factors.

\subsection{UnHiggsings of two U(1)'s to SU(3)}
\label{sec:SU3unHiggsing}

Next, we turn to F-theory compactifications with two U(1)'s that are embedded into the Cartan subalgebra of an SU(3) via
the adjoint Higgsing
\beq \label{eq:U1xU1->SU3}
		\xymatrixcolsep{4pc}\xymatrix{\text{U}(1)\times \text{U}(1)  \ar[r]^{\hspace{0.45cm}\text{adjoints}}_{\hspace{0.5cm}\parbox{1.9cm}{\centering\scriptsize$b_2,\, b_3\rightarrow 0$ or
		\\
	 $\Delta_{12},\,\Delta_{13}\rightarrow 0$}}& \ar[l] \text{SU}(3)\,.}
\eeq
We will demonstrate that this is only possible for models with either $[a_1]=0$ and $[s_8]\geq -K_B$ by tuning 
$b_2=b_3\equiv 0$ or with $[a_1],\, [b_1]> 0$ (where $[s_8]$ might or might not obey 
$[s_8]\geq -K_B$) by tuning $\Delta_{12}=\Delta_{13}\equiv 0$.\footnote{We note that models with $[a_1]>0$ and $[b_1]=0$, which implies $[s_8]<-K_B$ and $[s_3]>-K_B$, are equivalent upon $v\leftrightarrow w$ to models with $[a_1]=0$ 
and $[s_8]>-K_B$. Thus, we do not consider them here.} These conditions
guarantee the presence of adjoints of the SU(3). Additionally, we require $[a_2]=[a_3]=0$ in the first case and $[a_1]=[a_2]=[a_3]$ in the second case to ensure that no SU(2)'s are present.
We emphasize that the latter models admit SU(3)-divisors with ordinary double point singularities, giving rise to symmetric
matter representations.

\subsubsection{Geometric discussion}

We organize our discussion by analyzing the two types of models with unHiggsings to SU(3) separately.

\paragraph{Models with $[a_1]=0$ and $[s_8]\geq -K_B$:}
~\\
\indent 
For models with $[a_1]=0$ (which implies $[b_1]\geq 0$, yet allows us
to put $b_1\equiv 0$  by shifting $v\mapsto \frac1{a_1}(v-b_1w)$) we have to 
demand $[t]\equiv [s_8]\geq -K_B$  in order to have an SU(3) with adjoints. The absence of the SU(2) 
factors in \eqref{eq:unHiggsedGG} further requires to consider models $X^{(0)}_{n+1}$ with  
$[a_2]=[a_3]=0$, as follows from \eqref{eq:DivsUnHiggsedSpecial}. 
In addition, $[a_i]=0$ implies 
\beq
[b_i]=K_B+[s_8]\,,\qquad  i=2,3\,,
\eeq 
according to \eqref{eq:coeffsSectionsGen},  which implies that the number of VEVs of adjoints of the 
SU(3) agrees with the number of monomials in $b_2$, $b_3$ that are tuned to zero for the 
unHiggsing. 

Geometrically, as before the unHiggsing tranforms the generators of
the Mordell-Weil group of $X_{n+1}$ into the Cartan divisors of the
resolved $I_3^{\text{s}}$-singularity.  This is the analog of the
field theoretical statement that the two U(1)'s are embedded as the
Cartan generators of the SU(3) in the unHiggsed theory.

The Weierstrass form of the models at hand is given by
\eqref{eq:WSFSU2SU2SU3special} and precisely agrees with the general
Weierstrass model \eqref{eq:WSFSU2SU2SU3gen} with two $I_2$- and one
$I_3^{\text{s}}$-singularities.  If we set $[a_2]=[a_3]=0$ the SU(2)'s
disappear and we simply recover the most general model with one
$I_3^{\text{s}}$-singularity in \eqref{eq:WSFSU3} upon identifying
$t=s_8$.  Thus, by unHiggsing the considered Abelian models defined by
$X_{n+1}$ we reproduce the list of all F-theory models with one
$I^{\text{s}}_3$-singularity on a smooth divisor $t$.

\paragraph{Models with $[a_1],\,[b_1]> 0$:}
~\\
\indent We begin by noting that the class $[t]$ is automatically
effective
and satisfies $[t] \geq  [-K_B]$ as long as the term 
$a_1b_1 s_6\subset t$ is present, which is the case if both $a_1$ and $b_1$ are non-trivial 
sections\footnote{If {\it e.g.}~$a_1$ is 
constant the term involving $s_6$ in $t$ can be removed by redefining $s_8$ and we are 
back in a situation with $b_1=0$.}. Thus, we see that the SU(3) on
$t=0$ always has adjoints 
that can be used to Higgs the theory back to a model with $\text{U}(1)^2$ gauge 
symmetry.  We recall that the divisor $t=0$ has ordinary double point
singularities  that support
symmetric plus anti-symmetric representations, as shown in Section \ref{sec:NewWSFs}.

If we further assume that $[\lambda_1]=[\lambda_2]= 0$ are both trivial 
the gauge group of the tuned model $X_{n+1}^{(0)}$ is just to SU(3).  Then, we have a clear 
embedding of the two U(1)'s into the maximal torus of SU(3).
We observe that for 
$[\lambda_1]=[\lambda_2]=0$ we have $[a_1]=[a_2]=[a_3]$. Thus, the degrees of the sections
$\Delta_{12}$ and $\Delta_{13}$ defined in \eqref{eq:Delta12=0} and
\eqref{eq:FullTuning}, respectively, are equal. The number of monomials tuned to
zero in the unHiggsing by imposing $\Delta_{12},\,\Delta_{13}\rightarrow 0$ is given
{\it e.g.}~for $B_n=\mathbb{P}^2$ by
\beq \label{eq:dofunHiggsingSU3}
	([a_1]+2)([a_1]+1)+([a_1]+[s_8]-1)([a_1]+[s_8]-2)-2= 2 [a_1]([a_1]+[s_8])+[s_8]([s_8]-3)+2
\eeq
This agrees precisely with the number of VEVs of the 
$2x_{(\mathbf{1},\mathbf{1},\mathbf{8})}$ in Table \ref{tab:MultiesGenModel}, where the factor 
of $2$ is included due to the rank of SU(3). 

The Weierstrass model we obtain by unHiggsing is given by \eqref{eq:WSFSU2SU2SU3} for
$\lambda_1=\lambda_2=1$. As pointed out earlier, it deviates from the standard form of an 
$I^{\text{s}}_3$ singularity but still yields a model with SU(3) gauge group as shown in 
Section \ref{sec:NewWSFs}. We note that at the current stage it is unknown whether every
Calabi-Yau elliptic fibration with $I^{\text{s}}_3$ singularities over a divisor $t=0$ with ordinary 
double points can be written as the Weierstrass model \eqref{eq:WSFSU2SU2SU3} or whether 
unHiggsing $X_{n+1}$ reproduces every model of the form \eqref{eq:WSFSU2SU2SU3}. 

We emphasize that the unHiggsed theory has \textit{two} discrete degrees of freedom. These two 
parameters are the divisor classes  $[s_8]$ and $[a_1]=[a_2]=[a_3]$. The latter sets the 
number of ordinary double 
point singularities of $t=0$. Geometrically, the appearance of the extra parameter is expected due to 
the special structure of the Weierstrass model \eqref{eq:WSFSU2SU2SU3} which is key to the presence 
of an SU(3) gauge group. In contrast, if we remove the additional substructure of $t=0$ by imposing 
$[a_1]=0$ we are left with only one parameter $[s_8]$, which is the class of the $t$ supporting the SU(3), as expected.

\subsubsection{UnHiggsing from matching of spectra}

Again, we can rediscover those Calabi-Yau manifolds $X_{n+1}$
which arise from  unHiggsing to an SU(3) with adjoints
from the point of view of the matter spectrum of Abelian theory.

First we assume that there are no matter fields in the representation $(-2,-2)$ in Table 
\ref{tab:MultiesGenModel}. Then, we recall from Section \ref{sec:Higgsing-3} that an Abelian model with 
two U(1)'s which arises from
a broken SU(3) has to have the charge lattice shown in Figure \ref{f:3-weights}. Notably, matter with
charges $(2,0)$, $(0,2)$ has to be absent while matter with charges $(2,1)$, $(1,2)$ and $(-1,1)$
should generically be present. The unique solution to this is given by $[a_1]=[a_2]=[a_3]=0$, 
as follows from Table \ref{tab:MultiesGenModel}. We note that the remaining multiplicities
automatically assume the form
\bea \label{eq:relationsHiggsingSU3}
	&x_{(2,1)}=x_{(1,2)}=x_{(-1,1)}=[s_8]\cdot([s_8]+K_B)=2x_{\mathbf{8}}-2\,,&\nn\\ 
	&x_{(1,0)}=x_{(0,1)}=x_{(1,1)}=-3[s_8]\cdot(3K_B+[s_8])=x_{\mathbf{3}}\,,&
\eea
which implies that the matter spectrum falls into representations of SU(3) if we unHiggs by $b_2$, 
$b_3\rightarrow 0$. Here, we denote the multiplicity of an SU(3) representation $\mathbf{R}$
by $x_{\mathbf{R}}$. 
Indeed, the relations \eqref{eq:relationsHiggsingSU3} readily follow from the branching
rules of representations in an adjoint Higgsing given in Section \ref{sec:Higgsing-3} and a comparison
between Tables \ref{tab:spectrumUnHiggsedSpecial} and \ref{tab:Multies_Simpl}. Note that we have to 
subtract one adjoint as it is eaten up by the massive gauge bosons. 
We can also  obtain the relations \eqref{eq:relationsHiggsingSU3} formally from
the general relations \eqref{eq:relationsUnHiggsing} by dropping all multiplicities of 
SU$(2)\times$SU(2) representations.

In order to include matter fields in the representation $(-2,-2)$ all we have to do is to relax the 
condition of a vanishing multiplicity $x_{(-2,-2)}$. As the discussion in Section 
\ref{sec:Higgsing-singular}
shows, the only source for this representation is the symmetric representation
of SU(3). The branching  \eqref{eq:branching4} of the symmetric suggests that we have to demand
$x_{(-2,-2)}=x_{(-2,0)}=x_{(0,-2)}$. A unique solution  to this in a natural parametrization is given by
\beq \label{eq:sola1a2a3s8}
	[a_1]=[a_2]=[a_3]\,, \quad  [s_8]=[t]-2[a_1]
\eeq
for an effective class $[t]$.  Using
this relation we clearly reproduce the relations \eqref{eq:relationsUnHiggsing} after setting
the multiplicities of all SU(2)-representations to zero. Thus, in an unHiggsing $t$ plays the role of the 
divisor class supporting the SU(3) and $[a_1]$ is related to the number of symmetric representations. The entire spectrum of the Abelian theory falls into representations of SU(3).

\subsubsection{Examples: models on $\mathbb{P}^2$}

We conclude  by constructing explicit models for the base $B_2=\mathbb{P}^2$. 
We start with an Abelian model specified by the CY-threefold $X_{3}$, solve the effectiveness 
conditions \eqref{eq:coeffsSectionsGen} and impose the conditions for
an enhancement to SU(3). Then we tune the complex structure in order to unHiggs
to $X^{(0)}_{3}$. We obtain the following list of models:
\beq \label{eq:SU(3)modelsonP2}
\begin{tabular}{|c|c||c|c|c|c|c|c||c|} \hline
$[a_i]$ & $[s_8]$ & $b_i$& $s_1$ & $s_2$ & $s_3$ & $s_5$ & $s_6$ & $[t]$  \\ \hline
  0& 6& 3& 0& 0& 0& 3& 3& 6\\
  0& 5& 2&  3& 2& 1& 4& 3& 5\\
  0& 4& 1& 6& 4& 2& 5& 3& 4\\
0& 3&  0& 9& 6& 3& 6& 3& 3\\
  \hline \hline
 1& 4& 2& 0& 1& 2& 2& 3& 6 \\ 
  1& 3& 1& 3& 3& 3& 3& 3& 5\\
\hline
\end{tabular}
\eeq
We note that the first four lines reproduce precisely the list of
models with a standard $I_3^{\text{s}}$ singularity  
on a smooth divisor
found in \eqref{eq:HiggsableSU3onP2}, where we identify $t=s_8$. 
The last two lines are models corresponding to the two possible solutions to the 
condition 
that $[s_1]$ is effective when the
 divisor $t$ is of the form
\eqref{eq:tDivisor} with an ordinary double point. Indeed, for 
$[\lambda_1]=[\lambda_2]=0$ the effectiveness constraint
on $[s_1]$
reduces to
\beq
3[t]+[s_1]=18\,,
\eeq
which is solved only by $[t]=[a_1]+[b_1]+3=5,6$ with $[s_8]=3,4$ 
and $[a_1]=[b_1]=1$ or $[a_1]=[b_1]-1=1$, respectively.\footnote{Note that there exists a model with $[a_1]-1=[b_1]=1$ and 
$[s_8]=2$. It is mapped to the model $[a_1]=[b_1]-1=1$ and $[s_8]=4$ using the symmetry $v\leftrightarrow v$.}

Again, we emphasize that the new Calabi-Yau manifolds $X_{n+1}$ are needed to have a 
geometrical description in F-theory of the Higgs branch of an SU(3) gauge theory broken by adjoints 
to two U(1)'s. The relevant
Abelian theories can not be obtained from the $dP_2$-elliptic fibrations
in 
\cite{Borchmann:2013jwa,Cvetic:2013nia,Cvetic:2013uta,Borchmann:2013hta,Cvetic:2013jta}, unless $[s_8]=3$.

There are some caveats regarding the completeness of our construction
of models with an SU(3) on a divisor with double points in
$\mathbb{P}^2$.  As we have seen in \eqref{eq:SU(3)modelsonP2}, we
only obtain models with one or two double points on a quintic or a
sextic, respectively.  However, we expect that models on
$\mathbb{P}^2$ with more double points or an SU(3) divisor of a lesser
degree exist.  In particular, we expect the existence of an SU(3) on a
quartic with one adjoint and up to two double points, on a quintic
with 
an adjoint and
up to five double points and on a sextic with an adjoint and up
to  nine ordinary double points. On a technical level, the reason for
the limitations of our model is the term $b_1a_1s_6$ in $t$, which
imposes the condition $[t]-3=[a_1]+[b_1]$ fixing the number of double
points. Currently, we do not know of any geometrical or physical
reason for this constraint, which is intrinsic to our constructions,
to hold universally, or how to see such a constraint from the
low-energy theory.

\subsection{Unhiggsing of two U(1)'s to SU$(2)\times$SU$(2)\times$SU(3)}
\label{sec:SU2SU2SU3Unhiggsing}

Finally, we consider Abelian F-theory models with two U(1)'s that
unHiggs to a model $X^{(0)}_{n+1}$ in which the full non-Abelian gauge
group $G_{\text{uni}}=\text{SU}(2)\times \text{SU}(2)\times
\text{SU}(3)$ is realized. In this case, the two U(1)'s are embedded
diagonally into the Cartan subalgebra of $G_{\text{uni}}$.

Depending on the discrete parameters specifying $X_{n+1}$, the Abelian theory can be obtained
from the non-Abelian one by different Higgsings. While all Abelian
models considered here arise from  breaking  $G_{\text{uni}}$ by the bifundamentals
$(\mathbf{2},\mathbf{1},\mathbf{3})$ and $(\mathbf{1},\mathbf{2},\mathbf{3})$, 
\beq \label{eq:U1xU1->SU2SU2SU3}
		\xymatrixcolsep{4pc}\xymatrix{\text{U}(1)\times \text{U}(1)  \ar[r]^{\hspace{-1.1cm}\text{bifunds.}}& \ar[l] \text{SU}(2)\times\text{SU}(2)\times\text{SU}(3)\,,}
\eeq
certain models admit an alternative Higgsing via adjoints and  fundamentals. This alternative Higgsing
can be viewed as a two-stage version of the bifundamental Higgsing. It illuminates the geometrical 
process which turns a vertical into a horizontal divisor, {\it i.e.}~a section into a Cartan divisor. Thus, we
will discuss it in detail.

There are four different types of models $X_{n+1}$ that unHiggs to $G_{\text{uni}}$. They are
\begin{itemize}
	\item specialized models ($[a_1]=0$) with $[a_2],\,[a_3]\geq -K_B$, as in 
	Section \ref{sec:SU2SU2unHiggsing}, but with $[s_8]> 0$. 
	The  Higgsing alternative to \eqref{eq:U1xU1->SU2SU2SU3} is 
	by adjoints of the two SU(2)'s and a consecutive Higgsing
	by fundamentals of the residual SU(3) with appropriate U(1)$\times$U(1) charges.
	\item specialized models with  $[s_8]\geq -K_B$, as in Section \ref{sec:SU3unHiggsing}, but without $[a_2]=[a_3]=0$ and models with $[a_1],\,[b_1]>0$, 
	again as in Section \ref{sec:SU3unHiggsing}, but relaxing $[\lambda_1]=[\lambda_2]=0$.
	As an alternative to \eqref{eq:U1xU1->SU2SU2SU3} we can Higgs by adjoints of the SU(3) and
	break the residual SU$(2)\times$SU(2) by fundamentals with U(1)$\times$U(1) charges.
	\item specialized models with $[a_2]< -K_B$ but $[a_3]\geq-K_B$ or vice versa and 
	$[s_8]<-K_B$. In this case we
	can Higgs one SU(2) with adjoints, then Higgs the remaining SU$(2)\times$SU(3) by a bifundamental
	and finally Higgs the residual SU(2) by a fundamental with non-trivial U(1) charges.
	\item specialized models with  $[a_2],\,[a_3]<-K_B$ and $[s_8]< -K_B$. In this case, there is no
	alternative Higgsing to \eqref{eq:U1xU1->SU2SU2SU3}.
\end{itemize}

We note that some non-Abelian theories defined by $X^{(0)}_{n+1}$
have Higgs branches parameterized by Higgses in the bifundamental  representation
$(\mathbf{2},\mathbf{2},\mathbf{1})$, different fundamentals (with trivial U(1) charges) and with 
other final gauge symmetries. The analysis of these branches is beyond the scope of this article and is 
left for future works.

\subsubsection{Geometric discussion \& comparison to field theory}

We will analyze the four different types of models in the order outlined above.

\paragraph{Models with $[a_1]=0$, $[a_2],\,[a_3]\geq -K_B$ and $[s_8]>0$:}
~\\
\indent 
These models are analyzed similarly to those discussed in Section \ref{sec:SU2SU2unHiggsing}, 
however relaxing the condition
$[s_8]= 0$. In this case, the tuning $b_2$, $b_3\rightarrow 0$ unHiggsing the 
U(1)'s produces the model $X^{(0)}_{n+1}$. 
First, by tuning $b_2\equiv 0$ we get the gauge group 
$\text{SU}(2)\times\text{SU}(2)\times\text{U}(1)$, {\it cf.}~the discussion before 
\eqref{eq:GGafterfirstUnHiggs}. Then, we get the full  gauge group $G_{\text{uni}}$ if we also set 
$b_3\equiv 0$ as shown before \eqref{eq:GGUnHiggsed}. The 
Weierstrass model of $X^{(0)}_{n+1}$ is given by \eqref{eq:WSFSU2SU2SU3}.

In order to understand the embedding of
the U(1)'s into the Cartan subalgebra of $G_{\text{uni}}$ from a geometric point of view  we note
that $[s_8]>0$, using \eqref{eq:coeffsSections2}, implies
\beq 
	[b_i]\geq [s_8	]\,,\qquad i=2,3\,. 
\eeq
Thus, the degree of the $b_i$ is larger than that of $s_8$. Thus, we can tune the complex structure of 
$X_{n+1}$   so that
\beq \label{eq:SU(3)tuning}
	b_2=s_8 b'_2\,,\qquad b_3=s_8b_3'
\eeq
for appropriate sections $b'_2$ and $b'_3$. We obtain an F-theory model with two U(1)'s and an 
extra $I_3^{\text{s}}$-singularity at $s_8=0$, as can be seen by inspecting of the Weierstrass form 
\eqref{eq:WSFcubic} with \eqref{eq:fullf}, \eqref{eq:fullg} for $X_{n+1}$. 
Thus, the SU(3) gauge group at $s_8$ can already be induced prior to unHiggsing 
the U(1)'s in the Abelian model. The U(1)'s are unHiggsed into two SU(2)'s by tuning  
$b_2'=b_3'\equiv 0$  corresponding to an adjoint Higgsing as in Section \ref{sec:SU2SU2unHiggsing}.

The field theoretic Higgsing corresponding to the intermediate tuning 
\eqref{eq:SU(3)tuning} can be inferred by noting  that the U(1) charges of the theory are changed. 
Indeed, geometrically,  the Shioda map is altered due to the
presence of the Cartan divisors of the SU(3) \cite{shioda1990mordell,Park:2011ji}.  
This change shifts the U(1) charges of all matter fields in such a way that the new spectrum of the
theory after the tuning \eqref{eq:SU(3)tuning} is of the form of Figure \ref{f:2-2-weights}, 
while the original smooth model $X_{n+1}$ has a spectrum like in Figure \ref{f:223-weights}.
Thus, the field-theoretical analogue of \eqref{eq:SU(3)tuning} has to be a Higgsing
by fundamentals $\mathbf{3}_{(q_1,q_2)}$ with appropriate U(1) charges $(q_1,q_2)$. The correct 
U(1) charges are identified by starting with the spectrum in Table \ref{tab:spectrumUnHiggsedSpecial} 
of the unHiggsed model $X^{(0)}_{n+1}$. Then, we Higgs the two SU(2)'s by adjoints and identify 
the corresponding triplets $\mathbf{3}_{(q_1,q_2)}$ by
the requirement that the resulting spectrum matches Table \ref{tab:Multies_Simpl}. 
We find that
the correct states are precisely the triplets $\mathbf{3}_{(-1,0)}$ and $\mathbf{3}_{(0,1)}$ that
are inside the bifundamentals $(\mathbf{2},\mathbf{1},\mathbf{3})$ and 
$(\mathbf{1},\mathbf{2},\mathbf{3})$, respectively. Thus, the U(1)-generators of $X_{n+1}$
are embedded into the Cartans of $G_{\text{uni}}$ as in \eqref{eq:embeddingU1s}, {\it i.e.}~the 
bifundamental Higgsings.

We note that the spectrum of the Abelian
theory obeys  the relations \eqref{eq:relationsUnHiggsing} with $x_{(-2,-2)}=0$.
It is important to emphasize that these relations are simultaneously 
compatible with a Higgsing by bifundamentals as in \eqref{eq:U1xU1->SU2SU2SU3} and  by
a combination of adjoint and fundamental Higgses, as just described.

We summarize the chain of Higgsings that are possible if we invoke the tunings \eqref{eq:SU(3)tuning}  
in the following diagram: 
\begin{equation} \label{eq:223byadjoints+funds}
	\xymatrixcolsep{4pc}\xymatrix{
	\text{\phantom{aa....}U}(1)^2\phantom{aa...}\ar@{->}[r]^{\hspace{-0.8cm}b_2\rightarrow 0} &  \text{SU}(2)\times \text{U}(1)\times \text{SU}(2)\ar@{->}[r]^{b_3\rightarrow 0} &   \text{SU}(2)\times \text{SU}(2)\times \text{SU}(3)\ar@{<-}[ld]^{b_3'\rightarrow 0}\\
	\text{U}(1)^2\times \text{SU}(2)\ar@{->}[d]_{b_2=b_2's_8}\ar@{->}[r]^{\hspace{-0.8cm}b_2\rightarrow 0}\ar@{<-}[u]^{b_3=b_3's_8}& \text{SU}(2)\times \text{U}(1)\times \text{SU}(3)\ar@{<-}[u]^{b_3=b_3's_8}&\\
	\text{U}(1)^2\times \text{SU}(3)\ar@{->}[ru]_{b_2'\rightarrow 0}&&
	}
\end{equation}
Here we indicate the relevant tunings of the complex structure of the
Calabi-Yau manifold $X_{n+1}$ next to the respective arrow. 
The first line of \eqref{eq:223byadjoints+funds} is identical to 
\eqref{eq:U1xU1->SU2SU2SU3}.
Vertical arrows correspond to  a
Higgsing by a fundamental with non-trivial U(1) charges, horizontal arrows are bifundamental Higgsings  and diagonal arrows
are SU(2)
adjoint Higgsings. The latter correspond to tunings that unHiggs U(1)'s into Cartan generators of 
SU(2)'s. 

We note that starting from the non-Abelian theory $X^{(0)}_{n+1}$ there are other possible Higgs 
branches. Higgsing by the bifundamental $(\mathbf{2},\mathbf{2},\mathbf{1})$ leads to a theory with 
just one U(1), while Higgsings by fundamentals of SU(2)'s lead to a theory with no residual Abelian gauge 
group. These Higgsings are not considered here, as they are not related to the Abelian model $X_{n+1}$.

\paragraph{Models with $[a_1]=0$ and $[s_8]\geq -K_B$ or $[a_1],\,[b_1]>0$:}
~\\
\indent 
These models are analyzed in analogy to those in Section
\ref{sec:SU3unHiggsing}.  However, the full gauge group
$G_{\text{uni}}$ is realized as can be seen {\it e.g.}~from the
Weierstrass model given by \eqref{eq:WSFSU2SU2SU3special}.  Indeed,
models with $[a_1]=0$ are unHiggsed by tuning $b_2,\,b_3\rightarrow 0$
to $X_{n+1}^{(0)}$ also in the case if the conditions $[a_2]=[a_3]= 0$
are relaxed. Similarly, models with $[a_1],$ $[b_1]>0$ are unHiggsed
  by imposing \eqref{eq:Delta12=0} and \eqref{eq:FullTuning}. This is
  solved, assuming $[a_2]-[a_1]\geq 0$, $[a_3]-[a_1]\geq 0$, by
  \eqref{eq:tuningSU2} and \eqref{eq:secondtuning} with non-trivial
  $[\lambda_1]$ and $[\lambda_2]$.  In other cases, we are in
  situation \textbf{D)} in Section \ref{sec:TuningModel}, which can be
  made equivalent to a case with $[a_1]=0$ by setting $b_1 = 0$, as
  discussed there.
We assume in the rest of this section that we are in the situation
where $[a_2]-[a_1]\geq 0$, $[a_3]-[a_1]\geq 0$ applies.

Geometrically, we understand the embedding of the U(1)s into the maximal torus of $G_{\text{uni}}$ 
by noting that the condition $[s_8]\geq -K_B$ implies, by virtue of \eqref{eq:coeffsSections2}, that 
\beq
 		[b_i]\geq [a_i]\,,\qquad i=1,2,3\,,
\eeq
{\it i.e.}~the degrees of the $b_i$ are larger than that of the corresponding $a_i$. Thus, we can tune
the complex structure of $X_{n+1}$ so that
\beq \label{eq:generateSU2SU2}
	b_2=a_2b_2'\,,\qquad b_3=a_3b_3'\,,
\eeq
for appropriate sections $b_2'$, $b_3'$. The resulting F-theory models still have two rational
sections generating a rank two MW-group  but
also have $I_2$-singularities at $a_2=0$ and $a_3 = 0$.
In fact, the resulting elliptic fibration is described by
\beq
	p=uf_u+a_2a_3v(v+b_2'w)(v+b_3'w)\,,
\eeq
which clearly has $I_2$ singularities at $a_2=0$ and $a_3 = 0$, yet
has three distinct rational points. The  
relevant tunings that reduce the MW-rank are then given by $b_2',b_3'\rightarrow 0$, which 
creates the $I_3^{\text{s}}$ singularity at $s_8=0$. 

Analogously, a general model $X_{n+1}$ with $[a_1],$ $[b_1]>0$ can be tuned by setting
\beq \label{eq:generateSU2SU2gen}
	a_i=\lambda_{i-1} a_i'\,,\quad b_i=\lambda_{i-1} b_i'
\eeq
for appropriate sections $a_i'$, $b_i'$ for $i=2,3$ with $[a_2']=[a_1]$ and  $[a_3']=[a_1]$. 
The CY-constraint  \eqref{eq:cubicfactorized} then assumes the form
\beq \label{eq:intermediateSU2}
p = u f_u(u,v,w) +\lambda_1 \lambda_2 (a_1 v + b_1 w)(a_2'v+b_2'w) (a_3' v+ b_3' w )\,,
\eeq
so we see that the tuning \eqref{eq:generateSU2SU2gen} maintains all rational sections, but introduces 
already $I_2$ singularities corresponding to the SU(2)'s at $\lambda_1=0$ and 
$\lambda_2=0$. 
In this case, the tunings that unHiggs the U(1)'s to the SU(3) associated to an $I_3^{\text{s}}$ singularity at $t=0$, {\it cf.}~Section \ref{sec:NewWSFs}, are given by
\beq
	a_1b_i'-b_1a_i'\equiv 0
\eeq
for both $i=2$, $3$ yielding the CY-manifold $X_{n+1}^{(0)}$ with gauge group 
$G_{\text{uni}}$. 

In field theory, the intermediate tunings \eqref{eq:generateSU2SU2}, \eqref{eq:generateSU2SU2gen} 
corresponds to a Higgsing
by fundamentals of the SU(2)'s which carry non-trivial U(1)-charges. Again, this is clear from
the change in the Shioda map of the rational sections due to the SU(2)'s. The correct matter fields
are in the representations $(\mathbf{2},\mathbf{1})_{(1,0)}$ and
$(\mathbf{1},\mathbf{2})_{(0,-1)}$ that originate from the bifundamentals $(\mathbf{2},\mathbf{1},\mathbf{3})$ and 
$(\mathbf{1},\mathbf{2},\mathbf{3})$, respectively, so that the U(1)'s are embedded via 
\eqref{eq:embeddingU1s} into the Cartan generators of $G_{\text{uni}}$.

The spectrum of the Abelian theories obtained from models $X_{n+1}$ with $[a_1]=0$ obey the 
relations \eqref{eq:relationsUnHiggsing} with $x_{(-2,-2)}=0$ while models with $[a_1],\,[b_1]>0$ 
allow for a non-zero multiplicity $x_{(-2,-2)}$. We note again that the relations 
\eqref{eq:relationsUnHiggsing} are compatible both with a Higgsing
by bifundamentals or by adjoints and fundamentals, as the one just described. 

We summarize the chain of Higgsings we just discussed that also allows for the tunings 
\eqref{eq:generateSU2SU2} in the following diagram: 
\begin{equation} \label{eq:223bySU3adjoints+funds}
	\xymatrixcolsep{4pc}\xymatrix{
	\text{\phantom{aa....}U}(1)^2\phantom{aa...}\ar@{->}[r]^{\hspace{-0.8cm}b_2\rightarrow 0} &  \text{SU}(2)\times \text{U}(1)\times \text{SU}(2)\ar@{->}[r]^{b_3\rightarrow 0} &   \text{SU}(2)\times \text{SU}(2)\times \text{SU}(3)\ar@{<-}[ld]^{b_3'\rightarrow 0}\\
	\text{SU}(2)\times\text{U}(1)^2\ar@{->}[d]_{b_2=a_2b_2'}\ar@{->}[r]^{\hspace{-0.8cm}b_2\rightarrow 0}\ar@{<-}[u]^{b_3=a_3 b_3'}& \text{SU}(2)^2\times \text{U}(1)\times \text{SU}(2)\ar@{<-}[u]^{b_3=a_3b_3'}&\\
	\text{SU}(2)^2\times \text{U}(1)^2 \ar@{->}[ru]_{b_2'\rightarrow 0}&&
	}
\end{equation}
Here  the relevant tunings of the complex structure of the
Calabi-Yau manifold $X_{n+1}$ are indicated next to the respective arrow. 
The first line of \eqref{eq:223bySU3adjoints+funds} is the Higgsing in  
\eqref{eq:U1xU1->SU2SU2SU3}.
Vertical arrows correspond to  a
Higgsing by a fundamental with non-trivial U(1) charges, horizontal arrows are bifundamental Higgsings  
and diagonal arrows are adjoint Higgsings of the SU(3). The latter correspond to tunings that unHiggs 
U(1)'s into Cartan generators of SU(3)'s. 

We emphasize that starting from the non-Abelian theory $X^{(0)}_{n+1}$ there are other possible 
Higgs branches. Higgsing by the bifundamental $(\mathbf{2},\mathbf{2},\mathbf{1})$ and then by the 
adjoints of SU(3) leads to  a theory with three U(1)'s, while Higgsings by fundamentals of SU(3)'s and 
then by the bifundamental of the SU(2)'s leads to a theory with one Abelian gauge 
group. These Higgsings are not considered here as they are not relevant for obtaining the Abelian model 
$X_{n+1}$.

We conclude with the observation that the effectiveness condition \eqref{eq:masterCond} 
implies that whenever $[t]\geq -K_B$ the model can at most have one SU(2) with an adjoint. 
Consequently there are models with adjoints for one SU(2) and the SU(3). These models exhibit an 
adjoint Higgsing to a theory with U$(1)^3\times$SU(2) gauge group.

\paragraph{Models with $[a_1]=0$, $[a_2]< -K_B$, $[a_3]\geq -K_B$ and $[s_8]<-K_B$:}
~\\
\indent 
The main distinction of the following discussion from the above is that the CY-manifolds 
$X_{n+1}$ considered here do not yield non-Abelian theories with adjoints of rank two gauge 
groups. In fact, for models with $[a_1]=0$, $[s_8]<-K_B$ and $[a_2]< -K_B$, $[a_3]\geq -K_B$ only
one SU(2) in $X^{(0)}_{n+1}$ carries adjoints. Thus, we can Higgs the theory on this adjoint 
yielding one U(1). The remaining non-Abelian group SU$(2)\times$SU(3) is Higgsed on a bifundamental
to SU(2)$\times$U(1) and then on a charged fundamental to U(1).   As before the spectrum of the Abelian and the non-Abelian theory obey the relations \eqref{eq:relationsUnHiggsing}.
Geometrically,  there exists only one intermediate tuning
\beq
	b_3=b_3's_8\,,
\eeq
for appropriate $b_3'$ which creates an SU(2) on $s_8=0$ prior to unHiggsing any U(1).

The chain of (un-)Higgsings possible in this model is given by
\begin{equation} \label{eq:2byadjoints+funds}
	\xymatrixcolsep{4pc}\xymatrix{
	\text{\phantom{aa....}U}(1)^2\phantom{aa...}\ar@{->}[r]^{\hspace{-0.8cm}b_2\rightarrow 0} &  \text{SU}(2)\times \text{U}(1)\times \text{SU}(2)\ar@{->}[r]^{b_3\rightarrow 0} &   \text{SU}(2)\times \text{SU}(2)\times \text{SU}(3)\ar@{<-}[ld]^{b_3'\rightarrow 0}\\
	\text{U}(1)^2\times \text{SU}(2)\ar@{->}[r]^{\hspace{-0.8cm}b_2\rightarrow 0}\ar@{<-}[u]^{b_3=b_3's_8}& \text{SU}(2)\times \text{U}(1)\times \text{SU}(3)\ar@{<-}[u]^{b_3=b_3's_8}&
	}
\end{equation}
As before the relevant tunings of the complex structure of $X_{n+1}$ are indicated next to the 
respective arrow, vertical arrows correspond to Higgsings by charged fundamentals, horizontal arrows are Higgsings by bifundamentals and diagonal arrows are adjoint Higgsings.

We note that the number of alternative Higgsings of the non-Abelian 
theory defined by $X_{n+1}^{(0)}$ is limited. All Higgsings alternative to \eqref{eq:2byadjoints+funds}
yield theories with one or no U(1). 

\paragraph{Models with $[a_1]=0$ and $[a_2],\,[a_3],\,[s_8]<-K_B$:}
~\\
\indent 
For these models no intermediate tuning \eqref{eq:SU(3)tuning} of $X_{n+1}$ is possible.
Thus, the unHiggsing of the Abelian sector is induced by $b_2$, $b_3\rightarrow 0$ and cannot be 
split into steps. We obtain a non-Abelian model described by $X^{(0)}_{n+1}$ with gauge group 
$G_{\text{uni}}$ with a spectrum related to that of the Abelian model via the relations \eqref{eq:relationsUnHiggsing}. The only (un-)Higgsing chain possible is the one in 
\eqref{eq:U1xU1->SU2SU2SU3}:
\begin{equation} \label{eq:noadjoints}
	\xymatrixcolsep{4pc}\xymatrix{
	\text{U}(1)^2\ar@{->}[r]^{\hspace{-1.5cm}b_2\rightarrow 0} &  \text{SU}(2)\times \text{U}(1)\times \text{SU}(2)\ar@{->}[r]^{b_3\rightarrow 0} &   \text{SU}(2)\times \text{SU}(2)\times \text{SU}(3)
	}
\end{equation}

\subsubsection{Examples: models on $\mathbb{P}^2$}

We conclude the discussion of models with an unHiggsing to a
non-Abelian theory with the full gauge group $G_{\text{uni}}$ 
(or in special cases $SU(2) \times SU(3)$, when one $SU(2)$ is on a
trivial divisor)
by
constructing explicit models for the base $B_2=\mathbb{P}^2$.  As
before, we start with an Abelian F-theory model specified by the
threefold $X_{3}^{(2)}$ which is in one of the four classes of models
discussed in this section. Then we tune the model appropriately to
the threefold $X^{(0)}_{3}$.
 We discuss the four different classes of
models separately.
Note again that we assume in this discussion that we are in 
situation {\bf A)}, of the possibilities distinguished above.

The list of models $X_{3}$ on $B_2=\mathbb{P}^2$ with
$[a_1] = 0$,
 $[a_2],\,[a_3]\geq 3$ and 
$[s_8]>0$ is given by:
\beq
\begin{tabular}{|c|c||c|c|c|c|c|c|c|c|} \hline
$[a_2]$ & $[a_3]$& $b_2$ & $b_3$ & $s_1$ & $s_2$ & $s_3$ & $s_5$ & $s_6$ & $s_8$ \rule{0Em}{1Em}\\ \hline
3 & 3 & 1 & 1 & 3 & 4 & 5 & 2 & 3 & 1 \\
 3 & 3 & 2 & 2 & 0 & 2 & 4 & 1 & 3 & 2 \\
 3 & 4 & 1 & 2 & 1 & 3 & 5 & 1 & 3 & 1 \\
 \hline
\end{tabular}
\eeq 
These
three models
 are all the models in the list  \eqref{eq:SU2SUSU3OnP2} of theories with 
SU(2)$\times$SU(2)$\times$SU(3) gauge group over 
$B_2=\mathbb{P}^2$ that meet the requirement  $[a_2],\,[a_3]\geq 3$, {\it i.e.}~have SU$(2)\times$SU(2) with adjoints. 
Thus, we have reproduced the complete list of such models by unHiggsing the Abelian theories defined by $X_{3}$.

The list of models $X_{3}$ on $B_2=\mathbb{P}^2$ with $[s_8]\geq 3$ and 
$[a_2],$
or
$[a_3]\geq0$ reads:
\beq
\begin{tabular}{|c|c||c|c|c|c|c|c|c|c|} \hline
$[a_2]$ & $[a_3]$& $b_2$ & $b_3$ & $s_1$ & $s_2$ & $s_3$ & $s_5$ & $s_6$ & $s_8$ \rule{0Em}{1Em}\\ \hline
 0 & 1 & 2 & 3 & 1 & 1 & 1 & 3 & 3 & 5 \\
  0 & 1 & 1 & 2 & 4 & 3 & 2 & 4 & 3 & 4 \\
 0 & 1 & 0 & 1 & 7 & 5 & 3 & 5 & 3 & 3 \\
 0 & 3 & 1 & 4 & 0 & 1 & 2 & 2 & 3 & 4 \\
   0 & 3 & 0 & 3 & 3 & 3 & 3 & 3 & 3 & 3 \\
  0 & 2 & 1 & 3 & 2 & 2 & 2 & 3 & 3 & 4 \\
 0 & 2 & 0 & 2 & 5 & 4 & 3 & 4 & 3 & 3 \\
 0 & 4 & 0 & 4 & 1 & 2 & 3 & 2 & 3 & 3 \\
 1 & 1 & 1 & 1 & 5 & 4 & 3 & 4 & 3 & 3 \\
 1 & 1 & 2 & 2 & 2 & 2 & 2 & 3 & 3 & 4 \\
 1 & 2 & 1 & 2 & 3 & 3 & 3 & 3 & 3 & 3 \\
 1 & 2 & 2 & 3 & 0 & 1 & 2 & 2 & 3 & 4 \\
 1 & 3 & 1 & 3 & 1 & 2 & 3 & 2 & 3 & 3 \\
 2 & 2 & 2 & 2 & 1 & 2 & 3 & 2 & 3 & 3 \\
 \hline
\end{tabular}
\eeq 
Again,
 we reproduce all models in the list  in
\eqref{eq:SU2SUSU3OnP2} of theories 
with SU(2)$\times$SU(2)$\times$SU(3) gauge group over 
$B_2=\mathbb{P}^2$ that meet the requirement of having an SU(3) with adjoints.

The number of models $X_{3}$ on $B_2=\mathbb{P}^2$ with $[a_1],\,[b_1]>0$ is very limited.
Indeed, by inspection of the last term of $g$ in  \eqref{eq:WSFSU2SU2SU3}, we obtain
the equality
\beq \label{eq:conditionsSU223model}
	-6K_B-2[\lambda_1]-2[\lambda_2]-3[t]=[s_1]\geq 0\,,
\eeq
which agrees precisely with the condition \eqref{eq:masterCond} we have derived from the Abelian model $X_{n+1}$.
This condition is very constraining as both classes $[a_1]$ and $[b_1]$ have to be strictly non-zero so that 
$[t]=3+[a_1]+[b_1]\geq 5$ on $\mathbb{P}^2$. The only solutions to it are $[a_1]=[b_1]=1$ for $[t]=5$, which allows for 
$[\lambda_1]=[\lambda_2]=0$ or $[\lambda_1]=0$, $[\lambda_2]=1$, and $[a_1]=[b_1]-1=1$, which only allows for 
$[\lambda_1]=[\lambda_2]=0$. Thus, we only obtain one solution with one SU(2), in addition to the two solutions in 
\eqref{eq:SU(3)modelsonP2}:
\beq
\begin{tabular}{|c|c|c||c|c|c|c|c|c|c|c|c|} \hline
$[a_1]$ &$[a_2]$ & $[a_3]$& $b_1$& $b_2$ & $b_3$ & $s_1$ & $s_2$ & $s_3$ & $s_5$ & $s_6$ & $s_8$ \rule{0Em}{1Em}\\ \hline
 1 & 1 &2 & 1 & 1 & 2 & 1 & 2 & 3 & 2 &3&3 \\
 \hline
\end{tabular}
\eeq 
This is a model with an SU(3) on a quintic in $\mathbb{P}^2$ with one ordinary double point singularity at $a_1=b_1=0$ and 
an SU(2) on a line $\lambda_2=0$. As discussed earlier we reproduce the full list of models compatible with the structure of 
Weierstrass models of the form \eqref{eq:WSFSU2SU2SU3} and divisor $t=0$ of the form \eqref{eq:tDivisor}. However, we 
currently do not know whether this is the complete list of models with
an SU(3) on a divisor 
 in 
$\mathbb{P}^2$
with ordinary double points that carry symmetric representations.

The list of models $X_{3}$ on $B_2=\mathbb{P}^2$ with $[s_8]< 3$, 
$[a_3]\geq 3$ and $[a_2]<3$ together with models having $[a_2],\,[a_3],\,[s_8]< 3$ reads:
\beq
\begin{tabular}{|c|c||c|c|c|c|c|c|c|c|}
\hline
 \multicolumn{10}{|l|}{Models with $[s_8]< 3$, 
$[a_3]\geq 3$ and $[a_2]<3$ \rule{0pt}{1Em}}\\
 \hline
$[a_2]$ & $[a_3]$& $b_2$ & $b_3$ & $s_1$ & $s_2$ & $s_3$ & $s_5$ & $s_6$ & $s_8$ \rule{0Em}{1Em}\\ \hline
 1 & 3 & 0 & 2 & 4 & 4 & 4 & 3 & 3 & 2 \\
 1 & 4 & 0 & 3 & 2 & 3 & 4 & 2 & 3 & 2 \\
 1 & 5 & 0 & 4 & 0 & 2 & 4 & 1 & 3 & 2 \\
 2 & 3 & 0 & 1 & 5 & 5 & 5 & 3 & 3 & 1 \\
 2 & 3 & 1 & 2 & 2 & 3 & 4 & 2 & 3 & 2 \\
 2 & 4 & 0 & 2 & 3 & 4 & 5 & 2 & 3 & 1 \\
 2 & 4 & 1 & 3 & 0 & 2 & 4 & 1 & 3 & 2 \\
 2 & 5 & 0 & 3 & 1 & 3 & 5 & 1 & 3 & 1 \\
 \hline
 \multicolumn{10}{|l|}{Models with $[a_2],\,[a_3],\,[s_8]< 3$\rule{0pt}{1Em}}\\
 \hline
 1 & 1 & 0 & 0 & 8 & 6 & 4 & 5 & 3 & 2 \\
 1 & 2 & 0 & 1 & 6 & 5 & 4 & 4 & 3 & 2 \\
 2 & 2 & 0 & 0 & 7 & 6 & 5 & 4 & 3 & 1 \\
 2 & 2 & 1 & 1 & 4 & 4 & 4 & 3 & 3 & 2 \\
 \hline
\end{tabular}
\eeq 
We note that this covers all
remaining models in \eqref{eq:SU2SUSU3OnP2}  that can be Higgsed supersymmetrically 
to an Abelian theory with two U(1)'s. There are models that do not admit such Higgsings, either because the realized 
non-Abelian gauge group is too small or because there are less than two matter fields
in both bifundamental representations $(\mathbf{2},\mathbf{1},\mathbf{3})$ and $(\mathbf{1},\mathbf{2},\mathbf{3})$, 
respectively. In the latter case, a D-flat Higgsing is impossible. We note, though, that some of those models do have a 
Higgsing to a model with one U(1). It is satisfying that the effectiveness conditions \eqref{eq:coeffsSectionsGen} and 
\eqref{eq:coeffsSections2} required by the existence of a smooth and generic model $X_{n+1}$ reproduce these field 
theory results.

\section{Further research directions} 
\label{sec:further}

While we believe that the model constructed in this paper gives a
fairly complete picture of the most general classes of F-theory models
that can be constructed with two Abelian factors U(1)$\times$U(1),
many of the results we have found suggest generalizations that  may
have much broader consequences for our understanding of F-theory
models and the corresponding supergravity theories.
In particular, the general structure of the unHiggsing patterns
identified in the U(1)${}^2$ models considered here suggests that
similar patterns may govern F-theory models with arbitrary numbers of
Abelian factors U(1)${}^k$.  Our construction has also given rise to an
explicit realization of a class of models where a non-Abelian gauge
group is supported on a singular curve, giving rise to a non-generic
matter representation.  
In this section we look ahead to how these features may be generalized
in future work.

\subsection{More U(1) factors}

In the most general class of U(1) $\times$ U(1) models that we have
considered here, the charge spectrum (Figure~\ref{f:charges-general})
can be identified with the set of charges arising when a non-Abelian
group SU(2)$\times$SU(2)$\times$SU(3) is broken by Higgsing.  As we
have shown, this corresponds to the fact that the generic model can be
``unHiggsed'' by transforming the horizontal divisors (sections)
associated with the two U(1) factors into vertical divisors in the
total space of the Calabi-Yau that lie over divisors of the form $A C$
and $B C$ in the base of the elliptic fibration.  Here, $C$ is a
common factor of the two divisors, and picks up the rank  two SU(3)
non-Abelian group, while $A$ and $B$ support SU(2) gauge factors.

This suggests a natural structure for general models with more U(1)
factors.  For three U(1) factors U(1)$\times$U(1)$\times$U(1), for
example, we could imagine transforming all three horizontal divisors
into vertical divisors.  The most general structure for the divsors
in the base that would support the three non-Abelian factors would be
\begin{equation}
A_1 B_{12}B_{13}C, A_2 B_{12}B_{23}C, A_3 B_{13}B_{23}C \,.
\label{eq:}
\end{equation}
Each $A_i$
divisor would then support an SU(2) factor, each $B_{i j}$ divisor
would support an SU(3) factor, and $C$ would support a rank three
factor such as SU(4).   The general U(1)${}^3$ model\footnote{For a special construction of F-theory compactifications with U(1)${}^3$ see \cite{Cvetic:2013qsa}.} could then be
realized by Higgsing a non-Abelian theory with gauge group
SU(2)${}^3\times$SU(3)${}^3 \times$SU(4).  Of course, in many cases some
of the factors would be trivial, with many rank three Abelian groups
realized from Higgsings of rank three non-Abelian groups such as $\text{SU}(2)
\times \text{SU}(2) \times \text{SU}(2),\, \text{SU}(2) \times \text{SU}(3),$ and 
$\text{SU}(4)$. 
Degenerate divisors $A_i$, $B_{ij}$ and $C$ could also lead to larger product groups in an 
unHiggsing. As in
the case of one or two U(1) factors, the presence of additional
non-Abelian gauge groups in the original theory having U(1) factors,
and/or additional geometrical structure in the base could in some
situations make the unHiggsed non-Abelian theory singular, or bring the
theory to a superconformal fixed point where a geometric transition
would be necessary to give a smooth Calabi-Yau.

It is straightforward to generalize this kind of construction to an
arbitrary number of U(1) factors.  A very general class of $\text{U}(1)^k$
models could then be constructed from the Higgsing of a non-Abelian
theory with a gauge group $\text{SU}(2)^k \times \text{SU}(3)^{k (k -1)/2}\times
\cdots \times \text{SU}(k)^k \times \text{SU}(k + 1)=\prod_{i=1}^{k} \text{SU}(i+1)^{c_i}$,
where  the combinatoric  factor is
$c_i={k \choose i}$.
For larger ranks, we could also have other non-Abelian groups arising
in the unHiggsing.  For rank 2, we did not need to consider $G_2$
separately, since it can be Higgsed to SU(3) on the adjoint without
reducing the rank.  As the rank increases, however, we  can
realize U(1) factors by Higgsing more general groups including
$\text{SO}(N)$ factors, exceptional groups such as $E_6, E_7,  E_8$, etc.  This
will give rise to a rich collection of possibilities for higher rank
Abelian groups that can be realized from Higgsing higher rank
non-Abelian structures.

While this type of unHiggsing construction might be expected to give
the most general classes of models with any number of Abelian factors
there are a number of issues that would need to be addressed.  First,
while this story seems likely to hold in the low-energy supergravity
theory, the realization in F-theory through an explicit Weierstrass
model is quite nontrivial.  Even for two U(1) factors, the general
model with explicit formulae given in Appendix \ref{app:WSF}  are quite complex,
and do not follow in any simple way even though we can explicitly
construct Weierstrass models in principle for the unHiggsed $\text{SU}(2)
\times \text{SU}(2) \times \text{SU}(3)$ model more directly.  While as discussed
 in \cite{Morrison:2012ei, Morrison:2014era}, we have an explicit
formulation of the Higgsing of a single SU(2) factor in terms of
Weierstrass models, and as discussed here this can be extended to
theories with multiple non-Abelian factors, this construction becomes
progressively more complicated as the number of factors increases.
Another issue that would need to be addressed for any complete
understanding of models with more Abelian factors is the analysis of
special cases where the horizontal divisors cannot be unHiggsed to
vertical divisors of a non-Abelian theory with the same rank.  Given
that some examples of this are already encountered for two U(1)
factors, we expect an even more exotic set of special cases when the
number of U(1) factors is increased.
Despite these technical issues, however, it seems that exploring the
set of $\text{U}(1)^3,\, \text{U}(1)^4, \ldots$ models that can be realized in
F-theory through Higgsings of rank $k$ and higher non-Abelian groups
holds promise as a way of gaining some systematic understanding of
these rather difficult theories.

\subsection{Exotic matter representations}

Another issue that has arisen in this paper, in which we have likely
only touched on the tip of a rather complex iceberg of possibilities,
is the appearance of non-standard matter representations.
As discussed in Section \ref{sec:U1sandUnHiggsing}, an SU(2) or SU(3)
factor that is tuned in F-theory on a smooth divisor $D$ will carry
generically only matter in the fundamental and adjoint
representations.  In 6D models, the number of adjoints simply
corresponds to the genus $g$ of the curve $D$.  As discussed in 
\cite{Sadov:1996zm,Morrison:2011mb}, however, when the divisor $D$ is
itself singular, other representations of the gauge group can arise.
A simple formula was derived in \cite{Kumar:2010am}
from anomaly cancellation, which determines the contribution to the
arithmetic genus of a curve $D$ that should be associated with the
singularity carrying any given  matter representation.  For the
adjoint and symmetric representations, the contribution is 1.  Thus, in particular, as was originally
suggested by Sadov in \cite{Sadov:1996zm}
(see also \cite{Bershadsky:1996nh}),
we expect that tuning an SU(3) on certain divisors with double point
type singularities should give rise to matter in the symmetric
representation.  Indeed, a large class of the models we considered here can be
associated with the Higgsing of SU(2) $\times$SU(2)$\times$SU(3)
non-Abelian theories, where the SU(3) lies on a curve of the form
\begin{equation}
t = A x^2 + B x y + C y^2 \,,
\label{eq:singular-curve}
\end{equation}
where there are double point singularities at the points $x = y = 0$.
In the simplest case, a 6D model on $\P^2$ with $x, y$ defining lines
and $A, B, C$ defining cubics, there is a single double point
singularity at $x = 0$.  This corresponds to a symmetric
representation of SU(3), which carries charge $(2, 2)$ in the Higgsed
U(1)$\times$U(1) theory.  As we have discussed in earlier sections,
this double point singularity apparently cannot be removed without changing the
gauge group of the theory.

This is one of the first situations where we have an explicit example
of a non-generic matter representation arising on a singular curve.
At an algebraic level, Weierstrass models with this kind of structure
are rather subtle.  This can be illustrated clearly in a simple
example.  Consider tuning an $\text{SU}(N)$ on a generic divisor $\sigma(x,
y)$, where $x, y$ are coordinates on the base.  As analyzed in detail
in \cite{Morrison:2011mb}, this can be done systematically by tuning
the coefficients in an expansion $f = f_0+ f_1 \sigma + f_2 \sigma^2 +
\cdots, g = g_0+ g_1 \sigma + g_2 \sigma^2 + \cdots$, where $f_0 = -3u
(x, y)^2, g_0 = 2u (x, y)^{3}$, and similar conditions hold at higher
orders so that the discriminant cancels at leading orders in $\sigma,$
with {\it e.g.}, $4f_0^3 = -108u^6 = - 27g_0^2$.  If $\sigma$ is a
singular curve, however, such as for example $\sigma = 4 x^3+27y^2$,
we can have $f_0 = x, g_0 = y$, and the discriminant will vanish to
leading order.  This is the kind of condition that is automatically
satisfied by the form (\ref{eq:singular-curve})
in the context we have encountered it here.  An outstanding
challenge for F-theory is to develop tools and understanding for
systematically constructing such singular Weierstrass models to
realize arbitrary gauge groups and matter representations that are
allowed by low-energy supergravity consistency conditions such as
anomaly cancellation.

Within the context of the models developed here, one question is
whether there is a construction for models with U(1) charges
associated with arbitrary numbers of
symmetric representations of SU(3).  In particular, considered over
$\P^2$, the singular curve (\ref{eq:singular-curve}) must be of degree
at least five.  It should be possible, however, to construct F-theory
models with an SU(3) on a singular quartic over $\P^2$.  The quartic
has genus three, so we should be able to have one or two symmetric
representations, and then Higgs on the adjoint representation leaving
one or two charge $(2, 2)$ states in the low-energy U(1)$\times$U(1)
model.

More generally, we might hope to construct models with higher U(1)
charges from more exotic representations. Even for models with a
single U(1), we expect that there are solutions with matter having
charge $q=3$, see \cite{Klevers:2014bqa} for the Abelian model, associated with Higgsing of an SU(2) having matter in the
three-index symmetric representation.  At this time no explicit
geometric F-theory realization of such a matter representation is yet
known.  Even more exotic matter representations can in principle
arise when we
consider higher-rank groups.

\subsection{Matching 6D supergravity} 

It was conjectured in \cite{Kumar:2009us} that every low-energy 6D
supergravity theory can either be realized in string theory or suffers
from some quantum inconsistency.  The close connection between the
structure of 6D supergravity and F-theory \cite{Kumar:2009ac,Kumar:2010ru}
suggests that F-theory may provide a framework for systematically
constructing vacua in each branch of the complete moduli space of 6D
supergravity theories, with different branches associated with
different bases for the F-theory compactification connected through
tensionless string transitions.  At this point, however, ``string
universality'' for 6D supergravity theories has not been proven.
F-theory geometry places certain additional constraints on low-energy
physics.  While some such constraints have been understood as
consistency conditions that must be satisfied by the low-energy
supergravity theories \cite{Seiberg:2011dr}, other such constraints are
not yet understood from the low-energy point of view.  Theories with
Abelian factors provide an important regime for testing the
correspondence between the set of F-theory constructions and
consistent supergravity theories, in which there are many open
questions.  Some analysis of the set of consistent solutions to the 6D
anomaly equations in the presence of Abelian factors was carried out
in \cite{Erler:1993zy, Park:2011wv}.  In the context of the work in this paper,
the question is whether a correspondence can be found between the set
of apparently consistent 6D supergravity theories with two Abelian
factors and the set of F-theory constructions.

Two specific  examples of situations where this kind of question arise
have been encountered in this paper.  A first question is whether the
general model we have described here in fact captures all F-theory
constructions with two U(1) factors.  As we have noted, the $dP_2$
U(1)$\times$U(1) models identified in \cite{Borchmann:2013jwa,Cvetic:2013nia,Cvetic:2013uta,Borchmann:2013hta,Cvetic:2013jta} give rise to different 
charge spectra from the generic class of models we have constructed
here, though these models are also described in the framework given
here by relaxing the constraints on the divisor classes associated
with the relevant parameters.
It would be interesting to study whether there are other exceptional
cases such as this, some of which may go beyond the construction
presented here.

Another significant class of examples relates to the exotic matter
configurations described in the previous section.  Considering even
just symmetric matter representations of SU(3), there are a variety of
U(1)$\times$U(1) models that would seem natural from the low-energy
point of view that are not reproduced by our general class of
constructions.  In particular, since under anomaly cancellation an
adjoint behaves identically to a combination of asymmetric and a
fundamental representation of SU(3), we can consider a general model
where SU(3) is realized on a curve of genus $g$, and replace anywhere
from 1 to $g -1$ of the generic adjoint representations with a
symmetric + fundamental combination of representations.  Only a small
subset of these are realized through the model constructed here.  For
example, as mentioned above we could tune an SU(3) gauge group on the
base $\P^2$ on a curve $C$ of degree 4, which has genus $3$.
According to low-energy considerations only, it seems that it should
be possible to replace one or two adjoint representations with a
symmetric + fundamental, and Higgs on the remaining adjoint, which
would give a U(1)$\times$U(1) model with one or two charged matter
fields in the $(2, 2) + (-2, 0) + (0, -2)$ representations.  This is
impossible to realize, however, in the spectrum described in
Table \ref{tab:MultiesGenModel}.  Logically there are several possibilities.  It
may be that these theories are actually inconsistent due to as-yet not
understood compatibility conditions between low-energy field theory
and quantum gravity.  It may be that these theories are consistent but
cannot be realized in F-theory.  It may be that these theories can be
realized in F-theory but in a very different way than using the
construction of this paper.  Determining which of these possibilities
is correct may be a fruitful way of uncovering further the beautiful
and subtle correspondence between geometry and physics that F-theory
provides.


\section{Conclusions}
\label{conclusions}

We conclude by  summarizing  the key results of the paper:
\begin{itemize}

\item

We have given what we believe is the most general construction of a globally
defined F-theory compactication with U(1)$\times$U(1) gauge factors.
Specifically, we have constructed a general class of elliptically fibered
Calabi-Yau manifolds with rank two Mordell-Weil group.  The elliptic
fibration is based on the elliptic curve $\mathcal{E}$ with three
(non-toric) rational points at general positions, which is realized as
a specialized cubic in $\mathbb{P}^2$.  We have obtained the Weierstrass form
for this general class of models.  Though the focus is on Calabi-Yau
threefolds, the construction is general and is valid for an arbitrary
F-theory base manifold.  The work is a generalization of the earlier
constructions
\cite{Borchmann:2013jwa,Cvetic:2013nia,Cvetic:2013uta,Borchmann:2013hta,Cvetic:2013jta,Klevers:2014bqa}
where the rational points in in $\mathcal{E}$ were chosen at specific
positions and assumed to be toric points in $\mathbb{P}^2$.

\item 
We have provided a detailed analysis of the codimension two singularities
of these elliptically fibered Calabi-Yau manifolds, and have
provided their global
resolution, which is represented as a complete intersection in an
$n+4$-dimensional ambient space, constructed by two blow-ups at two non-toric
points in the ambient space of the elliptic fiber.  This allows for
the full determination of charges and multiplicities for the matter
spectrum of general global 6D U(1)$\times$U(1) F-theory models.  We
also explicitly checked that the matter spectra cancel all 6D
anomalies.

The 
matter spectra of the models constructed here are more general and
contain additional matter representations, compared
to those of specialized constructions
\cite{Borchmann:2013jwa,Cvetic:2013nia,Cvetic:2013uta,Borchmann:2013hta,Cvetic:2013jta,Klevers:2014bqa}.
In particular, they  include representations with $(-2,-2)$, $(-2,1)$
and $(2,0)$ charges under U(1)$\times$U(1).  Representations with
higher charges could emerge in special cases where the elliptic curve
is further specialized.

\item 
We have provided a detailed geometric and field theoretical study of the
general unHiggsing mechanism in the moduli space of globally defined
F-theory models with U(1)$\times$U(1) symmetry.  
These transitions have a geometric
interpretation as the transformation of a rational section to a
Cartan divisor of a non-Abelian gauge symmetry; in such an unHiggsing
process, the nonlocal horizontal divisors, associated with sections in
the Mordell-Weil group, become local vertical divisors, associated
with the Cartan generators of non-Abelian gauge groups of Kodaira
singularities in the elliptic fibration over the base.  One of the
principal results is that these models can be unHiggsed to a
non-Abelian gauge group SU(2)$\times$SU(2)$\times$SU(3).  The
structure of the non-Abelian group can be understood geometrically by
identifying the general form of the vertical divisors, associated with
unHiggsing the two U(1) factors, to be $AC$ and $BC$, where $C$ is a
common factor.  The unHiggsing process leads to SU(2) factors on $A$
and $B$, while on $C$ the two elements of the Cartan generators
combine, resulting in SU(3).  We provide the geometric and the field
theory analysis of the non-Abelian matter spectra that after Higgsing
result in the spectra of general U(1)$\times$U(1) models.  In
particular, we note that the appearance of matter with $(-2,-2)$
charges originates from the symmetric representation of SU(3).

We also note that in some cases the divisors $A$, $B$, and  $C$
can be reducible; for example  $A$
 can support SU(2)$\times$SU(2)
or more SU(2) factors.  In the most  generic such case,  the unHiggsed
symmetry results in SU(2)$^3\times$SU(3). In these cases the Higgsing
to a general U(1)$\times$U(1) is due to bi-fundamental non-Abelian
matter only.

\item 
An important new result that has emerged from the geometric analysis
of these models and the associated unHiggsing mechanism is the first
explicit geometric construction of a symmetric representation of the
unitary gauge symmetry $SU(N)$ for $N > 2$, in this case SU(3).  The associated
Weierstrass models have an intricate and nontrivial algebraic
structure. Specifically, the appearance of a symmetric representation
is associated with the tuning of an SU(3) divisor with a double point
type singularity that cannot be removed while retaining the $I_3$
Kodaira singularity type. It is expected that tuning of more intricate singular divisors
supporting non-Abelian groups could result in other higher-dimensional
representations, such as three-index symmetric tensor representations.
These higher-dimensional representations are expected to play a role
in the unHiggsing of Abelian models with larger Abelian charges.

\item The U(1)$\times$U(1) construction suggests a generalization to
  theories with any number of U(1) factors; for a model with U(1)$^k$
  factors with $k >2$, we expect that an unHiggsing of $k$-horizontal
 divisors into vertical divisors would result in an unHiggsed
 phase with a gauge group like $\prod_{i=1}^k \text{SU}(i+1)^{c_i}$,
 where $c_i={k \choose i}$.

\end{itemize}

\subsubsection*{Acknowledgments}
We would like to thank Kang Sin Choi, Antonella Grassi, Thomas Grimm, James Halverson,  Andreas Kapfer, Albrecht 
Klemm, Sven Krippendorf, Craig Lawrie, Wolfgang Lerche, David Morrison, Daniel Park,
Maximilian Poretschkin,  Nikhil Raghuram, Soo-Jong Rey, Sakura Sch\"afer-Nameki and Peng Song  for helpful discussions and comments. MC would like to thank the theory group at CERN and DK the EWHA Womans University
for hospitality during the final stages of this project. 
The work is supported in part by the DOE through (HEP) Award DE-SC0013528
(MC,HP)
and (HEP) Award DE-SC00012567 (WT), in part by
the Fay R. and Eugene L. Langberg Endowed Chair (MC), and in part by the Slovenian Research Agency (ARRS) (MC).


\appendix


\section{The Weierstrass form of $X_{n+1}$}
\label{app:WSF}

The Weierstrass form of the elliptic curve \eqref{eq:cubicfactorized} can be computed by first mapping to the cubic in $dP_2$ 
and then using the results of  \cite{Borchmann:2013jwa,Cvetic:2013nia} for its Weierstrass form. Alternatively, one can use 
Delign\'{e}'s approach of constructing 
the Weierstrass coordinates $[z:x:y]$ as sections of appropriate line bundles on $\mathcal{E}$. Either way, we find the 
following expressions for $f$ and $g$:
\begin{footnotesize}
\bea \label{eq:fullf}
f\!\!&\!\!\!=\!\!\!&\!\!-\frac{1}{48} \left(s_6^2\!-\!4 s_3 s_8\right)^2+\frac{1}{2} b_1 b_2 b_3 s_3 \left(2 s_3 s_5\!-\!s_2 s_6\right)+\frac{1}{6}(a_1b_2 b_3 +a_2b_1 b_3+a_3 b_1 b_2)\left(s_2 s_6^2+s_3 \left(2 s_2 s_8\!-\!3 s_5 s_6\right)\right)\nn\\
\!\!\!&\!\!\!-\!\!\!&\!\!\frac{1}{3} (a_3^2b_1^2 b_2^2 +a_2^2b_1^2 b_3^2+a_1^2b_2^2 b_3^2)\left(s_2^2\!-\!3 s_1 s_3\right)+\frac{1}{6} (b_1a_2a_3+b_2a_1a_3+b_3a_1a_2) \left(2 b_1 b_2 b_3 \left(s_2^2\!-\!3 s_1 s_3\right)\!-\!3 s_2 s_6 s_8\right.\nn\\
\!\!\!&\!\!\!+\!\!\!&\!\!\left. s_5 \left(s_6^2+2 s_3 s_8\right)\right)
\!+\!\frac{1}{6} (a_2a_3^2b_1^2 b_2+a_2^2 a_3 b_1^2 b_3+a_1a_3^2b_1 b_2^2+a_1^2a_3b_2^2 b_3+a_1a_2^2b_1 b_3^2+a_1^2a_2b_2 b_3^2 )\left(2 s_2 s_5\!-\!3 s_1 s_6\right)\nn\\
\!\!\!&\!\!\!-\!\!\!&\!\!\frac{1}{3}(a_2^2a_3^2 b_1^2+a_1^2a_2^2b_3^2+b_2^2a_1^2a_3^2-a_1a_2a_3^2b_1 b_2-a_1a_2^2a_3b_1 b_3-a_1^2a_2a_3b_2 b_3) \left(s_5^2-3 s_1 s_8\right)\nn\\
\!\!\!&\!\!\!+\!\!\!&\!\!a_1a_2a_3\big(b_1 b_2 b_3 \left(3 s_1 s_6-2 s_2 s_5\right)+s_2 s_8^2-\frac{1}{2} s_5 s_6 s_8\big)\,,
\eea
\end{footnotesize}
\begin{footnotesize}
\bea \label{eq:fullg}
g\!\!&\!\!\!=\!\!\!&\!\!\frac{1}{864} \left(s_6^2-4 s_3 s_8\right)^3 +\frac{1}{24} b_1 b_2 b_3 s_3 \left(\left(s_2 s_6-2 s_3 s_5\right) \left(s_6^2-4 s_3 s_8\right)+6 b_1 b_2 b_3 s_3 \left(s_2^2-4 s_1 s_3\right)\right)\nn\\
&\!\!-\!\!\!&\!\!\frac{1}{72}(b_2 b_3a_1+b_3 b_1a_2+ b_1 b_2a_3) \left(12 b_1 b_2 b_3 s_3 \left(s_2 s_3 s_5+\left(s_2^2-6 s_1 s_3\right) s_6\right)+\left(s_6^2-4 s_3 s_8\right) \left(s_2 s_6^2+s_3 \left(2 s_2 s_8-3 s_5 s_6\right)\right)\right)\nn\\
&\!\!+\!\!\!&\!\!\frac{1}{36}( b_1^2 b_2^2a_3^2+b_1^2 b_3^2a_2^2+b_2^2 b_3^2a_1^2) \left(3 \left(3 s_5^2-8 s_1 s_8\right) s_3^2+\left(4 s_2^2 s_8-3 s_6 \left(2 s_2 s_5+s_1 s_6\right)\right) s_3+2 s_2^2 s_6^2\right)\nn\\
&\!\!+\!\!\!&\!\!\frac{1}{72} (b_1a_2a_3+b_2a_1a_3+b_3a_1a_2) \left(2 b_1 b_2 b_3 \left(-6 \left(s_5^2+2 s_1 s_8\right) s_3^2+\left(2 s_8 s_2^2+18 s_5 s_6 s_2-33 s_1 s_6^2\right) s_3+s_2^2 s_6^2\right)\right.\nn\\
&\!\!-\!\!&\!\!\left.\left(s_6^2-4 s_3 s_8\right) \left(s_5 \left(s_6^2+2 s_3 s_8\right)-3 s_2 s_6 s_8\right)\right)+\frac{1}{27} (b_1^3 b_2^3a_3^3+b_1^3 b_3^3a_2^3+b_2^3 b_3^3a_1^3) s_2 \left(9 s_1 s_3-2 s_2^2\right)\nn\\
&\!\!+\!\!\!&\!\!\frac{1}{72} (b_1^2 b_2a_2a_3^2+b_1^2 b_3a_2^2a_3+a_1a_2^2b_1 b_3^2+a_1^2a_2b_2 b_3^2+a_1^2a_3b_2^2 b_3+a_1a_3^2b_1 b_2^2) \left(4 b_1 b_2 b_3 s_2 \left(2 s_2^2-9 s_1 s_3\right)\right.\nn\\
&\!\!+\!\!&\!\!\left. s_6 \left(s_6 \left(2 s_2 s_5+3 s_1 s_6\right)-12 s_3 s_5^2\right)+4 \left(s_2 s_3 s_5-3 \left(s_2^2-5 s_1 s_3\right) s_6\right) s_8\right)\nn\\
&\!\!+\!\!\!&\!\!\frac{1}{72}a_1a_2a_3 \left(16 b_1^2 b_2^2 b_3^2 s_2 \left(9 s_1 s_3-2 s_2^2\right)+6 b_1 b_2 b_3 \left(s_6 \left(6 s_3 s_5^2+s_6 \left(9 s_1 s_6-8 s_2 s_5\right)\right)+2 \left(3 \left(s_2^2+2 s_1 s_3\right) s_6-8 s_2 s_3 s_5\right) s_8\right)\right.\nn\\
&\!\!+\!\!&\!\!\left.3 s_8 \left(s_5 s_6-2 s_2 s_8\right) \left(s_6^2-4 s_3 s_8\right)\right)\nn\\
&\!\!+\!\!&\!\!\frac{1}{18} (a_2^3a_3b_1^3 b_3^2+a_3^3a_2b_1^3 b_2^2+a_3^3a_1b_1^2 b_2^3+a_1^3a_3b_2^3 b_3^2+a_1^3a_2b_2^2 b_3^3+a_2^3a_1b_1^2 b_3^3) \left(\left(2 s_2^2-3 s_1 s_3\right) s_5-3 s_1 s_2 s_6\right)\nn\\
&\!\!+\!\!&\!\!\frac{1}{36} (b_1^2a_2^2a_3^2+b_2^2a_1^2a_3^2+b_3^2a_1^2a_2^2) \left(8 b_1 b_2 b_3 \left(-2 s_5 s_2^2+3 s_1 s_6 s_2+3 s_1 s_3 s_5\right)+2 \left(s_6^2+2 s_3 s_8\right) s_5^2-6 s_2 s_6 s_8 s_5\right.\nn\\
&\!\!-\!\!&\!\!\left.3 s_8 \left(s_1 \left(s_6^2+8 s_3 s_8\right)-3 s_2^2 s_8\right)\right)+\frac{1}{36} (a_1^2a_2a_3b_2 b_3 +a_2^2a_1a_3b_1 b_3+a_3^2a_1a_2b_1b_2)\left(4 b_1 b_2 b_3 \left(\left(2 s_2^2-3 s_1 s_3\right) s_5-3 s_1 s_2 s_6\right)\right.\nn\\
&\!\!+\!\!&\!\!\left.\left(s_6^2+2 s_3 s_8\right) s_5^2+18 s_2 s_6 s_8 s_5-6 \left(s_2^2+2 s_1 s_3\right) s_8^2-33 s_1 s_6^2 s_8\right)\nn\\
&\!\!+\!\!&\!\!\frac{1}{18} (b_2 b_3^3a_1^3a_2^2+b_1 b_3^3a_2^3a_1^2+b_2^3 b_3a_1^3a_3^2+b_1 b_2^3a_3^3a_1^2+b_1^3 b_3a_2^3a_3^2+b_1^3 b_2a_3^3a_2^2) \left(s_2 \left(2 s_5^2-3 s_1 s_8\right)-3 s_1 s_5 s_6\right)\nn\\
&\!\!+\!\!&\!\!\frac{2}{9}( b_1^2 b_2^2a_1a_2a_3^3+ b_1^2 b_3^2a_1a_3a_2^3+ b_2^2 b_3^2a_2a_3a_1^3) \left(3 s_1 s_5 s_6+s_2 \left(3 s_1 s_8-2 s_5^2\right)\right)\nn\\
&\!\!+\!\!&\!\!\frac{1}{18} (b_1a_1a_2^2a_3^2+b_2a_2a_1^2a_3^2+b_3a_3a_1^2a_2^2) \left(2 b_1 b_2 b_3 \left(s_2 \left(2 s_5^2-3 s_1 s_8\right)-3 s_1 s_5 s_6\right)-3 s_8 \left(s_6 s_5^2+\left(s_2 s_5-6 s_1 s_6\right) s_8\right)\right)\nn\\
&\!\!+\!\!&\!\!\frac{1}{4} (b_1^4 b_2^2a_2^2a_3^4+b_1^4 b_3^2a_2^4a_3^2+b_2^2 b_3^4a_1^4a_2^2+b_1^2 b_3^4a_2^4a_1^2+b_2^4 b_3^2a_1^4a_3^2+b_1^2 b_2^4a_3^4a_1^2) s_1^2\nn\\
&\!\!-\!\!&\!\!\frac{1}{2} (b_1^3 b_2^3a_1a_2a_3^4+b_2^3 b_3^3a_2a_3a_1^4+b_1^3 b_3^3a_1a_3a_2^4) s_1^2-\frac{1}{54} (a_1^3a_2^3b_3^3+a_2^3a_3^3b_1^3+a_1^3a_3^3b_2^3) \left(9 s_1 \left(3 b_1 b_2 b_3 s_1-2 s_5 s_8\right)+4 s_5^3\right)\nn\\
&\!\!+\!\!&\!\!\frac{1}{18} (b_1^2 b_2a_1a_2^2a_3^3+b_1 b_2^2a_2a_1^2a_3^3+b_1^2 b_3a_1a_3^2a_2^3+b_3 b_1^2a_1a_3^2a_2^3+b_2^2 b_3a_2a_3^2a_1^3+b_2 b_3^2a_3a_2^2a_1^3) \left(9 b_1 b_2 b_3 s_1^2+2 s_5^3-9 s_1 s_8 s_5\right)\nn\\
&\!\!+\!\!&\!\!(a_1a_2a_3)^2(-\frac{3}{2}  b_1^2 b_2^2 b_3^2 s_1^2+\frac{2}{9} b_1 b_2 b_3 s_5 \left(9 s_1 s_8-2 s_5^2\right)+\frac{1}{4} s_8^2 \left(s_5^2-4 s_1 s_8\right))\,.
\,
\eea
\end{footnotesize}

We  can use Weierstrass coordinates of the rational points of the elliptic curve in $dP_2$ in \cite{Cvetic:2013nia}  to obtain the Weierstrass coordinates of the rational points \eqref{eq:coordsPQR}. This yields
\begin{footnotesize}
\bea \label{eq:WScoordsQ}
	z_Q\!\!&\!\!=\!\!&\!\!a_1 b_2-a_2 b_1\,,\nn\\
	 x_Q\!\!&\!\!=\!\!&\!\!b_1^2 b_2^2 s_3^2\!-\!b_1 b_2 \left(a_2 b_1\!+\!a_1 b_2\right) s_3 s_6\!+\!\frac{1}{12} (a_2^2 b_1^2\!+\!a_1^2 b_2^2) \left(s_6^2+8 s_3 s_8\right)\!+\!\frac{1}{6} a_1 a_2 b_1 b_2 \left(5 s_6^2\!+\!4 s_3 s_8\right)\nn\\
	 &\!\!+\!\!&\!\!\frac{1}{3} \left(a_2 b_1-a_1 b_2\right)^2 \left(\left(2 a_3 b_1 b_2-a_2  b_1b_3-a_1 b_2b_3 \right)s_2+ \left(2 a_1a_2 b_3-a_1 a_3 b_2-a_2a_3 b_1\right)s_5\right)\nn\\
	 &\!\!-\!\!&\!\!a_1 a_2 \left(a_2 b_1+a_1 b_2\right)  s_6 s_8 +a_1^2 a_2^2 s_8^2\,,\nn\\
\	 y_Q\!\!&\!\!=\!\!&\!\!-b_1^3 b_2^3 s_3^3+\frac{1}{2} \left(a_2 b_1-a_1 b_2\right){}^4 \left(a_3 b_1-a_1 b_3\right) \left(a_2 b_3-a_3 b_2\right)s_1
-\frac{1}{2} b_2^4 b_3 s_3 s_5 a_1^4-\frac{1}{2} a_3 b_2^4 s_2 s_8 a_1^4\nn\\
&+&\frac{1}{2} a_2 b_2^3 b_3 \left(s_5 s_6+s_2 s_8\right) a_1^4+\frac{1}{2} a_2 a_3 b_2^3 s_5 s_8 a_1^4-a_2^2 b_2^2 b_3 s_5 s_8 a_1^4+\frac{1}{2} a_3 b_1 b_2^4 \left(s_3 s_5+s_2 s_6\right) a_1^3\nn\\
&+&a_2 a_3 b_1 b_2^3 \left(s_2 s_8-s_5 s_6\right) a_1^3+\frac{1}{2} b_2^3 s_3 \left(b_1 b_2 b_3 s_2+s_6 s_8\right) a_1^3-\frac{1}{2} a_2^2 a_3 b_1 b_2^2 s_5 s_8a_1^3\nn\\
&+&a_2^3 \left(-s_8^3+2 b_1 b_2 b_3 s_5 s_8\right) a_1^3-\frac{1}{2} a_2^2 b_2 \left(b_1 b_2 b_3 \left(s_5 s_6+s_2 s_8\right)-3 s_6 s_8^2\right) a_1^3\nn\\
&+&\frac{1}{2} a_2 b_2^2 \left(2 b_1 b_2 b_3 \left(s_3 s_5-s_2 s_6\right)-s_8 \left(s_6^2+2 s_3 s_8\right)\right) a_1^3-a_3 b_1^2 b_2^4 s_2 s_3 a_1^2-\frac{1}{2} a_2 a_3 b_1^2 b_2^3 \left(s_3 s_5+s_2 s_6\right) a_1^2\nn\\
&-&\frac{1}{2} b_1 b_2^3 s_3 \left(s_6^2+2 s_3 s_8\right) a_1^2+\frac{1}{2} a_2 b_1 b_2^2 \left(s_6^3+5 s_3 s_8 s_6-b_1 b_2 b_3 s_2 s_3\right) a_1^2+a_2^2 a_3 b_1^2 b_2^2 \left(2 s_5 s_6-s_2 s_8\right) a_1^2\nn\\
&+&\frac{1}{2} a_2^3 a_3 b_1^2 b_2 \left(2 b_1 b_2 b_3 s_1-s_5 s_8\right) a_1^2+a_2^4 b_1^2 b_3 \left(-s_5 s_8\right) a_1^2-\frac{1}{2} a_2^3 b_1 \left(b_1 b_2 b_3 \left(s_5 s_6+s_2 s_8\right)-3 s_6 s_8^2\right) a_1^2\nn\\
&-&a_2^2 b_1 b_2 \left(b_1 b_2 b_3 \left(s_3 s_5-2 s_2 s_6\right)+s_8 \left(2 s_6^2+s_3 s_8\right)\right) a_1^2+2 a_2 a_3 b_1^3 b_2^3 s_2 s_3 a_1+\frac{3}{2} b_1^2 b_2^3 s_3^2 s_6 a_1\nn\\
&-&\frac{1}{2} a_2^2 a_3 b_1^3 b_2^2 \left(s_3 s_5+s_2 s_6\right) a_1+a_2^3 a_3 b_1^3 b_2 \left(s_2 s_8-s_5 s_6\right) a_1+\frac{1}{2} a_2^4 b_1^3 b_3 \left(s_5 s_6+s_2 s_8\right) a_1\nn\\
&-&a_2 b_1^2 b_2^2 s_3 \left(2 s_6^2+s_3 s_8\right) a_1+\frac{1}{2} a_2^4 a_3 b_1^3 \left(s_5 s_8\right) a_1+\frac{1}{2} a_2^2 b_1^2 b_2 \left(s_6^3+5 s_3 s_8 s_6-b_1 b_2 b_3 s_2 s_3\right) a_1\nn\\
&+&\frac{1}{2} a_2^3 b_1^2 \left(2 b_1 b_2 b_3 \left(s_3 s_5-s_2 s_6\right)-s_8 \left(s_6^2+2 s_3 s_8\right)\right) a_1-\frac{1}{2} a_2^4 a_3^2 b_1^5 b_2 s_1-a_2^2 a_3 b_1^4 b_2^2 s_2 s_3\nn\\
&-&\frac{1}{2} a_2^4 b_1^4 b_3 s_3 s_5+\frac{3}{2} a_2 b_1^3 b_2^2 s_3^2 s_6+\frac{1}{2} a_2^3 a_3 b_1^4 b_2 \left(s_3 s_5+s_2 s_6\right)-\frac{1}{2} a_2^4 a_3 b_1^4 s_2 s_8-\frac{1}{2} a_2^2 b_1^3 b_2 s_3 \left(s_6^2+2 s_3 s_8\right)\nn\\
&+&\frac{1}{2} a_2^3 b_1^3 s_3 \left(b_1 b_2 b_3 s_2+s_6 s_8\right)\,,
\eea
\end{footnotesize}
and, using the symmetry $(a_2,b_2)\leftrightarrow (a_3,b_3)$ of $\mathcal{E}$,
\begin{footnotesize}
\bea \label{eq:WScoordsR}
	z_R\!\!&\!\!=\!\!&\!\!a_1 b_3-a_3 b_1\,,\nn\\
	 x_R\!\!&\!\!=\!\!&\!\!b_1^2 b_3^2 s_3^2\!-\!b_1 b_3 \left(a_3 b_1\!+\!a_1 b_3\right) s_3 s_6\!+\!\frac{1}{12} (a_3^2 b_1^2\!+\!a_1^2 b_3^2) \left(s_6^2+8 s_3 s_8\right)\!+\!\frac{1}{6} a_1 a_3 b_1 b_3 \left(5 s_6^2\!+\!4 s_3 s_8\right)\nn\\
	 &\!\!+\!\!&\!\!\frac{1}{3} \left(a_3 b_1-a_1 b_3\right)^2 \left(\left(2 a_2 b_1 b_3-a_3  b_1b_2-a_1 b_2b_3 \right)s_2+ \left(2 a_1a_3 b_2-a_1 a_2 b_3-a_2a_3 b_1\right)s_5\right)\nn\\
	 &\!\!-\!\!&\!\!a_1 a_3 \left(a_3 b_1+a_1 b_3\right)  s_6 s_8 +a_1^2 a_3^2 s_8^2\,,\nn\\
\	 y_Q\!\!&\!\!=\!\!&\!\!-b_1^3 b_3^3 s_3^3+\frac{1}{2} \left(a_3 b_1-a_1 b_3\right){}^4 \left(a_2 b_1-a_1 b_2\right) \left(a_3 b_2-a_2 b_3\right)s_1
-\frac{1}{2} b_3^4 b_2 s_3 s_5 a_1^4-\frac{1}{2} a_2 b_3^4 s_2 s_8 a_1^4\nn\\
&+&\frac{1}{2} a_3 b_3^3 b_2 \left(s_5 s_6+s_2 s_8\right) a_1^4+\frac{1}{2} a_2 a_3 b_3^3 s_5 s_8 a_1^4-a_3^2 b_3^2 b_2 s_5 s_8 a_1^4+\frac{1}{2} a_2 b_1 b_3^4 \left(s_3 s_5+s_2 s_6\right) a_1^3\nn\\
&+&a_2 a_3 b_1 b_3^3 \left(s_2 s_8-s_5 s_6\right) a_1^3+\frac{1}{2} b_3^3 s_3 \left(b_1 b_2 b_3 s_2+s_6 s_8\right) a_1^3-\frac{1}{2} a_3^2 a_2 b_1 b_3^2 s_5 s_8a_1^3\nn\\
&+&a_3^3 \left(-s_8^3+2 b_1 b_2 b_3 s_5 s_8\right) a_1^3-\frac{1}{2} a_3^2 b_3 \left(b_1 b_2 b_3 \left(s_5 s_6+s_2 s_8\right)-3 s_6 s_8^2\right) a_1^3\nn\\
&+&\frac{1}{2} a_3 b_3^2 \left(2 b_1 b_2 b_3 \left(s_3 s_5-s_2 s_6\right)-s_8 \left(s_6^2+2 s_3 s_8\right)\right) a_1^3-a_2 b_1^2 b_3^4 s_2 s_3 a_1^2-\frac{1}{2} a_2 a_3 b_1^2 b_3^3 \left(s_3 s_5+s_2 s_6\right) a_1^2\nn\\
&-&\frac{1}{2} b_1 b_3^3 s_3 \left(s_6^2+2 s_3 s_8\right) a_1^2+\frac{1}{2} a_2 b_1 b_3^2 \left(s_6^3+5 s_3 s_8 s_6-b_1 b_2 b_3 s_2 s_3\right) a_1^2+a_3^2 a_2 b_1^2 b_3^2 \left(2 s_5 s_6-s_2 s_8\right) a_1^2\nn\\
&+&\frac{1}{2} a_3^3 a_3 b_1^2 b_3 \left(2 b_1 b_2 b_3 s_1-s_5 s_8\right) a_1^2+a_3^4 b_1^2 b_2 \left(-s_5 s_8\right) a_1^2-\frac{1}{2} a_3^3 b_1 \left(b_1 b_2 b_3 \left(s_5 s_6+s_2 s_8\right)-3 s_6 s_8^2\right) a_1^2\nn\\
&-&a_3^2 b_1 b_3 \left(b_1 b_2 b_3 \left(s_3 s_5-2 s_2 s_6\right)+s_8 \left(2 s_6^2+s_3 s_8\right)\right) a_1^2+2 a_2 a_3 b_1^3 b_3^3 s_2 s_3 a_1+\frac{3}{2} b_1^2 b_3^3 s_3^2 s_6 a_1\nn\\
&-&\frac{1}{2} a_3^2 a_2 b_1^3 b_3^2 \left(s_3 s_5+s_2 s_6\right) a_1+a_3^3 a_2 b_1^3 b_3 \left(s_2 s_8-s_5 s_6\right) a_1+\frac{1}{2} a_3^4 b_1^3 b_2 \left(s_5 s_6+s_2 s_8\right) a_1\nn\\
&-&a_3 b_1^2 b_2^2 s_3 \left(2 s_6^2+s_3 s_8\right) a_1+\frac{1}{2} a_3^4 a_2 b_1^3 \left(s_5 s_8\right) a_1+\frac{1}{2} a_3^2 b_1^2 b_3 \left(s_6^3+5 s_3 s_8 s_6-b_1 b_2 b_3 s_2 s_3\right) a_1\nn\\
&+&\frac{1}{2} a_3^3 b_1^2 \left(2 b_1 b_2 b_3 \left(s_3 s_5-s_2 s_6\right)-s_8 \left(s_6^2+2 s_3 s_8\right)\right) a_1-\frac{1}{2} a_3^4 a_2^2 b_1^5 b_2 s_1-a_3^2 a_2 b_1^4 b_2^2 s_2 s_3\nn\\
&-&\frac{1}{2} a_3^4 b_1^4 b_3 s_3 s_5+\frac{3}{2} a_3 b_1^3 b_3^2 s_3^2 s_6+\frac{1}{2} a_3^3 a_2 b_1^4 b_3 \left(s_3 s_5+s_2 s_6\right)-\frac{1}{2} a_3^4 a_3 b_1^4 s_2 s_8-\frac{1}{2} a_3^2 b_1^3 b_3 s_3 \left(s_6^2+2 s_3 s_8\right)\nn\\
&+&\frac{1}{2} a_3^3 b_1^3 s_3 \left(b_1 b_2 b_3 s_2+s_6 s_8\right)\,.
\eea
\end{footnotesize}

\section{Weierstrass models with $I_2$  and $I_3$ singularities}
\label{app:GeneralWSFrk2}

In this appendix, we summarize and study the general Weierstrass models of Calabi-Yau elliptic fibrations 
with $I_2$  and $I_3$ singularities (mostly the split case) at codimension one in the base $B_n$.  
These are the simplest singularities used to engineer F-theory models with gauge algebras su(2) and 
su(3). Here, we consider two models with rank two gauge algebra, namely su(2)$\oplus$su(2) and 
su(3), and one model with an su(2)$\oplus$su(2)$\oplus$su(3) gauge algebra. 
The corresponding Weierstrass models serve as reference points
for comparison with the Weierstrass models obtained  in the main text by unHiggsing
the Calabi-Yau manifolds $X_{n+1}$ with rank two Mordell-Weil group. 

As we are interested in the most generic Weierstrass forms, {\it i.e.}~those with the most complex structure 
moduli, we only consider $I_n$ singularities. Other 
realizations of rank two gauge algebras, such as Kodaira fibers of type $III$ and $IV$, can be obtained 
form these models via additional tunings as analyzed in Tate's algorithm, so that {\it e.g.}~$f$ and $g$ vanish to higher order. 

\subsection{The SU(2)$\times$SU(2) Weierstrass model}
\label{app:SU2SU2WSF}

An F-theory model with two su(2) gauge algebras on two distinct divisors $\sigma=0$, 
$\tau=0$ in $B_{n}$, which we simply denote $\sigma$ and $\tau$ by abuse of notation, is
specified by a Weierstrass model with two $I_2$ singularities. It can be constructed using Tate's 
algorithm yielding 
\beq \label{eq:WSFSU2SU2}
	y^2=x^3+\big( -\tfrac{1}{3}\tilde{a}_2^2+\tilde{a}_4 \sigma  \tau\big)xz^4+\big(\tfrac{2}{27}\tilde{a}_2^3-\tfrac{1}{3} \tilde{a}_2 \tilde{a}_4 \sigma  \tau+\tilde{a}_6 \sigma ^2 \tau ^2\big)z^6\,,
\eeq
where the coefficients $\tilde{a}_i$ are sections of the line bundles 
\beq \label{eq:sectionsSU2SU2}
	\tilde{a}_2\in \mathcal{O}(-2K_B)\,,\qquad \tilde{a}_4\in \mathcal{O}(-4K_B-[\sigma]-[\tau])\,,\qquad  \tilde{a}_6\in \mathcal{O}(-6K_B-2[\sigma]-2[\tau])\,
\eeq
on $B_{n}$ by the requirement that $f\in\mathcal{O}(-4K_B)$ and $g\in \mathcal{O}(-6K_B)$. These coefficients are expressed by the leading Tate coefficients 
$\mathfrak{a}^{(k)}_i$, $i=1,2,3,4,6$,\footnote{We expand the Tate coefficients
as $\mathfrak{a}_i=(\sigma\tau)^k \mathfrak{a}_i^{(k)}$ where $k$ is according to Tate's algorithm  $k= 0,0,1,1,2$ for an $I_2$ singularity.} in a Tate form as
\beq \label{eq:TateSU2SU2}
	 \tilde{a}_2=\tfrac14\big(4\mathfrak{a}^{(0)}_2+(\mathfrak{a}^{(0)}_1)^2\big)\,, \qquad \tilde{a}_4=\mathfrak{a}_4^{(1)}+\tfrac12\mathfrak{a}_1^{(0)}\mathfrak{a}_3^{(1)}\,,\qquad\tilde{a}_6=\mathfrak{a}_6^{(2)}
	+\tfrac14 (\mathfrak{a}_3^{(1)})^2\,.
\eeq
Any model with two $I_2$ singularities
on smooth divisors can be shown to assume this form 
\cite{Morrison:2011mb}.

We are interested in models where both SU(2)'s carry adjoints. In 6D,  the number of adjoints is 
computed by the topological genus $g_1$, $g_2$ of the curves $\sigma$, $\tau$ 
which agrees with the arithmetic genus  for smooth curves. The latter is  computed as
\beq \label{eq:arithmeticGenus}
	g_i=1+\tfrac12[\Sigma]\cdot ([\Sigma]+K_B)\,,\qquad i=1,2\,,
\eeq
with $\Sigma=\sigma$, $\tau$.
Thus, demanding $g_i\geq 1$ implies 
\beq
	[\sigma]=-K_B+Y\,,\qquad [\tau]=-K_B+Z\,,
\eeq
for effective classes $Y$, $Z$. Thus, effectiveness of the sections in \eqref{eq:sectionsSU2SU2} implies upper bounds on both $Y$ and $Z$,
\beq\label{eq:UpperBoundXY}
	Y+Z\leq -K_B\,,
\eeq
for the model to be generic.

For a concrete base $B_2=\mathbb{P}^2$ for which 
$\mathcal{O}(-K_B)=\mathcal{O}_{\mathbb{P}^2}(3)$ this condition yields the 
following list of allowed values for the degree of the divisors $\sigma$ and $\tau$, together 
with the corresponding degrees of the $\tilde{a}_i$:
\beq \label{eq:HiggsableSU2OnP2}
\begin{tabular}{|c|c||c|c|c|} \hline
$[\sigma]$ & $[\tau]$ & $\tilde{a}_2$ & $\tilde{a}_4$ & $\tilde{a}_6$  \\ \hline
  6 & 3 & 6 & 3 & 0 \\
  5 & 4 & 6 & 3 & 0  \\
  5 & 3 & 6 & 4 & 2 \\
   4 & 5 & 6 & 3 & 0 \\
   4 & 4 & 6 & 4 & 2 \\
    4 & 3 & 6 & 5 & 4\\
    3& 6 &6 & 3&0\\
    3& 5 & 6& 4&2\\
    3& 4 & 6&5&4\\
    3& 3 & 6& 6&6\\
    \hline
\end{tabular}
\eeq

The generic matter spectrum of the theory can be extracted directly from its Weierstrass model \eqref{eq:WSFSU2SU2}. By 
analyzing the orders of vanishing of $f$, $g$ and the discriminant
\beq
\Delta=4f^3+27g^2 =\sigma^2 \tau^2(4 \tilde{a}_2^3 \tilde{a}_6-\tilde{a}_2^2 \tilde{a}_4^2+\sigma \tau(4 \tilde{a}_4^3-18 \tilde{a}_2 \tilde{a}_4 \tilde{a}_6 )+27 \tilde{a}_6^2 \sigma^2 \tau^2)\,,
\eeq
we find the following  singular fibers and corresponding 
local (non-adjoint)
matter representations:
\beq \label{eq:Multies_SU2SU2}
\text{
\begin{tabular}{|c|c|c|} \hline
Representation &Multiplicity  & Fiber  \rule{0pt}{14pt} \\ \hline
$(\mathbf{2},\mathbf{1})$ &  $-2[\sigma]\cdot (4[K_B]+[\sigma]+[\tau])$ \rule{0pt}{14pt} & $I_3$\\ \hline
$(\mathbf{1},\mathbf{2})$ &  $-2[\tau]\cdot (4[K_B]+[\sigma]+[\tau])$ \rule{0pt}{14pt} & $I_3$\\ \hline
$(\mathbf{2},\mathbf{2})$ & $[\sigma]\cdot  [\tau]$ \rule{0pt}{14pt} & $I_4$ \\ \hline
\end{tabular}
}
\eeq
We note that there are additional singular fibers of Kodaira type $III$ at the codimension two loci 
$\tilde{a}_2=0$ and $\sigma=0$ or $\tau=0$, respectively. As the number of fiber components does not 
increase at these loci, there is no matter supported at these loci.

We readily check using the data in \eqref{eq:arithmeticGenus}, \eqref{eq:Multies_SU2SU2} and the anomaly coefficients $[\sigma]$, $[\tau]$ for the two SU(2) groups that all 6D anomalies are canceled.

\subsection{The $I_3$ Weierstrass model}
\label{app:SU3WSF}

The canonical ansatz for an F-theory model with su(3) gauge algebra is a Weierstrass model with an 
$I_3$-singularity at codimension one.
It is well-know that one has to distinguish between the split, $I_3^s$, and non-split, $I_3^{ns}$, 
case. Given the Weierstrass form of an elliptic fibration
with $I_3$-singularity at $t=0$ in the base $B_{n}$, the split condition is expressed as the condition that the following monodromy
cover is reducible, {\it i.e.}~has two distinct rational roots \cite{Grassi:2011hq}:
\beq 
	\Psi^2+\left.\frac{9g}{2f}\right\vert_{t=0}=0\,.
\eeq
This is the case, if and only if $\left.\frac{9g}{2f}\right\vert_{t=0}$ is a perfect square.  As shown in
Section \ref{sec:NewWSFs} it is sufficient if this condition holds locally around $t=0$ requiring, however, a 
novel structure of the Weierstrass model.

The Weierstrass form in the non-split case 
is discussed in \cite{Katz:2011qp}. In this case the gauge algebra is generically 
$\text{sp}(1)=\text{su}(2)$ due to the monodromy action on the fiber around the branch points of the 
monodromy cover \eqref{eq:I3monoCover}. Thus, we disregard the non-split case in the following.
 
In the split case the gauge algebra is $\text{su}(3)$ and the Weierstrass model assumes,  
if $\frac{9g}{2f}\vert_{t=0}$ is \textit{globally} a square, the standard 
form  that follows from Tate's algorithm \cite{Katz:2011qp}:
\beq \label{eq:WSFSU3}
	y^2=x^3+(-\tfrac{1}{48}\tilde{a}_1^4+\tfrac12 \tilde{a}_1 \tilde{a}_3t+\tilde{a}_4 t^2)xz^4+(\tfrac1{864}\tilde{a}_1^6-\tfrac{1}{24} \tilde{a}_3 \tilde{a}_1^3 t+\tfrac{1}{12}  \left(3 \tilde{a}_3^2-\tilde{a}_1^2 \tilde{a}_4\right)t^2+\tilde{a}_6 t^3)z^6\,.
\eeq
This is the form that is appropriate when $t$ is a smooth divisor,
which supports adjoint but not symmetric representations of SU(3).
Denoting the divisor $t=0$ supporting the $\text{su}(3)$ by $t$ by abuse of notation, we find 
that the coefficients 
$\tilde{a}_i$ have to be sections of the following line bundles on $B_{n}$:
\beq \label{eq:sectionsSU3}
	\tilde{a}_1\in \mathcal{O}(-K_B)\,,\quad \tilde{a}_3\in \mathcal{O}(-3K_B-[t])\,,\quad  
	\tilde{a}_4\in 
	\mathcal{O}(-4K_B-2[t])\,,\quad
	\tilde{a}_6\in\mathcal{O}(-6K_B-3[t])\,.
\eeq
In terms of the leading Tate coefficients $\mathfrak{a}_i^{(k)}$ in a Tate model\footnote{Again 
we expand the Tate coefficients
as $\mathfrak{a}_i=(t)^k \mathfrak{a}_i^{(k)}$ where $k$ is according to Tate's algorithm  $k= 0,1,1,2,3$ for an $I_3^{\text{s}}$-singularity.}, they can 
be expressed as
\beq
\label{eq:TateSU3}	
 	 \tilde{a}_1=\mathfrak{a}_1^{(0)}\,,\quad   \tilde{a}_3=
 	 \tfrac13(3\mathfrak{a}_3^{(1)}-\mathfrak{a}_1^{(0)}\mathfrak{a}_2^{(1)})\,,\quad \tilde{a}_4=\mathfrak{a}_4^{(2)}-\tfrac13(\mathfrak{a}_2^{(1)})^2\,,\quad \tilde{a}_6=\mathfrak{a}_6^{(3)}-\tfrac13 \mathfrak{a}_2^{(1)}\mathfrak{a}_4^{(2)}+\tfrac2{27}(\mathfrak{a}_2^{(1)})^3\,.
\eeq

In 6D, the genus  of the curve $t$ counts the number of matter fields in the adjoint representation
of the su(3). As before, we have at least one adjoint if 
\beq
	[t]=-K_B+Z\,,
\eeq
for an effective divisor $Z$. Effectiveness of all coefficients in \eqref{eq:sectionsSU3} imposes an upper 
bound on $Z$ of the form
\beq\label{eq:UpperBoundZ}
	Z\leq -K_B\,,
\eeq
so that the model \eqref{eq:WSFSU3} is generic.

For the concrete base $B=\mathbb{P}^2$ this condition is solved for the following choices of the 
degree of $[t]$, implying corresponding classes for the coefficients $\tilde{a}_i$:
\beq \label{eq:HiggsableSU3onP2}
\begin{tabular}{|c||c|c|c|c|} \hline
$[t]$ & $\tilde{a}_1$ & $\tilde{a}_3$ & $\tilde{a}_4$ & $\tilde{a}_6$  \\ \hline
  6 & 3 & 3 & 0 & 0 \\
  5 & 3 & 4 & 2 & 3  \\
  4 & 3 & 5 & 4 & 6 \\
   3 & 3 & 6 & 6 & 9 \\
    \hline
\end{tabular}
\eeq

As before, we can extract the generic matter spectrum of the theory directly from its Weierstrass 
model \eqref{eq:WSFSU3}. By analyzing the orders of vanishing of $f$, $g$ and the discriminant
\beq
\Delta=4f^3+27g^2 =t^3(\tfrac1{16} \tilde{a}_1^3 (\tilde{a}_1^3 \tilde{a}_6-\tilde{a}_3^3-\tilde{a}_1^2 \tilde{a}_3 \tilde{a}_4)+\mathcal{O}(t))\,,
\eeq
we find the following  singular fibers and corresponding
local 
matter representations:
\beq \label{eq:Multies_SU3}
\text{
\begin{tabular}{|c|c|c|} \hline
Representation &Multiplicity  & Fiber  \rule{0pt}{14pt} \\ \hline
$\mathbf{3}$ &  $-3[t]\cdot (3K_B+[t])$ \rule{0pt}{14pt} & $I_4$\\ \hline
\end{tabular}
}
\eeq
There are additional singular fibers of Kodaira type $IV$ at the codimension two loci $t=\tilde{a}_1=0$. 
As the number of fiber components does not increase, there is no matter supported at these loci.

We readily check using the data in \eqref{eq:Multies_SU3} and the anomaly coefficient $[t]$ that all 6D 
anomalies are canceled.

\subsection{The SU(2)$\times$SU(2)$\times$SU(3) Weierstrass model}
\label{app:SU2SU2SU3WSF}

Finally, we combine the previous two subsections to construct the Weierstrass model of
an elliptic fibration with two $I_2$  and one $I_3^s$ singularity.  The Weierstrass form we obtain
agrees precisely with \eqref{eq:WSFSU2SU2SU3special} of the unHiggsed specialized model 
$X_{n+1}$.

All we have to do is to combine the forms \eqref{eq:WSFSU2SU2} and \eqref{eq:WSFSU3}, where the 
appropriate way to specialize coefficients is determined by Tate's algorithm. For example, we can start 
with the Weierstrass form  
\eqref{eq:WSFSU2SU2}  in the parametrization \eqref{eq:TateSU2SU2} for two $I_2$ singularities and 
impose the orders of vanishing  $(0,1,1,2,3)$ of the Tate coefficients $\mathfrak{a}_i$ on the SU(3) divisor $t=0$ for an $I_3^s$ singularity \cite{tate1975algorithm}. We expand 
the Tate coefficients as $\mathfrak{a}_i=\mathfrak{a}_i^{(k,l)}t^k(\sigma\tau)^l$.
Suppressing the superscripts on the $\mathfrak{a}_i^{(k,l)}$ to unclutter our notation  
we thus obtain the Weierstrass form 
appropriate for smooth $\sigma, \tau, t$
\bea \label{eq:WSFSU2SU2SU3gen}
	y^2&=&x^3+\big(-\tfrac{1}{48}(4\mathfrak{a}_2 t+\mathfrak{a}_1^2)^2+(\mathfrak{a}_4t+\tfrac12\mathfrak{a}_1\mathfrak{a}_3) \sigma  \tau t\big)xz^4\\&+&\big(\tfrac{1}{864}(4\mathfrak{a}_2t+\mathfrak{a}_1^2)^3-\tfrac{1}{12} (4\mathfrak{a}_2 t+\mathfrak{a}_1^2) (\mathfrak{a}_4t+\tfrac12\mathfrak{a}_1\mathfrak{a}_3) \sigma  \tau t+(\mathfrak{a}_6t
	+\tfrac14 \mathfrak{a}_3^2) \sigma ^2 \tau ^2t^2\big)z^6\,,\nn
\eea
where the leading Tate coefficients $\mathfrak{a}_i$  are sections of the line bundles
\bea \label{eq:sectionsSU2SU2SU3}
	&\mathfrak{a}_1\in \mathcal{O}(-K_B)\,,\qquad \mathfrak{a}_2\in \mathcal{O}(-2K_B-[t])\,,\qquad
	\mathfrak{a}_3\in \mathcal{O}(-3K_B-[\sigma]-[\tau]-[t])\,,&\nn\\ 
	&\mathfrak{a}_4\in \mathcal{O}(-4K_B-[\sigma]-[\tau]-2[t])\,,\qquad  \mathfrak{a}_6\in \mathcal{O}(-6K_B-2[\sigma]-2[\tau]-3[t])\,.&
\eea

Since we have analyzed the case of SU(2)$\times$SU(2) or SU(3) gauge groups already in the previous 
two subsections, we focus here only on the cases where $[t]$ is a non-trivial class and not both
$[\sigma]$ and $[\tau]$ are trivial simultaneously. 
(We thus include not only models with gauge group $SU(2) \times SU(2)
\times SU(3)$, but also with $SU(2) \times SU(3)$).
For the concrete base $B_2=\mathbb{P}^2$ for which 
$\mathcal{O}(-K_B)=\mathcal{O}_{\mathbb{P}^2}(3)$ the effectiveness conditions implied by 
\eqref{eq:sectionsSU2SU2SU3} yield the following list of allowed values for the degrees of the divisors $[\sigma]$, $[\tau]$ and $[t]$:
\beq \label{eq:SU2SUSU3OnP2}
\begin{array}{|c||c|c|c|c|c|c|c|c|c|c|c|c|c|c|c|}
\hline
 [\sigma] &1  & 1  & 2         & 2   & 3   & 3    & 3      & 4  &   4  & 4 & 5  & 5   & 6  & 6 & 7 \\
 \hline
[\tau] &0   & 1 & \leq 1  & 2   & 0   & 1  & \leq 3  & 0  &  \leq 2 & 3 &  \leq 1  & 2 & 0  & 1 & 0 \\\hline
[t] &\leq 5 &  \leq 4 & \leq 4 & \leq 3 & \leq 4 & \leq 3 & \leq 2 & \leq 3 & \leq 2 & 1 &  \leq 2 & 1 & \leq 2 & 1 & 1 \\
\hline
\end{array}
\eeq
Here we indicate by $\leq k$ that all integers less or equal to $k$ are allowed. In addition, we 
obtain valid models by exchanging the degrees of $[\sigma]$ and $[\tau]$.

The generic
local matter spectrum of the theory can be extracted directly from its Weierstrass model 
\eqref{eq:WSFSU2SU2SU3}. By 
analyzing the orders of vanishing of $f$, $g$ and the discriminant,
we find the following  singular fibers and corresponding 
matter representations: 
\beq \label{eq:Multies_SU2SU2SU3}
\text{
\begin{tabular}{|c|c|c|} \hline
Representation &Multiplicity    \rule{0pt}{14pt} \\ \hline
$(\mathbf{2},\mathbf{2},\mathbf{1})$ &  $[\sigma]\cdot[\tau]$ \rule{0pt}{14pt} \\ \hline
$(\mathbf{2},\mathbf{1},\mathbf{3})$ &  $[\sigma]\cdot [t]$ \rule{0pt}{14pt} \\ \hline
$(\mathbf{1},\mathbf{2},\mathbf{3})$ & $[\tau]\cdot [t]$ \rule{0pt}{14pt} \\ \hline
$(\mathbf{2},\mathbf{1},\mathbf{1})$ & $[\sigma]\cdot  (-8K_B-2[\sigma]-2[\tau]-3[t])$  \rule{0pt}{14pt} \\  \hline
$(\mathbf{1},\mathbf{2},\mathbf{1})$ & $[\tau]\cdot  (-8K_B-2[\sigma]-2[\tau]-3[t]) $ \rule{0pt}{14pt} \\ \hline
$(\mathbf{1},\mathbf{1},\mathbf{3})$ & $[t]\cdot (-9K_B-2[\sigma]-2[\tau]-3[t])$ \rule{0pt}{14pt} \\ \hline
\end{tabular}
}
\eeq
We note that there are additional singular fibers of Kodaira type $III$ and $IV$ at the codimension two 
loci  $\sigma=(\mathfrak{a}_1^2 + 4 \mathfrak{a}_2 t)=0$ or 
$\tau=(\mathfrak{a}_1^2 + 4 \mathfrak{a}_2 t)=0$ and
$\mathfrak{a}_1=t=0$ , respectively, that do not  is support additional matter.

We readily check using the data in \eqref{eq:arithmeticGenus}, \eqref{eq:Multies_SU2SU2SU3} and the anomaly coefficients that all 6D anomalies are canceled.

\section{Map of $X_{n+1}$ to the quartic in $\text{Bl}_1\mathbb{P}^2(1,1,2)$}
\label{app:quarticinP112}

In this appendix, we argue that the unHiggsing of the F-theory model defined by $X_{n+1}$ can be interpreted as an
unHiggsing of the two  U(1)'s to two SU(2)'s on reducible divisors $X=AC$ and $Y=BC$. 
To this end, we use the natural presentation of \cite{Morrison:2012ei} for a model with one U(1) as  quartic in 
$\text{Bl}_1\mathbb{P}^2(1,1,2)$, which will be non-generic 
with an additional non-toric rational section in the case of $X_{n+1}$.
Most notably, due to this non-generic form the singularity on the 
common component $C$ enhances just to $I_3$, corresponding to an SU(3) gauge group. This is in contrast to the 
expected enhancement of two $I_2$ singularities to an $I_4$.  

It was argued in \cite{Morrison:2012ei} that every model with at least one 
rational section can be brought into the form of the quartic in 
$\text{Bl}_1\mathbb{P}^2(1,1,2)$. In particular, this implies that there exists a birational map 
of the model \eqref{eq:cubicfactorized} to the quartic. In the following, as an illustration 
we construct this birational map for the specialized model \eqref{eq:abcdModel}.

We recall from \cite{Morrison:2012ei} that the quartic elliptic curve in $\text{Bl}_1\mathbb{P}^2(1,1,2)$ with homogeneous coordinates $[U:V:W]$ (we set the coordinate of the exceptional divisor to one)
reads
\beq \label{eq:quartic}
 W(W+bV^2) = U(e_0 U^3+e_1 U^2 V + e_2 U V^2 + e_3 V^3)\,.
\eeq
It has a rational point $\tilde{Q}$ at $[U:V:W]=[0:1:-b]$ and its zero point is at $[0:1:0]$.

There are two possible presentations of the specialized model
as the quartic \eqref{eq:quartic}, 
corresponding to the two possible choices which rational 
point in \eqref{eq:abcdModel} is mapped to the rational point $\tilde{Q}$ in 
$\text{Bl}_1\mathbb{P}^2(1,1,2)$.  Here, we choose to map
$Q$ in  \eqref{eq:abcdModel} to $\tilde{Q}$, noting that the quartic for the choice of 
$R\mapsto\tilde{Q}$ is obtained by exchanging $(a_2,b_2)\leftrightarrow(a_3,b_3)$.

One way to find the quartic is to construct sections in the bundles $\mathcal{O}(k(P+Q))$ for 
$k=1,2,3,4$. (This requires moving the point $R$ out of the plane $u=0$, which can be achieved
by a variable transformation on \eqref{eq:abcdModel}.) Here, we construct the map indirectly
by matching Weierstrass models. The Weierstrass model for the specialized model is given
in Appendix \ref{app:WSF} for $a_1=1$, $b_1=0$. It has to agree with the Weierstrass form
\eqref{eq:Abelian-1} upon appropriate identifications of the coefficients $b$ and $e_i$, 
$i=0,1,2,3$, with the coefficients in \eqref{eq:abcdModel}. Clearly, the system of equations
resulting from this is under-determined. We have to specify in addition that the rational point 
$Q$ maps to the rational point $\tilde{Q}$ in \eqref{eq:quartic}. We recall that the Weierstrasss
coordinates of $\tilde{Q}$ read \cite{Morrison:2012ei}
\beq
	[x:y:z]=[e_3^2-\tfrac23 b^2 e_2:- e_3^3+ b^2 e_2 e_3 -\tfrac12b^4 e_1:b ]\,.
\eeq
Comparing to the Weierstrass coordinates \eqref{eq:WScoordsQ} for $Q$ with $a_1=1$ and 
$b_1=0$ and demanding equality of the Weierstrass models, we obtain
\bea \label{eq:quarticcoeffs}
b&=&b_2\,,\nn\\
e_0 &=& \tfrac{1}{4} (b_3^2 (s_2^2-4 s_1 s_3)-2 a_3 b_3 (s_2 s_5-2 s_1 s_6)+a_3^2 (s_5^2-4 s_1 s_8)), \nn \\
   e_1 &=& \tfrac{1}{2} b_3 (2 s_3 s_5-s_2 s_6+2 a_2 b_3 s_1)+a_3 (s_2 s_8-b_2 b_3 s_1-\tfrac12 s_5 s_6)\,, \nn \\
   e_2 &=& \tfrac{1}{4}(s_6^2-4 s_3 s_8+2 b_2 (a_3 s_5+b_3 s_2)-4 a_2 b_3 s_5)\,, \nn \\
   e_3 &=& a_2 s_8-\tfrac12b_2 s_6\,. 
\eea 
The coordinates of the rational point $R$ in the quartic \eqref{eq:quartic} are given  by
\bea \label{eq:quarticR}
	[U:V:W]\!\!&\!\!=\!\!&\!\![b_3 (a_2 b_3-a_3 b_2):a_3^2 s_8-a_3 b_3 s_6+b_3^2 s_3:-\tfrac{1}{2} b_3 (b_3^3 \left(a_2^2 b_3 s_2-a_2 s_3 s_6+2 b_2 s_3^2\right)\nn\\
	&+\!\!&\!\!\left.a_3^2 b_3 \left(b_2 \left(s_6^2+2 s_3 s_8+2 a_2 b_3 s_5\right)\!-\!3 a_2 s_6 s_8+b_3 b_2^2 s_2\right)\!-\!a_3 b_3^2 \left(a_2^2 b_3 s_5+3 b_2 s_3 s_6\right.\right.\nn\\
	&-\!\!&\!\!\left.\left.a_2 \left(s_6^2+2 s_3 s_8-2 b_2 b_3 s_2\right)\right)-a_3^3 \left(b_3 b_2^2 s_5+b_2 s_6 s_8-2 a_2 s_8^2\right)\right)]\,.
\eea
We note that it is a non-toric rational point.

The coefficient $e_3$ is the locus of SU(2) enhancement if the section U(1) corresponding to 
the section $\hat{s}_{\tilde{Q}}$ is unHigssed. This unHiggsing is triggered by 
$b_2\rightarrow 0$ in which case we obtain $e_3= a_2 s_8$ according to 
\eqref{eq:quarticcoeffs}. Thus after unHiggsing we obtain two SU(2)'s on the two components 
$a_2=0$ and $s_8=0$. In contrast, if we set $b_3\rightarrow 0$ we unHiggs  the U(1) 
associated to the non-toric section $\hat{s}_R$ as its coordinates in
\eqref{eq:quarticR} then coincide
with those of the zero point $[0:1:0]$. Furthermore, we 
observe from \eqref{eq:quarticcoeffs} that $e_0\sim a_3^2$, $e_1\sim a_3$, inducing an 
$I_2$ singularity at $a_3=0$. In addition, we obtain an $I_2$ at $s_8=0$ as can be seen from 
the discriminant. Thus we can interpret the tuning $b_3\rightarrow 0$ again as an unHiggsing to an
SU(2) on a reducible divisor $e'_3=a_3s_8$. In fact, $e'_3$ is the coefficient in the second \eqref{eq:quartic} 
we obtain by mapping $R\mapsto \tilde{Q}$.

We emphasize that in the simultaneous tuning $b_2,\, b_3\rightarrow 0$ the $I_2$ singularity on
$s_8$ only enhances to an $I_3$. Thus, we confirm 
the picture of an SU(2) on two divisor $X=AC$, $Y=BC$ with $A=a_2$, $B=a_3$ and $C=s_8$ in the unHiggsed theory, as claimed.

\bibliographystyle{utphys}	
\bibliography{ref}

\end{document}